\documentclass{aa}  
\usepackage{natbib}
\usepackage{graphicx}
\usepackage{txfonts}
\usepackage{xcolor}
\begin{document} 

   \title{LLAMA: The $M_{BH}$ - $\sigma_{\star}$ Relation of the most luminous local AGNs}

   \author{Turgay Caglar\inst{1}\fnmsep\thanks{caglar@strw.leidenuniv.nl}
            \and
            Leonard Burtscher\inst{1}
            \and
            Bernhard Brandl\inst{1}
            \and
            Jarle Brinchmann\inst{1,2}
            \and
            Richard I. Davies\inst{3}
            \and
            Erin K. S. Hicks\inst{4}
            \and
            Michael Koss\inst{5}
            \and
            Ming-Yi Lin\inst{6}
            \and
            Witold Maciejewski\inst{7}
            \and
            Francisco M\"{u}ller-S\'{a}nchez\inst{8}
            \and
            Rogemar A. Riffel\inst{9}
            \and
            Rog\'{e}rio Riffel\inst{10}
            \and
            David J. Rosario\inst{11}
            \and
            Marc Schartmann\inst{3,12}
            \and
            Allan Schnorr-M\"{u}ller\inst{10}
            \and
            T. Taro Shimizu\inst{3}
            \and
            Thaisa Storchi-Bergmann\inst{10,13}
            \and
            Sylvain Veilleux\inst{14,15}
            \and
            Gilles O. de Xivry\inst{16}
            \and
            Vardha N. Bennert\inst{17}
            }

    \institute{Leiden Observatory, PO Box 9513, 2300 RA, Leiden,  \\ 
                the Netherlands \email{caglar@strw.leidenuniv.nl}
                \and  Instituto de Astrofísica e Ciências do Espaço, Universidade do Porto, CAUP, Rua das Estrelas, 4150-762 Porto, Portugal
                \and Max-Planck-Institut für extraterrestrische Physik, Postfach 1312, D-85741, Garching, Germany 
                \and Department of Physics \& Astronomy, University of Alaska Anchorage, AK 99508-4664, USA
                \and Eureka Scientific Inc, Oakland, CA, USA
                \and Institute of Astronomy and Astrophysics, Academia Sinica, 11F of AS/NTU Astronomy-Mathematics Building, No.1, Sec. 4, Roosevelt Rd, Taipei 10617, Taiwan
                \and Astrophysics Research Institute, Liverpool John Moores University, IC2 Liverpool Science Park, 146 Brownlow Hill, L3 5RF, UK
                \and Center for Astrophysics and Space Astronomy, University of Colorado, Boulder, CO 80309-0389, USA
                \and Universidade Federal de Santa Maria, Departamento de F\'{i}sica/CCNE, 97105-900, Santa Maria, RS, Brazil
                \and Departamento de Astronomia, Universidade Federal do Rio Grande do Sul, IF, CP 15051, 91501-970 Porto Alegre, RS, Brazil 
                \and Centre for Extragalactic Astronomy, Department of Physics, Durham University, South Road, Durham DH1 3LE, UK
                \and Universit\"{a}ts-Sternwarte München, Scheinerstraße 1, 81679 München, Germany
                \and Harvard-Smithsonian Center for Astrophysics, 60 Garden St., Cambridge, MA 02138, USA
                \and Department of Astronomy and Joint Space-Science Institute, University of Maryland, College Park, Maryland 20742 USA
                \and Institute of Astronomy and Kavli Institute for Cosmology Cambridge, University of Cambridge, Cambridge CB3 0HA, United Kingdom
                \and Space Sciences, Technologies, and Astrophysics Research Institute, Universit\'{e} de Li\'{e}ge, 4000 Sart Tilman, Belgium
                \and Department of Physics, California Polytechnic State University, San Luis Obispo, CA 93407, USA
            }
\authorrunning{Caglar et al. 2018}

   \date{}

 
  \abstract
   {The $M_{BH}$ - $\sigma_{\star}$ relation is considered a result of co-evolution between the host galaxies and their super-massive black holes. For elliptical bulge hosting inactive galaxies, this relation is well established, but there is still discussion whether active galaxies follow the same relation.}
   {In this paper, we estimate black hole masses for a sample of 19 local luminous AGNs (LLAMA) in order to test their location on the $M_{BH}$ - $\sigma_{\star}$ relation. In addition, we test how robustly we can determine the stellar velocity dispersion in the presence of an AGN continuum, AGN emission lines and as a function of signal/noise ratio.}
   {Super-massive black hole masses ($M_{BH}$) were derived from the broad-line based relations for H$\alpha$, H$\beta$ and Pa$\beta$ emission line profiles for the Type 1 AGNs. We compare the bulge stellar velocity dispersion ($\sigma_{\star}$) as determined from the Ca II triplet (CaT) with the dispersion measured from the near-infrared CO (2-0) absorption features for each AGN and find them to be consistent with each other. We apply an extinction correction to the observed broad line fluxes and we correct the stellar velocity dispersion by an average rotation contribution as determined from spatially resolved stellar kinematic maps.}
   {The H$\alpha$-based black hole masses of our sample of AGNs were estimated in the range 6.34 $\leq$ $\log{M_{BH}}$ $\leq$ 7.75 M$_\odot$ and the $\sigma_{\star CaT}$ estimates range between 73 $\leq$ $\sigma_{\star CaT}$ $\leq$ 227 km s$^{-1}$. From the so-constructed $M_{BH}$ - $\sigma_{\star}$ relation for our Type 1 AGNs, we estimate the black hole masses for the Type 2 AGNs and the inactive galaxies in our sample.}
   {In conclusion, we find that our sample of local luminous AGNs is consistent with the $M_{BH}$ - $\sigma_{\star}$ relation of lower luminosity AGNs and inactive galaxies, after correcting for dust extinction and the rotational contribution to the stellar velocity dispersion.}


   \maketitle
%
\section{INTRODUCTION}
Theoretical and observational evidence in the last decade has shown that super-massive black holes (SMBHs) reside in the majority of galaxy nuclei and play a substantial role in the evolution of galaxies. \citet{Lynden} recognized that SMBHs primarily grow via mass accretion, during which an extreme amount of energy is released. Nowadays, it is widely accepted that Active Galactic Nuclei (AGNs) are powered by mass accretion onto SMBHs via the conversion of gravitational energy into radiation through accretion disks \citep[e.g.,][and references therein]{Padovani}. The feeding of SMBHs begins with materials accretion at extragalactic scales, which subsequently passes through galactic and nuclear scales to the broad-line region (BLR) and accretion disk before falling into the black hole or being ejected by jets or winds \citep{TSB}. The materials in the host galaxy residing near the nucleus can be ionized by radiation  \citep[e.g.,][]{Davidson,Netzer1985}. Spectral studies have confirmed the existence of two distinct regions of excited gas clouds near the nucleus, referred as the broad-line region and the narrow-line region ( NLR). The BLR gas resides at sub-parsec scales, whereas NLR gas can be found up to a few kpc from the central black hole (BH) \citep{Netzer}. Studying the characteristics of these gas clouds is crucial for understanding AGN emission lines. 

Detailed investigations of the BLR became possible in the last few decades due to large dedicated observing campaigns \citep[e.g.,][]{BLMC,Peterson1993,ONPE2002,Denney2006,Bentz2006,Bentz2009,Denney2010,Grier2012,Grier2013,Bentz2016}. They have allowed the interaction between the SMBH and surrounding gas clouds to be characterised in detail. Under virial equilibrium, it is possible to use the BLR gas as an estimator for SMBH mass using the line widths of rotation-broadened emission lines. Even though virial black hole masses ($M_{BH}$) are roughly consistent with masses derived from other methods \citep[e.g.,][]{Peterson2004,Peterson2007}, there are a few complications, namely the structure, kinematics, and orientation of the BLR. To obtain accurate black hole masses, it is fundamental to know these BLR properties. Application of the virial theorem allows one to use the emission line width of the BLR gas as a tracer of BLR rotational velocity. While the radius of the BLR is inferred from Reverberation Mapping (RM), other efforts to resolve the structure, kinematics and orientation of the BLR have been limited so far \citep{Pancoast,Grier2017}, but new instrumentation developments have allowed recent progress to directly resolve the BLR \citep{GRAVITY2}. Correspondingly, these parameters have been used for estimating black hole masses of AGNs.

A growing body of evidence suggests a tight connection between the evolution and formation of SMBHs and host galaxies \citep[e.g.,][]{FeMe, Tremaine, MeFe, Geb, FeFo, Gultekin, Beifiori, McMa, KoHo}. This tight connection suggests that host galaxy properties, such as stellar velocity dispersion and/or bulge mass, can be used a proxy for black hole mass. The observational present-day black hole mass-galaxy comparisons, i.e. black hole mass - stellar velocity dispersion ($M_{BH}$ - $\sigma_{\star}$), show a very strong correlations for inactive galaxies, which are hosting elliptical bulges \citep[e.g.,][hereafter MM13, KH13, respectively]{McMa, KoHo}. This tight relation is usually attributed as evidence that feedback mechanisms must be responsible for linking the growth of galaxy bulges to accretion, although the exact feedback mechanism is still under debate. Using the observational data, the $M_{BH}$ - $\sigma_{\star}$ relation has been parameterized as a power-law function with index $\alpha$ ($M_{BH}$ $\propto$ $\sigma^{\alpha}$), where $\alpha$ was found to be between 3 and 6. From a theoretical concept, the difference between the power-law index is attributable to different feedback models: momentum-driven or energy-driven winds, which expects an $\alpha$ = 4 \citep{King} and $\alpha$ = 5 \citep{SiRe} relation, respectively. In these models, shocked shells of matter are driven outwards by winds; correspondingly, the galaxy bulges grow via the central star-formation. In both models, AGN accretion must approach the Eddington limit in order to form winds that can blow gas out of the host galaxy. In case of major mergers, a larger amount of gas can be driven onto the SMBH, and fuelling of black holes can lead to a coupled BH-bulge growth. But, co-evolution can occur relatively slow in the case of secular evolution, which results in the formation of pseudo-bulges. Even though the $M_{BH}$ - $\sigma_{\star}$ correlation is very tight for the galaxies hosting elliptical bulges, galaxies with pseudo-bulges are reported to lie below the $M_{BH}$ - $\sigma_{\star}$ relation \citep[e.g.,][]{Greene2010,Korm2011,KoHo}.

The assumption that AGNs and inactive galaxies follow the same $M_{BH}$ - $\sigma_{\star}$ relation is still under debate. In previous studies, \citet{Nelson}, \citet{Onken} and \citet{YuLu} investigated the $M_{BH}$ - $\sigma_{\star}$ relation of AGNs; unfortunately, their measurements suffered from low-quality data and an unreliable $M_{BH}$ - $\sigma_{\star}$ relation for inactive galaxies. Afterwards, \citet{GreeneHo2006a} found an intrinsic scatter of 0.61 dex from the $M_{BH}$ - $\sigma_{\star}$ relation for local AGNs using the RM and single-epoch black hole masses. Accordingly, \citet{Woo2010,Woo2013,Woo2015,Graham,Park} and \citet{Batiste} reported shallower $M_{BH}$ - $\sigma_{\star}$ relations for reverberation-mapped AGNs. But, the resulting discrepancy between active and inactive galaxies was assumed to be related to unreliable $\sigma_{\star}$ calculations of AGNs and/or the lack of AGNs in the high SMBH mass regime. Unfortunately, the number of high SMBH masses ($M_{BH}$ $>$ 10$^{8}$ M$_{\odot}$) from reverberation-mapped AGNs was too low to make a direct comparison with the inactive sample. To increase the number of the AGNs, other studies concentrated on single-epoch SMBH mass estimations, but a few large offsets ($>$ 0.5 dex) from the inactive $M_{BH}$ - $\sigma_{\star}$ relation were also reported from the single-epoch based investigations \citep{Bartho,GreeneHo2006a,Shen,Subramanian,Koss}. Thus, the intrinsic scatter from inactive $M_{BH}$ - $\sigma_{\star}$ relation remains highly uncertain for AGNs.

To calibrate the $M_{BH}$ - $\sigma_{\star}$ scaling relation, black hole masses are mostly determined by modelling stellar kinematics or spatially resolving gas for galaxies in the local universe. On the other hand, black hole masses are determined via RM or megamaser disks for AGNs. In RM-based estimations, a dimensionless scale factor $f$ is required to convert the virial product into $M_{BH}$s, and it is estimated assuming an average multiplicative offset from the $M_{BH}$ - $\sigma_{\star}$ relation for AGN-hosting galaxies \citep{Onken}. Although the $M_{BH}$ - $\sigma_{\star}$ relation appears to be tight, the slope of the relation remains uncertain (i.e. the slope of both AGN and/or inactive samples). Previous studies reported significantly different slopes of the $M_{BH}$ - $\sigma_{\star}$ relation for AGNs with respect to the $M_{BH}$ - $\sigma_{\star}$ relation for inactive galaxies \citep{Woo2010,Woo2013,Woo2015,Graham,Park,Vanden,Shankar2016,Shankar2019,Batiste}. However, these authors noted that the discrepancy between AGNs and inactive galaxies may be due to sample selection bias. 

In order to overcome selection biases in the studies of local AGNs, the Local Luminous AGNs with Matched Analogues (LLAMA) sample was created \citep{Davies}. The AGNs in this sample are selected in the ultra-hard X-rays, avoiding issues with obscuration for all but the most Compton-thick galaxies. As the name implies it comes with a sample of (stellar mass, distance, inclination, Hubble type) matched inactive galaxies to be able to compare galaxy properties among AGNs and similar inactive host galaxies. Over the last five years, this sample has been observed with VLT/X-SHOOTER, VLT/SINFONI, APEX and HST, and more observations are planned or proposed. These observations have so far been used to study the environmental dependence of AGN activity \citep{davies2017}, nuclear stellar kinematics \citep{Lin}, the gas content and star formation efficiencies \citep{rosario2018} as well as the nuclear star formation histories (Burtscher et al., in prep). In addition several single-object studies have been performed with this rich data set, e.g. on NGC 2110 \citep{rosario2019} and on NGC 5728 \citep{shimizu2019}.

In this paper, we present stellar velocity dispersions ($\sigma_{\star}$) calculated from the Ca II triplet (CaT) and the CO (2-0) absorption features and the broad-line based single-epoch black hole mass estimates for the hard X-ray selected Local Luminous AGN with Matched Analogues sample using the available X-SHOOTER and SINFONI data. We present a comparison of our results with the $M_{BH}$ - $\sigma_{\star}$ plane. We aim to understand the physical properties of the LLAMA sample of AGNs, and we also aim to test the robustness of the parameters which are used for the AGN $M_{BH}$ - $\sigma_{\star}$ relation. The paper is organized as follows: Section 2 reviews sample selection, observation and data reduction processes. Section 3 describes our estimation methods and the tests we performed for studying the robustness of $M_{BH}$ - $\sigma_{\star}$ parameters. In Section 4, we discuss our results. Finally, we conclude the paper in Section 5.


\begin{table*}
\caption{Galaxy properties, X-SHOOTER and SINFONI observation lists of our sample of galaxies. Sector 1 (top): the LLAMA AGNs, Sector 2 (bottom): The LLAMA Inactive Galaxies. 1) Object Name, 2) Distance, 3) Galaxy Morphology, 4) (a) Logarithmic X-ray luminosity, (b) integrated H-band luminosity in logarithm in solar unit, 5) X-SHOOTER observation date, 6) Airmass during the observation, 7) Seeing, 8) SINFONI observation date, 9) Airmass during the observation, 10) Seeing. Galaxy morphologies and distances are taken from the NASA Extragalactic database. B and AB indicates the existence and absence of bar, respectively. The hard X-ray luminosities (14 - 195 keV) are taken from the SWIFT-BAT 70 months survey \citep{Baumgartner}, where X-ray luminosities were corrected for absorption based on X-ray fittings by \citet{Ricci2017}. The list of abbreviation : Distance (Dist) Observation (Obs), Morphology (Morph), Air Mass (AirM), Peculiar (p). Seyfert types of the LLAMA AGNs are presented in Table $\ref{catvscotable}$.}
\vspace{2.5mm}
\centering
\begin{tabular}{cccc|ccc|cccc}
\hline
\hline
 		&  Properties  &   &  	 &  X-SHOOTER	& 	 & & SINFONI	& \\
 \hline
 Object Name & Dist & Morph & log $L$ & Obs. Date  & AirM & Seeing & Obs. Date  & AirM & Seeing   \\
 		& (Mpc)   &	  & erg s$^{-1}$	  &	DD/MM/YY	&  		& $\arcsec$	 &		DD/MM/YY  & 	& $\arcsec$  \\
\hline
1 & 2 & 3 & 4A & 5 & 6 & 7 & 8 & 9 & 10 \\
\hline
ESO 137-G034 & 35 & S0a(AB) & 42.76 & 19/05/19 & 1.2 & 0.78 & 18/04/14 & 1.2 & 0.75  \\
ESO 021-G004 & 39 & SA(s)0/a & 42.70 &02/08/16 & 1.8 & 0.83 & - & - & -  \\
MCG-05-14-12 & 41 & S0 & 42.65  & 11/12/13  & 1.0 & 0.61 & - & - & -  \\
MCG-05-23-16 & 35 &  S0  & 43.50 & 22/01/14 & 1.1 & 1.21 & 14/01/17 & 1.1 &  1.00  \\
MCG-06-30-15 &27 & S?  & 42.91 & 16/01/15  & 1.1 & 0.83 & 04/06/14 & 1.1 & 1.08 \\
NCG 1365 & 18 &  Sb (B)  & 42.60 & 10/12/13 & 1.0 & 1.34 & 18/11/10 & 1.1 & 0.78 \\
NGC 2110 &27 & S? (AB) & 43.63 &16/01/15  & 1.1  & 0.59 & 15/01/11 & 1.1 & 0.83 \\ 
NGC 2992 & 36 & Sa  & 42.52 & 26/02/14  & 1.3 & 0.72 & 05/02/17 & 1.0 & 0.85 \\
NGC 3081 & 34 & (R)SAB(r)0/a & 43.29 & 20/02/14  & 1.2 & 0.82 & 14/03/17 & 1.2 & 0.76 \\
NGC 3783 & 38 & Sb (B)  & 43.58 & 11/03/14  & 1.4 & 0.81 & 16/02/15 & 1.2 & 1.04 \\
NGC 4235 &  37 &  Sa & 42.64 & 13/05/15  & 1.2 & 0.73 & - & - & - \\
NGC 4388 & 39  & SA(s)b (B) & 43.70 & -  & -  & -   & 24/02/15 & 1.5 & 0.35   \\
NGC 4593 & 37 & Sb (B)  & 43.20 & 10/03/14 & 1.3 & 0.80 & 23/01/15 & 1.1 & 0.88 \\
NGC 5128 & 3.8 & S0 p  & 43.02 & 21/05/15 & 1.1 & 0.76 & - & - & - \\
NGC 5506 & 27 & Sa p & 43.30 & 03/03/16 & 1.1 & 0.64 & 12/03/15 & 1.098 & 0.72 \\
NGC 5728 & 39 & SAB(r)a: & 43.36 & 13/05/15  & 1.0 & 0.81 & 25/06/15 & 1.3 & 0.75  \\
NGC 6814 &  23 & SAB(rs)bc  & 42.75 & 13/05/15  & 1.1 & 0.86 & 05/06/14 &  1.0 & 0.83 \\ 
NGC 7172 & 37 & Sa & 43.32 & 12/08/15 & 1.0  & 1.6   & 20/07/14 & 1.0  &  0.77 \\
NGC 7213 & 22 &   Sa(s) &  42.49 & 13/07/16  & 1.3 & 0.47 & 16/07/14 & 1.1 & 0.83 \\ 
NGC 7582 & 22 & (R')SB(s)ab (B) & 43.29 & 27/07/17 & 1.2 & 0.69  &  14/07/14 & 1.1 & 0.91 \\
\hline
1 & 2 & 3 & 4B & 5 & 6 & 7 & 8 & 9 & 10 \\
\hline
ESO 093-G003 & 22 & SAB(r)0/a? & 9.86 &  22/01/14 & 1.3 & 0.98 & 06/04/17 & 1.4 & 0.86 \\
ESO 208-G021 & 17 & SAB0 & 10.88 & 12/12/13 & 1.1 & 0.95 & 14/03/17 & 1.2 & 1.02 \\
NGC 718 & 23 &  SAB(s)a	 & 9.89 & 05/12/15 & 1.2 & 0.61 & 13/08/14 & 1.2 & 0.82 \\
NGC 1079 & 19 & (R)SAB(rs)0/a & 9.91 & 23/11/13 & 1.0 & 1.12 & 17/11/051 & 1.1 & 0.88 \\ 
NGC 1315 & 21 & SB0?	 & 10.07 & 11/12/13 & 1.0 & 0.83 & - & - & - \\
NGC 1947 & 19 & S0 p & 10.45 & 23/12/13 & 1.4 & 0.77 & - & - & - \\
NGC 2775 & 21 & SA(r)ab	 & 9.84 &   15/11/15 & 1.5 & 0.74 & - & - & - \\
NGC 3175 & 14 & SAB(s)a? & 10.07 &   09/03/14 & 1.2 & 1.13 & 06/04/17 & 1.0 & 0.88 \\
NGC 3351 & 11 & SB(r)b & 10.39 &   21/02/14 & 1.3 & 1.04 & 27/01/15 & 1.3 & 0.89 \\
NGC 3717 & 24 & SAb & 10.40 &   22/03/14 & 1.2 & 1.34 &  - & - & - \\
NGC 3749 & 42 & SA(s)a & 10.48 &   22/03/14 & 1.0 & 0.93 &  - & - & - \\
NGC 4224 & 41 & SA(s)a & 10.22 & 13/05/15 & 1.2 & 0.66 & 24/02/15 & 1.2 & 0.91 \\
NGC 4254 & 15 & SA(s)c	 & 10.22 &   02/06/16 & 1.3 & 0.77 & 09/03/15 & 1.5 & 0.84 \\
NGC 4260 & 31 & SB(s)a	 & 10.25 & - & - & - & - & - & - \\ 
NGC 5037 & 35 & SA(s)a & 10.30 & 13/05/15 & 1.0 & 0.70 &  - & - & - \\
NGC 5845 & 25 & E & 10.46 & 16/03/16 & 1.2 & 0.69 & 14/03/17 & 1.2 & 0.61 \\
NGC 5921 &21 & SB(r)bc & 10.08 & 16/06/15 & 1.2 & 0.71 &  - & - & - \\
NGC 7727 & 26 & SAB(s)a p & 10.41 & 25/08/15 & 1.0 & 0.68 & 21/07/14 & 1.0 & 0.89 \\
IC 4653 &  26 & SB0/a(r) p &  9.48 & 19/05/2015 & 1.2 & 0.79 & 25/07/2017 & 1.6 & 1.11 \\
\hline
\vspace{0.5mm}
\label{t1}
\end{tabular}
\end{table*}

\section{SAMPLE SELECTION, OBSERVATION and DATA REDUCTION}

\subsection{Sample Selection}
A complete volume-limited sample of the most luminous X-ray-selected local AGNs in the Southern Hemisphere was compiled by \citet{Davies} as a the Local Luminous AGN with Matched Analogues project. The AGN sample was selected from the SWIFT-BAT 58 months survey \citep{Baumgartner} using the following three criteria: \\
\\
1. High X-ray luminosity ($\log{L_{14-195 keV}}$ $\geq$ 42.5 erg s$^{-1}$), to select bona-fide AGNs.  \\
2. Low-redshift AGNs ( $z$ < 0.01 ), to spatially resolve the nuclear regions.  \\ 
3. Observable from VLT ($\delta$ $<$ 15$^{\circ}$ ). \\ 

The LLAMA AGN sample comprises ten Type 1 and ten Type 2 AGNs \citep{Davies}. They were selected to be the most luminous local AGNs and are sufficiently powerful to sustain a BLR.

The matching inactive galaxy sample was selected by \citet{Davies} based on the following criteria: H-band luminosity (as a proxy of stellar mass), redshift, distance, inclination and host galaxy morphology. Due to these criteria, 19 inactive galaxies comprise the LLAMA inactive galaxy sample.

Here, we compare the physical properties of both sample. The mean H-band luminosities are $\log{L_{H}}$[$L_{\odot}$] = 10.3 $\pm{0.3}$ for AGN sample and $\log{L_{H}}$[$L_{\odot}$] = 10.2 $\pm{0.4}$ for inactive galaxy sample. The LLAMA inactive galaxies are also selected within the same redshift cut-off as active galaxy sample, which is  z < 0.01. The active and inactive galaxy sample have redshift-independent mean distances 31 and 24 Mpc, respectively. The average inclinations for each sample are found to be $\sim$ 45$^{\circ}$. Both active and inactive samples have a wide variety of galaxy morphologies with a peak distribution around early-disk types (S0 and Sa). Finally, also where possible, presence/absence of a bar is matched for both sample.

\subsection{Observations and Data Reduction}
The medium-resolution spectrograph X-SHOOTER on the Very Large Telescope (VLT), covering 0.3-2.3 $\mu$m, was used to observe the LLAMA sample. The X-SHOOTER observations were performed between November 2013 and June 2015, using the IFU-offset mode with a Field of View (FOV) of 1$\arcsec$.8 $\times$ 4$\arcsec$ Spectroscopic standard star observations were performed on the same nights with similar atmospheric conditions, and telluric standard stars were observed before and after the target. 
Data were obtained with resolution R $\sim$ 8400, 13200, 8300 for the ultraviolet (UVB), visual (VIS) and Near-infrared (NIR) arms, respectively. The X-SHOOTER data cubes were obtained using the ESO X-SHOOTER pipeline v2.6.0 \citep{Modigliani} within the ESO Reflex environment \citep{Freudling}. Finally, the spectra were corrected for telluric absorption using telluric standard stars. The data analysis of the X-SHOOTER observations was performed by \citet{Allan} and included most notably a correction for the [Fe II] multiplets in the 4000 -- 5600 \AA{} wavelength range. A more detailed description of the X-SHOOTER data processing will be given in Burtscher et al. (in prep.)  .

The SINFONI observations were performed between 2014 April and 2018 March with the H+K grating at a spectral resolution R $\sim$ 1500 for each 0$\arcsec$.05$\times$0$\arcsec$.1 spatial pixel leading to a total FOV of 3$\arcsec$.0$\times$3$\arcsec$.0. The observations were performed in adaptive optics (AO) mode and a standard near-infrared nodding technique was used. The telluric standard stars were observed before and after the target observations to obtain similar atmospheric conditions. SINFONI data were reduced using the SINFONI custom reduction package SPRED \citep{Abuter}. Further details about observation and data reduction are described by \citet{Lin}. 

Here, we note that the majority of XSHOOTER and SINFONI observations were performed for both active and inactive galaxy sample and the same data reduction approach was used for them. In Table \ref{t1}, we present the observation lists and basic properties of the LLAMA AGN and inactive galaxy sample.


\section{METHODS AND MODELS}

We performed the spectral analysis for 20 AGNs in our sample. In the first step, the AGN continuum was modelled and extracted from the spectra using additive polynomials in the form of power-law functions. We fit the spectra of each AGN using stellar templates to determine stellar velocity dispersions (see Section \ref{sec:stellar-kinematics}). The resulting  stellar velocity dispersion estimates are presented in Table \ref{catvscotable}. The emission lines from BLR and NLR were fit by applying multiple Gaussian models (Section \ref{sec:BLR}). Finally, black hole masses were obtained through virial `single-epoch' empirical correlations (Section \ref{sec:black-hole-mass}). The results are presented in Table \ref{Results}.  

\subsection{Velocity Dispersion Calculations} {\label{sec:stellar-kinematics}}

We obtained stellar velocity dispersions from the Ca II triplet (8498, 8552, 8662 $\AA$), where the AGN contamination is typically weaker than in the Mg b triplet (5069, 5154, 5160 $\AA$) \citep{GreeneHo2006b,Harris}. We also estimated stellar velocity dispersions from the CO (2-0) absorption at 2.2935 $\mu$m, since it is less affected by dust extinction. \citet{Riffel} reports that giant and super-giant stars are the dominant contributor for CaT and CO regions, respectively. To estimate stellar velocity dispersions, we used the penalised pixel-fitting (pPXF) method \citep{CapEms,Cap} adopting the X-SHOOTER G, M, K stellar population spectral library (127 stars) of \citet{Chen} for fitting the CaT absorption lines and the GEMINI NIR stellar library with spectral types ranging from F7 III to M5 III (60 stars) \citep{Winge} for fitting the CO (2-0) absorption lines.

pPXF adopts the Gauss-Hermite parametrisation for the line-of-sight velocity distribution in the pixel space, where bad pixels and emission lines can be easily excluded from the spectra, and continuum matching can be performed directly using additive polynomials. pPXF measures stellar velocity dispersions by making initial guesses using a broadening function for stellar templates. The fit parameters ($V$, $\sigma$, $h_{3}$, ..., $h_{m}$), where $h_{i}$ is the Hermite polynomial for the i-th parameter, are fitted simultaneously using pPXF, but it adds an adjustable penalty term to the $\chi^{2}$ to optimize the fit. In this way, the best fitting parameters of the Gauss-Hermite series can be estimated, and the lowest $\chi^{2}$ are provided by the definition of this method \citep[e.g.,][and references therein]{MaFr}. The uncertainties of stellar velocity dispersion estimates were obtained via bootstrapping by randomly resampling the residuals of the best-fit of pPXF, and repeating pPXF fitting 100 times.

To match the spectral resolutions of galaxy and template spectra, the template spectra were convolved with the line spread function of $\sim$ 70 km s$^{-1}$ for SINFONI data, while the XSHOOTER template spectra were convolved by $\sim$ 5 km s$^{-1}$. Since the CO absorption lines in the near-infrared tend to have lower signal to noise ratio (S/N $\sim$ 10) relative to the CaT absorption lines (S/N $\sim$ 50), we did not use $h_{3}$ and $h_{4}$ higher order moments for the CO (2-0) absorption lines fitting. The fitting procedure for the CO (2-0) absorption is explained in detail by \citet{Lin}. We note that the AGN emission lines (e.g., O I 4998 \AA , Fe II 8616 \AA) are masked to increase the accuracy of stellar velocity dispersion calculations. We fit the integrated spectrum from the X-SHOOTER within 1$\arcsec$.8 $\times$ 1$\arcsec$.8 radius for CaT, whereas the integrated spectrum withing 3$\arcsec$.0 $\times$ 3$\arcsec$.0 radius was used for fitting CO (2-0). Finally, the resulting $\sigma_{\star}$ estimates are corrected for the instrumental broadening.

We then corrected $\sigma_{\star}$ estimates from the 1$\arcsec$.8 slit-width to an effective radius using the following power-law function in the form:

\begin{equation}
\sigma_{re} = \sigma_{ap} \left( \frac{r_{ap}}{r_{re}} \right)^{\alpha}
\end{equation}
where $\alpha$ is the slope, $r_{e}$ is effective radius. Since $\log{L_{H}}$[$L_{\odot}$] = 10.3 $\pm{0.3}$, which is assumed to be a proxy of stellar mass, for the LLAMA AGN sample , we adopt $\alpha$ = 0.077$\pm{0.012}$ for late-type galaxies within 10 < $\log{M_{\star}}$ < 11 $M_{\odot}$ \citep{Falcon}. We note that we only present the resulting best-fitting $\sigma_{\star}$ values obtained within instrument aperture in Table \ref{catvscotable}. But, we note that effective radius-corrected $\sigma_{\star}$ values are used in our $M_{BH}$ - $\sigma_{\star}$ relation investigations. We note that the effective radius correction changes the LLAMA $\sigma_{\star}$ estimates from 2\% to 18\% with a mean of $\sim$ 10\%.

\subsection{Bulge properties of the LLAMA sample}

In this paragraph, we explain our method to identify the bulge properties of the LLAMA sample. \citet{FiDr} list a few major indicators for identifying pseudo-bulges. However, none of these diagnostics can be used alone to identify pseudo-bulges. In the same work, the authors also claim that pseudo-bulge hosting galaxies tend to have S\'{e}rsic index $n$ < 2, bulge to total mass ratio $B/T$ $\leq$ 0.35 and $\sigma_{\star}$ < 130 km s$^{-1}$. Even though there are some exceptional cases, these three diagnostics are the best indicators for pseudo-bulges. Correspondingly, we collected $n$ and $B/T$ estimates from the literature. The collected diagnostic bulge type indicators are presented in Table \ref{catvscotable}. These diagnostic parameters for pseudo-bulge identification demonstrate that the majority of the LLAMA AGN sample hosts pseudo-bulges ($\sim$ 65\%).

\subsection{Emission Line Fitting} {\label{sec:BLR}}	 

We fit the spectra of our sample by adopting Astropy fitting routines \citep{astropy1,astropy2}. The broad-line emission can often be fit sufficiently well using a single Gaussian profile, but sometimes more complex approaches are required \citep[e.g. double peak BLR emissions, extended wings][]{Peterson2004,SBT}. H$\beta$ profiles were fit within a rest-frame range of 4700 - 5100 $\AA$, whereas H$\alpha$ profiles were fit within a rest-frame range of 6400 - 6800 $\AA$. First, the AGN continuum of each AGN was modelled using a power-law function for H$\beta$, H$\alpha$ and Pa$\beta$ region. We then describe narrow-emission lines using single Gaussian profile for each AGN. For H$\beta$ spectral region, we fit narrow H$\beta$, [O III] (4959 $\AA$), and [O III] (5007 $\AA$) lines using single Gaussian profile for each narrow component.  For H$\alpha$ region, we fit narrow H$\alpha$, [N II] (6548  $\AA$), [N II] (6583 $\AA$), [S II] (6718.3 $\AA$) and [S II] (6732.7 $\AA$) lines using single Gaussian profile for each narrow component. However, since H$\alpha$ is blended with two [N II] lines (6548 and 6583 $\AA$), we adopted $F_{[N II]}^{6583 \AA}$ = 2.96 $\times$ $F_{[N II]}^{6548 \AA}$ \citep{OsFe} and equal velocity dispersions for the [N II] lines in our calculations. Finally, Pa$\beta$ emission lines were fitted within the rest-frame range of 12200 - 13200  $\AA$, where we used a single Gaussian profile to describe the narrow component of Pa$\beta$ emission-line.

For fitting the BLR profiles, we used a single Gaussian model for some of AGN, but a second Gaussian profile was required to characterize the BLR profile for the following galaxies MCG-05-14-12, MCG-06-30-15, NGC 3783, NGC 4593, NGC 4235, NGC 6814 and NGC 7213.  For the broad-line profiles that required double Gaussian models, both Gaussian profiles are combined with each other, and the resulting FWHM is estimated from the new, combined profile. Uncertainties of the FWHM estimates are derived from the fit residuals. Here, we empathize that the narrow-emission line components and the AGN continuum were extracted, before we estimate the width of broad-emission line profiles. To test the reliability of the H$\alpha$ based calculations, we additionally studied the H$\beta$ and Pa$\beta$ (when H$\beta$ is not available) emission profiles for comparison. The resulting FWHM differences between H$\alpha$, H$\beta$ and Pa$\beta$ emission-line profiles of our sample are found to be less than 20\%, and this result is consistent with other observational results from different sample \citep[][]{GreeneHo2005,ShLi,MeRe16,Ricci}. For consistency, we used a same number of Gaussian models for fitting H$\alpha$, H$\beta$ and Pa$\beta$ emission-line profiles of each AGN. The resulting parameters are presented in Table \ref{Results}. 

In the case of MCG-05-14-12, NGC 1365 and NGC 2992 we detected blue-shifted emission lines in the spectra (> 500 km s$^{-1}$), which were also fitted with additional single Gaussian models. We excluded these blue-shifted emission lines, when we estimate our final BLR profiles of the LLAMA AGNs. We present the emission line fitting of our type 1 AGN sample in the Appendix (see \ref{appendix1}, \ref{appendix2}, \ref{appendix3}, \ref{appendix4}). 

MCG-05-14-12 and MCG-06-30-15 both show low emission line widths (FWHM $<$ 1700 km s$^{-1}$) and low [O III]/H$\beta$ ratios (0.2 and 0.9, respectively). According to the definition of narrow-line Seyfert 1 (NLS1) galaxies (FWHM $<$ 2000 km s$^{-1}$ and [O III]/H$\beta$ $<$ 3) reported by \citet{OsPo}, we classify them as such.

\subsection{Black Hole Mass Estimations} {\label{sec:black-hole-mass}}

By assuming gravitationally dominated, virialized, rotating gas in the BLR, black hole masses can be obtained by: 

\begin{equation}
M_{BH} = f \left( \frac{\Delta V^{2} R}{G} \right), 
\end{equation}
where $f$ is a factor that depends on the unknown structure, kinematics, and orientation of the BLR, $\Delta$ $V$ is the velocity dispersion of the broad emission line, $G$ is the gravitational constant and $R$ is the BLR radius \citep[e.g.,][]{Peterson2004}. In this equation, the $f$ factor converts the observed virial product into black hole masses.

From the RM studies, a strong correlation between the AGN continuum luminosity ($\lambda$ $L_{5100}$) and the radius of the BLR ($R_{BLR}$) have been determined \citep{Kaspi,Bentz09,Bentz13}. By adopting the $R_{BLR}$ - $\lambda$ $L_{5100}$ relation, black hole masses based on virial `single-epoch' empirical correlations can be obtained. The tight empirical correlations between $M_{BH}$ and emission from BLR regions can be expressed as: 

\begin{equation}
 M_{BH} = 10^{\alpha} \times \left( \frac{L_{H\alpha}}{10^{42} \: erg \: s^{-1}} \right)^{\beta}  \times \left( \frac{FWHM_{H\alpha}}{10^{3} \: km \: s^{-1}} \right)^{\gamma}  \times f_{FWHM}   \:   \:  \: \:   M_{\odot}
\end{equation}

\begin{equation}
 M_{BH} = 10^{\alpha} \times \left( \frac{\lambda L_{5100}}{10^{44} \: erg \: s^{-1}} \right)^{\beta}  \times \left( \frac{\sigma_{H\beta}}{10^{3} \: km \: s^{-1}} \right)^{\gamma} \times f_{\sigma}   \:  \:  \:  \:    M_{\odot}
\end{equation}

\begin{equation}
 M_{BH} = 10^{\alpha} \times \left( \frac{L_{Pa\beta}}{10^{42} \:erg \: s^{-1}} \right)^{\beta}  \times ,\left( \frac{FWHM_{Pa\beta}}{10^{4} \:km \: s^{-1}} \right)^{\gamma}  \times  \left( \frac{f_{\sigma}}{4.31} \right)   \: \:  \: \:      M_{\odot}
\end{equation}
where we adopt the $\alpha$, $\beta$, $\gamma$ values 6.544, 0.46, 2.06 for the $L_{H\alpha }$ - $FWHM_{H\alpha}$, 6.819, 0.533, 2.0 for the $L_{5100}$ - $\sigma_{H\beta}$ calibration \citep[][hereafter W15]{Woo2015}, and 7.834, 0.46, 1.88 for $L_{Pa\beta}$ - $FWHM_{Pa\beta}$ calibration reported by \citet{Lafranca}.Since some studies suggest that the line profile of H$\beta$ is not universal, and the second moment ($\sigma_{Line}$) of H$\beta$ profile gives more accurate H$\beta$-based $M_{BH}$ estimates \citep{Peterson2004,Collin}, we used $\sigma_{Line}$ for our H$\beta$-based $M_{BH}$ investigations. This effect will be discussed in Section \ref{sec:robo}.

The observed flux of broad H$\beta$ emissions weakens with the decrease of the inclination angle of AGN structure, and becomes undetectable for Sy 1.9 galaxies \citep[e.g.][]{Allan}. However, broad H$\alpha$ can be observed even in these moderately obscured AGNs. Therefore, we estimate black hole masses of our sample using broad H$\alpha$ emission lines for the entire sample, whereas we present the black hole masses obtained from H$\beta$ or Pa$\beta$ for comparison. 

Furthermore, we adopted $M_{BH}$ estimates of NGC 4388 and NGC 5728 obtained by \citet{Greene2016} and \citet{Braatz}, respectively. Finally, the $M_{BH}$ of NGC 5128 is adopted from \citet{Cap2009}, in which the authors used stellar kinematics to obtain $M_{BH}$ value. Therefore, we have thirteen $M_{BH}$ estimates in total for ten type 1 and three type 2 AGNs , which will be further used in our $M_{BH}$ and $\sigma_{\star}$ investigations.

\subsection{The $f$ Factor}
The black hole masses are estimated for our sample using the broad-line based single-epoch scaling relations. In the broad-line based black hole mass estimations, the dimensionless $f$ factor is an important parameter that can change the $M_{BH}$ estimates by an order of magnitude. The obscurity of geometry, kinematics and orientation of the BLR constitute systematic uncertainties encapsulated in the f factor. Although there was no precise method to obtain the $f$ factor, it is determined in the literature by assuming AGN-hosting galaxies follow the inactive $M_{BH}$ - $\sigma_{\star}$ relation \citep[e.g.,][]{Onken}. A mean value of $f$ $\sim$ 5 was reported for $\sigma_{\star}$-based $M_{BH}$ estimations with an intrinsic scatter of 0.35 dex, whereas the $f$ factor was found to be $\sim$ 1 for FWHM-based $M_{BH}$ estimations \citep[e.g.,][]{Woo2015,Grier2017}. 

Interestingly, \citet{SBT} and \citet{MeRe} show an anti-correlation between the FWHM$_{obs}$ and the $f$ factor, and \citet{MeRe} provided a relation for the $f$ factor calculations: $f$ = (FWHM$_{obs (line)}$/FWHM$^{0}_{obs}$)$^{\beta}$, where $\beta$ and FWHM$_{obs}^{0}$ values are -1.0$\pm{0.10}$, 4000$\pm{700}$ km s$^{-1}$ for H$\alpha$ and -1.17$\pm{0.11}$, 4550$\pm{1000}$ km s$^{-1}$ for H$\beta$, respectively. This formula is roughly consistent with the $f$ factor of 1.12 (W15), $f$ factor of 1.51 \citep{Grier} for both H$\alpha$ and H$\beta$ BLR gas with a FWHM range 2000-4000 km s$^{-1}$, whereas the difference between calibrations significantly increases for the BLR gas with FWHM $<$ 2000 km s$^{-1}$. Accordingly, the $f$ factor is reported to be different for every AGN \citep{Pancoast}.

Until recently, there was no direct method to obtain the $f$ factor, but interestingly the \citet{GRAVITY2} resolved the BLR region of 3C 273 using observational data from VLTI/GRAVITY.  In the same work, the authors reported an $f_{FWHM}$ = 1.3 $\pm{0.2}$ and $f_{\sigma}$ = 4.7 $\pm{1.4}$ for 3C 273. The \citet{GRAVITY2} noted that a comparison between RM and interferometry in the same objects can be very efficient for understanding the characteristics of BLRs and for increasing the accuracy of $M_{BH}$ estimations. Even though the $f$ factor remains as an uncertainty of $M_{BH}$ estimations of Type 1 AGNs for now, the $f$ factor of $\sim$ 1 and 5 are expected to represent the BLR structure for FWHM and $\sigma_{Line}$ estimations, respectively. Further investigations with VLTI/GRAVITY are required to resolve the BLR structures for each AGNs. 

The latest single-epoch RM based calibrations are presented by \citet{Woo2015}, and we use these for the further analysis: we adopt an $f$ factor of 4.47 ($\log{f = 0.65\pm{0.12}}$) for estimates based on $\sigma_{Line}$ of H$\beta$ and 1.12 ($\log{f = 0.05\pm{0.12}}$) for estimates based on the modeled FWHM of H$\alpha$, respectively. For the black hole mass estimates based on the Paschen-$\beta$ line, we re-calibrate the \citet{Lafranca} calibration adopting the same $f$ factor as for the H$\beta$ estimate.

\subsection{The Dust Extinction}

In the single-epoch reverberation mapping calibration, the luminosity is usually not corrected for extinction since the objects studied there are essentially unobscured (Type 1) AGNs. Since we also have moderately obscured Type 1 objects in our sample, an extinction correction must be applied to these objects to have accurate $M_{BH}$ estimations. In a previous LLAMA project, \citet{Allan} used the line ratios of various hydrogen recombination lines from the UV to the near-infrared to derive both the excitation conditions and the optical extinction to the BLR for 9 objects. We have adopted $A_{V}$(BLR) estimates from \citet{Allan} for nine of type 1 AGNs in our sample. We note that $A_{V}$(BLR) of NGC 7213 is obtained in this study using the same approach provided by \citet{Allan}. This method can only be used to type 1 AGNs, and a more detailed explanation for extinction calculation is given by \citet{Allan}.

Here, it is worth to mention that \citet{Leo} and \citet{Taro} also estimated the extinction in the BLR by comparing X-ray absorption and optical obscuration for some AGNs in our sample. The estimated $A_{V}$(BLR)s are found to be consistent with the ones reported by \citet{Allan}. Since the method from \citet{Allan} is a more direct method for obtaining the BLR extinction, we have used their $A_{V}$(BLR) estimates.

In order to convert from $A_{V}$ to the extinction at a any wavelength ($A_{\lambda}$), we employ the extinction law presented by \citet{Wild}:

\begin{equation}
A_{\lambda} / A_{V} = 0.6(\lambda/5500)^{-1.3} + 0.4(\lambda/5500)^{-0.7} . 
\label{extinctionwave}
\end{equation}
In this equation, the first term describes the dust extinction along the line of sight (assuming Milky Way dust), whereas the second term provides the dust extinction caused by the diffuse interstellar medium. \citet{Wild} reports that this equation provides a good correction for AGNs with a large dust reservoir. We use the Equation \ref{extinctionwave} to convert the BLR extinction in V-band to the BLR extinction in H$\alpha$ (6562.8 $\AA$), H$\beta$ (4861.4 $\AA$) and Pa$\beta$ (1281.8 $\AA$). As mention in \citet{Allan}, this relation gives a good correction for both the NLR and BLR of the LLAMA AGNs.

The resulting $A_{\lambda}$ (BLR) values are used to correct the extinguished BLR flux (S) of H$\alpha$ and the continuum flux of $L_{5100}$ using the following equation.

\begin{equation}
S_{corrected} = S_{observed} \times 10^{0.4 A_{\lambda}} 
\end{equation}

For highly obscured sources in our sample (NGC 1365, NGC 2992, MCG-05-23-16), we used Pa$\beta$ $M_{BH}$ calibration reported by \citet{Lafranca} (see Equation 4) for obtaining $M_{BH}$ values, since the broad H$\beta$ cannot be detected for these sources. Even though the near-infrared band suffers less from the dust extinction \citep{Landt}, we have also corrected the slightly extinguished BLR flux of Pa$\beta$ using the resulting $A_{\lambda}$(BLR) in our calculations.

\subsection{Accretion Rate}
In this section, we explain the method for estimating the Eddington ratios and accretion rates of our sample by adopting the following empirical relations. First, we obtain the bolometric luminosities by \citep{Winter}:  

 \begin{equation}
 \label{eq:lbol}
 \log{L_{Bol}} =  1.12 \log{L_{14-195 keV}} - 4.23  \:   \:  \: \: erg s^{-1}.
\end{equation}

Then, the Eddington luminosity ($L_{Edd}$) can be written as $L_{Edd}$ = 1.26 $\times$ 10$^{38}$ $M_{BH}$/M$_{\odot}$ \citep{Rybicki}. We used our single-epoch $M_{BH}$ values from H$\alpha$ to estimate the Eddington luminosities for the Type 1 sources.  To obtain Eddington luminosities for the LLAMA Type 2 sources, black hole masses that are calculated from the LLAMA M$_{BH}$ - $\sigma_{\star}$ relation (see Section \ref{sec:llama_m_sigma}) are used, whereas  we collected the megamaser black hole masses for NGC 4388 \citet{Greene2016} and NGC 5728 \citep{Braatz} , respectively. The Eddington ratio ($\lambda_{Edd}$) can be computed by: 

 \begin{equation}
 \label{eq:ledd}
 \lambda_{Edd} =  \left( \frac{L_{Bol}}{L_{Edd}} \right) .
\end{equation}

Finally, the mass accretion rate ($\dot{M}$) onto the black hole can be estimated by assuming a steady radiative efficiency $\epsilon$ = 0.1 \citep{CoHu}: 

 \begin{equation}
 \label{eq:macc}
\dot{M} =  \left( \frac{L_{Bol}}{\epsilon c^{2}} \right) .
\end{equation}

We note that the main contribution to uncertainty on the Eddington ratios and accretion rates originate from the uncertainty in bolometric luminosity, accretion efficiency and $M_{BH}$, which corresponds to an uncertainty of $\sim$ 0.4 - 0.5 dex \citep{Bian,Marinucci}. This uncertainty range is roughly consistent with the median value of our estimates. The resulting Eddington and mass accretion rates can be found in Table $\ref{Results}$.

\subsection{The Statistical Fitting Procedure}

The FITEXY, an IDL-based tool, developed by \citet{Press} and modified by \citet{Tremaine}, is an effective tool for estimating fit parameters for a linear regression model. The original idea of the FITEXY method is based on a modified version of Bivariate Correlated Errors and Intrinsic Scatter proposed by \citet{AkBe}. The FITEXY method minimizes the $\chi^{2}$ statistic and takes into account the measurement error for both dependent or independent variables for X and Y axes. In this method, $\chi^{2}$ is minimized by 

\begin{equation}
\chi^{2} = \sum_{i=1}^{N} \frac{(\mu_{i} - \alpha - \beta s_{i})^{2}}{\sigma^{2}_{\mu,i} + \beta \sigma^{2}_{s,i} + \epsilon^{2}_{0}} ,
\end{equation}  
where $\mu$ is $\log{(M_{BH}/M_{\odot})}$, s is $\log{(\sigma_{\star}/\sigma_{0}}$) where $\sigma_{0}$ is 200 km s$^{-1}$, $\sigma_{\mu}$ and $\sigma_{s}$ are measurement uncertainties in both variables and $\epsilon_{0}$ is the intrinsic scatter. 

To fit the $M_{BH}$ - $\sigma_{\star}$ relation, we used a single power law as expressed in the following equation: 

\begin{equation}
\log{(M_{BH}/M_{\odot})} = \alpha + \beta \log{\left( \frac{\sigma_{\star}}{\sigma_{0}} \right)},
\end{equation}  
where $\alpha$ is the intercept, $\beta$ is the slope of the single power law fit. Here, we emphasize that both $M_{BH}$ and $\sigma_{\star}$ parameters will be estimated using the data obtained from the same spectra for the LLAMA type 1 sources.

\section{RESULTS and DISCUSSION} {\label{sec:discussion}}

In this section we first study and discuss the robustness of the observables and assumptions involved in constructing the $M_{BH}$ - $\sigma_{\star}$ relation, before presenting the $M_{BH}$ - $\sigma_{\star}$ relation for our sample.

\subsection{Stellar velocity dispersion estimates: Optical versus Near-Infrared}

We provide stellar velocity dispersion estimates of the CaT absorption lines, results in the range of 73 $\leq$ $\sigma_{\star CaT}$ $\leq$ 227 km s$^{-1}$ for our sample of AGNs (see Table \ref{catvscotable}). Besides, the estimated $\sigma_{\star CaT}$ values for the LLAMA inactive sample are found to be 64 $\leq$ $\sigma_{\star CaT}$ $\leq$ 262 km s$^{-1}$. This shows that the LLAMA active and inactive sub-samples, which are matched on total stellar mass ($H$ band luminosity), have also comparable bulge stellar masses.

\begin{figure}[hbt!]
\centering
\includegraphics[width=9.0cm]{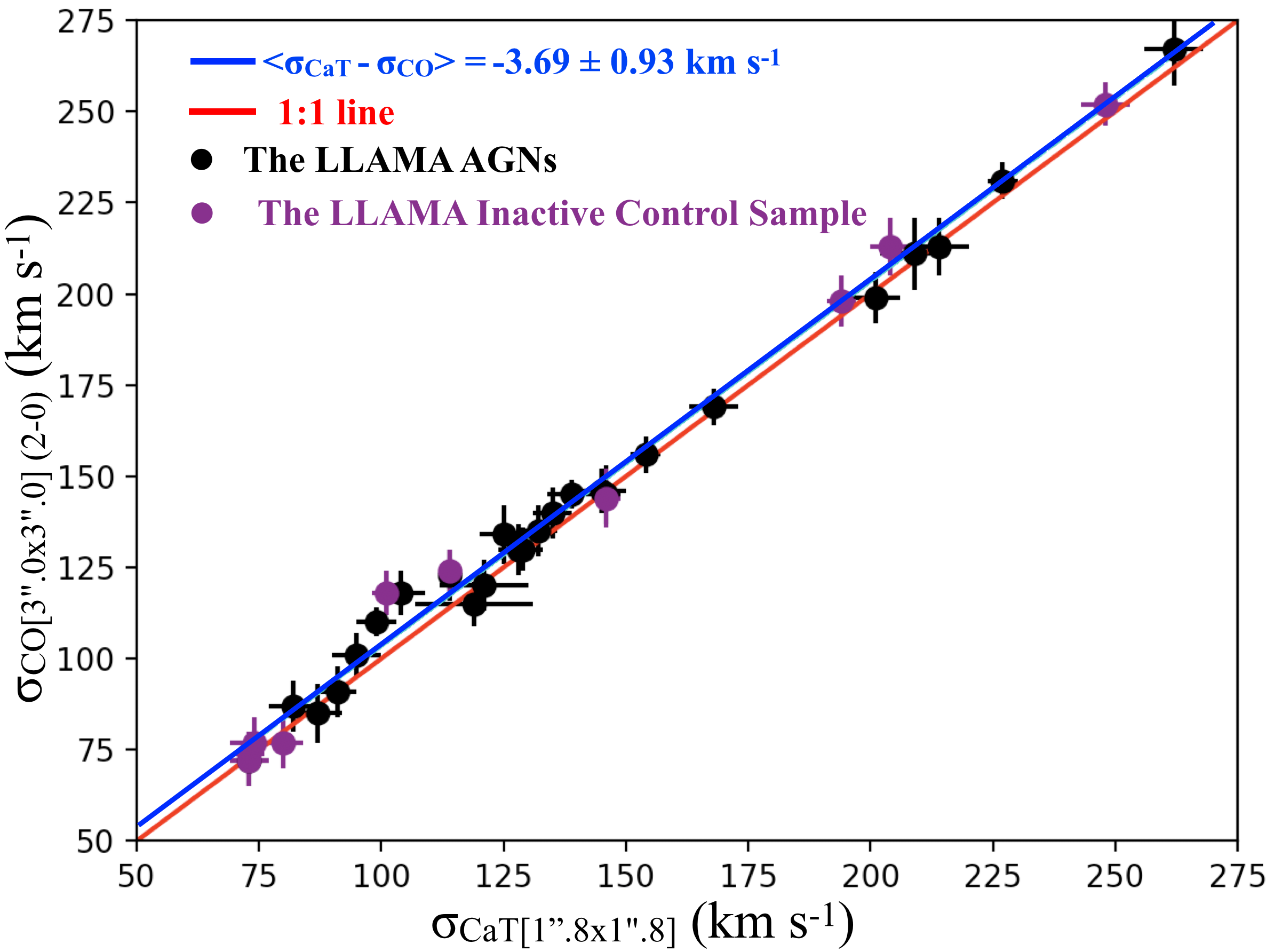}
\caption{The stellar velocity dispersion results, which are calculated from the CaT and CO (2-0) absorption features. We note that some of sources are not still observed for our entire sample, therefore, we compared the sources that we have both $\sigma_{CaT}$ and $\sigma_{CO (2-0)}$ estimates. The red solid line represents 1:1 line, whereas the blue solid line shows the offset between the $\sigma_{CaT}$ and $\sigma_{CO (2-0)}$ estimates of our data}. The LLAMA AGNs and inactive galaxies are presented as black and purple, respectively.
\label{catvsco}%
\end{figure}

Alternatively, we estimated the stellar velocity dispersion from the near-infrared CO (2-0) absorption band-head using the SINFONI data for a comparison. The  $\sigma_{\star CO (2-0)}$ values are found to be slightly higher ($\sim$ 3.69 $\pm{0.93}$ km s$^{-1}$) than the $\sigma_{\star CaT}$. The most likely explanation for this is that the near-infrared CO feature probes more deeply embedded (and therefore higher velocity dispersion) stellar populations than the optical CaT. Interestingly, our result shows a different trend than the results from \citet{Riffel}, where the authors claims that the discrepancy between  $\sigma_{\star CO (2-0)}$ and $\sigma_{\star CaT}$ is higher (<$\sigma_{\star CO (2-0)}$> - < $\sigma_{\star CaT}$ >  = 19$\pm{6}$ km s$^{-1}$). The  $\sigma_{\star CO (2-0)}$ versus $\sigma_{\star CaT}$ comparison and the resulting parameters are presented in Fig. \ref{catvsco} and Table \ref{catvscotable}.  

\begin{table*}[hbt!]
\begin{center}
\caption{Stellar velocity dispersion comparison between the estimates from CaT and CO (2-0) absorption lines. Sector 1 (top): the LLAMA AGNs, Sector 2 (bottom): The LLAMA inactive Galaxies. Columns are from left to right as follows: 1) Object name, 2) Bulge effective radius, 3) Stellar velocity dispersion estimates from the CaT absorption lines, 4) Stellar velocity dispersion estimates from the CO (2-0) transmission, 5) The rotation contribution in percentage, $\star$ : the assumed rotation contribution, which is the the average rotation contribution of LLAMA sample, 6) S\'{e}rsic index, 7) Bulge to total mass ratio (B/T), 8) Bulge type, where ($\it{L}$) is LINER and ($\it{n}$) indicates narrow-line Seyfert 1 galaxies according to our spectral investigations. Reference (Ref) for Seyfert activity in the literature; 0: This work, I: \citet{Allan}, II: \citet {Gu2006}, III: \citet{MaiRie} IV: \citet{Veron}, V: \citet{Gonza}. }

\begin{tabular}{ccccccccc} 
\hline
\hline
Object & $r_{e}$ &  $\sigma_{\star CaT}$ & $\sigma_{\star CO_{(2-0)}}$ & Correction & S\'{e}rsic index & B/T & Bulge Type & Seyfert Activity  \\ 
& arcsec & km s$^{-1}$ & km s$^{-1}$ & \% & & & \\
\hline
1 & 2 & 3 & 4 & 5 & 6 & 7 & 8\\
\hline
ESO 137-G034 & 6.94 (a)	& 128$\pm{4}$	& 130$\pm{7}$ &	10 ($\star$) & 2.13 (a) & 0.22 (a) & PB? & Sy 2 (II) \\

ESO 021-G004 & 16.7 (r) & 178$\pm{3}$	& - & 10 ($\star$) &	- & - & - & Sy 2 (0) \\

MCG-05-14-12 & 4.41 (r) &73$\pm{5}$ 	&	-  & 10 ($\star$) & - & - &	 - & Sy 1.0 (0 \it{n})  \\
MCG-05-23-16 & 9.37 (c) &	135$\pm{4}$ 	& 	140$\pm{7}$ & 11.8  &	3.20 (c) &	-  & CB?	& Sy 1.9 (I) \\
MCG-06-30-15 & 0.63 (d) &	95$\pm{5}$	&  101$\pm{6}$ & 10 ($\star$)   & 1.29 (d)	& 0.06 (d) & PB	& Sy 1.2 (0 \it{n})  \\ 

NGC 1365 & 12.8 (e) &	121$\pm{5}$	&	120$\pm{6}$ & 20 &	 0.86 (e)	&  0.25 (e)	& PB	& Sy 1.8 (I) \\
NGC 2110 & 6.80 (f) & 227$\pm{3}$ 	& 	231$\pm{5}$ & 10 ($\star$) &	2.70 (f)	& 0.39 (f)	& CB	 & Sy 2 (II) \\
NGC 2992 & 14.2 (r) &154$\pm{3}$ 	& 	156$\pm{5}$ & 12.2 & - &	- & -	& Sy 1.8 (I) \\
NGC 3081 &1.34 (g) & 132$\pm{4}$ & 135$\pm{7}$ & 10 ($\star$) & 2.10 (g)	& 0.10 (g) & PB? & Sy 2(II) \\
NGC 3783 &  1.45 (a) &	125$\pm{5}$	& 134$\pm{8}$ & 10 ($\star$)	& 1.24 (a)	& 0.21 (a)	& PB	& Sy 1.2 (I) \\ 
NGC 4235 & 2.70 (o) &	142$\pm{5}$ &	- & 10 ($\star$) & 6.00 (h)	& 0.50 (i)	& CB 	& Sy 1.2 (I) \\
NGC 4388 &  5.62 (p)	& -	& 117$\pm{6}$ & 18.8 &0.50 (j) & - & - & Sy 2 (II) \\
NGC 4593 & 6.21 (b) &  139$\pm{5}$ & 145$\pm{4}$ & 1.4	& 1.37 (b)	& 0.18 (b)	& PB &  Sy 1.2 (1) \\
NGC 5128 & 8.62 (k)	&	199$\pm{8}$	&	- & 10 ($\star$)	& 	2.63 (k)	&	1.00 (l)	& CB & Sy 2 (III) \\
NGC 5506 & 2.06 (m)    &   -           &   118$\pm{47}$ & 10 ($\star$)    & 0.50 (m) & 0.06 (m) & PB & Sy 1i (IV) \\
NGC 5728 & 4.02 (a)	&168$\pm{7}$	& 	169$\pm{9}$	& 2.8 & 1.10 (a)	& 0.23 (a)	& PB? & Sy 2 (II) \\
NGC 6814 & 1.08 (a) &	99$\pm{4}$ & 110$\pm{4}$ & 0 & 1.08 (a)	& 0.09 (a)	& PB	&	Sy 1.2 (I) \\
NGC 7172 & 1.16 (a)	& 145$\pm{5}$	& 146$\pm{6}$ & 10 ($\star$) & 1.16 (a)	& 0.25 (a) & PB?	&	Sy 2 (II) \\
NGC 7213 & 13.7 (a)	& 209$\pm{7}$	&	211$\pm{10}$ & 0	& 2.57 (a)	& 0.70 (a) & CB	&	Sy 1.0 (V $L$) \\
NGC 7582 & 1.99 (a)	& 129$\pm{4}$	& 130$\pm{6}$	& 10 ($\star$)	& 2.72	(a) & 0.28 (a) & PB?	&	 Sy 2 (II) \\
\hline
ESO 093-G003 & 11.5 (r) & 87$\pm{5}$ & 85$\pm{8}$ & - & - & - & - & -  \\
ESO 208-G021 & 7.47 (g) & 214$\pm{6}$ & 213$\pm{9}$ & - & 4.20 (g) & 0.97 (g) & CB & - \\
NGC 718 & 2.09 (a) &	104$\pm{5}$ & 118$\pm{7}$ & - & 1.32 (a) & 0.28 (a) & PB & - \\
NGC 1079 & 4.94 (g) & 114$\pm{2}$ & 123$\pm{7}$ & - & 2.20 (g) & 0.25 (g) & PB? & - \\ 
NGC 1315 & 16.1 (r) & 77$\pm{3}$ & -  & -  & - & - & -  \\
NGC 1947 & 30.1 (b) & 147$\pm{3}$  & - & - & 2.51 (b) & 0.68 (b) & CB & -  \\
NGC 2775 & 63.2 (h) & 175$\pm{6}$  & - & - & 3.49 (h) & 0.75 (i) & CB & -  \\
NGC 3175 & 40.1 (r) & 73$\pm{5}$ & 72$\pm{7}$ & - & - & - & - & - \\
NGC 3351 & 6.95 (a) & 91$\pm{4}$ & 91$\pm{7}$ & - & 0.80 (a) & 0.22 (a) & PB & - \\
NGC 3717 & 32.5 (r) & 137$\pm{5}$ & - & - & - & - & - & -  \\
NGC 4224 & 5.01 (a)	& 146$\pm{3}$ & 145$\pm{8}$ & - & 2.53 (a) & 0.29 (a) & CB? & -  \\
NGC 4254 & 12.59 (a) &	82$\pm{5}$ & 87$\pm{7}$ & - & 1.99 (a) & 0.19 (a) & PB? & -  \\
NGC 5037 & 23.2 (r) & 168$\pm{3}$ & - & - & - & - & - & -  \\
NGC 5845 & 0.49 (p) & 262$\pm{6}$ & 267$\pm{10}$ & - & - & 1.0 (i) & CB & -  \\
NGC 5921 & 3.59 (n) & 80$\pm{2}$ & - & - & 1.60 (n) & 0.50 (i) & PB? & - \\
NGC 7727 &  5.07 (a) & 201$\pm{5}$ & 199$\pm{7}$ & - & 1.68 (a) & 0.36 (a) & CB? & -  \\
IC 4653 & 17.0 (r) & 64$\pm{5}$ & - & - & - & - & - & -  \\
\hline
\end{tabular}
 \vspace{1ex}

     {\raggedright Notes: Bulge properties are taken from: (a) \citet{Lin}, (b) \citet{Gao}, (c) \citet{CapBal}, (d) \citet{ChenHu}, (e) \citet{Combes}, (f) \citet{Gadotti}, (g) \citet{Laurikainen}, (h) \citet{Heikki}, (i) \citet{deLap},  (j) \citet{Greene2010}, (k) \citet{FiDr2010}, (l) \citet{Korm2010}, (m) \citet{Yoshino}, (n) \citet{Knapen}, (o) \citet{Baggett} , (p) \citet{Vanden2016}, (r) \citet{Skrutskie}. The CaT region of NGC 5506 is highly contaminated by AGN emission lines, therefore the $\sigma_{CaT}$ is not presented in our study. There is no available X-SHOOTER observation for NGC 4388, but there are three available $\sigma_{\star CaT}$ estimates from the literature. The reported $\sigma_{\star CaT}$ values differ significantly: $\sigma_{\star CaT}$ = 119 km s$^{-1}$ \citep{Terlevich}, $\sigma_{\star CaT}$ = 165$\pm{21}$ km s$^{-1}$ \citep{Riffel} and $\sigma_{\star CaT}$ = 76 km s$^{-1}$ \citep{Greene2010}. But, the central velocity dispersion measurements of this galaxy, as reported by \citet{Greene2010,Saglia,Vanden2016}, are in the range of $\sim$ 100-120 km s$^{-1}$, which are consistent with our $\sigma_{CO (2-0)}$ estimate. Therefore, we used our $\sigma_{CO (2-0)}$ estimate as a surrogate for $\sigma_{\star CaT}$ for NGC 4388. \par}
\label{catvscotable}
\end{center}
\end{table*}

\subsection{The Robustness of Stellar Velocity Dispersion Estimations}

\begin{figure}[hbt!]
   \centering
   	\includegraphics[width=8.5cm]{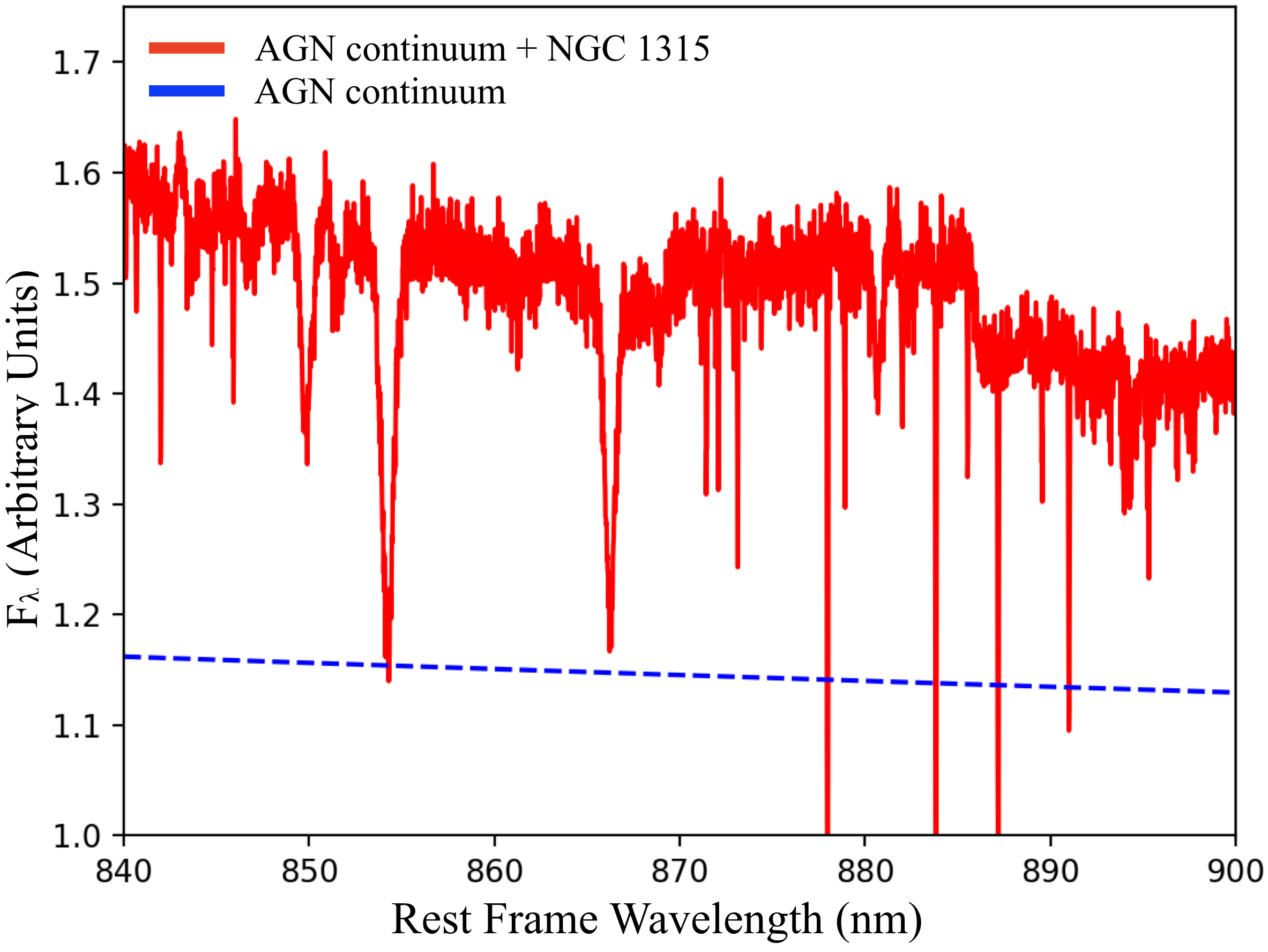}
	\includegraphics[width=8.5cm]{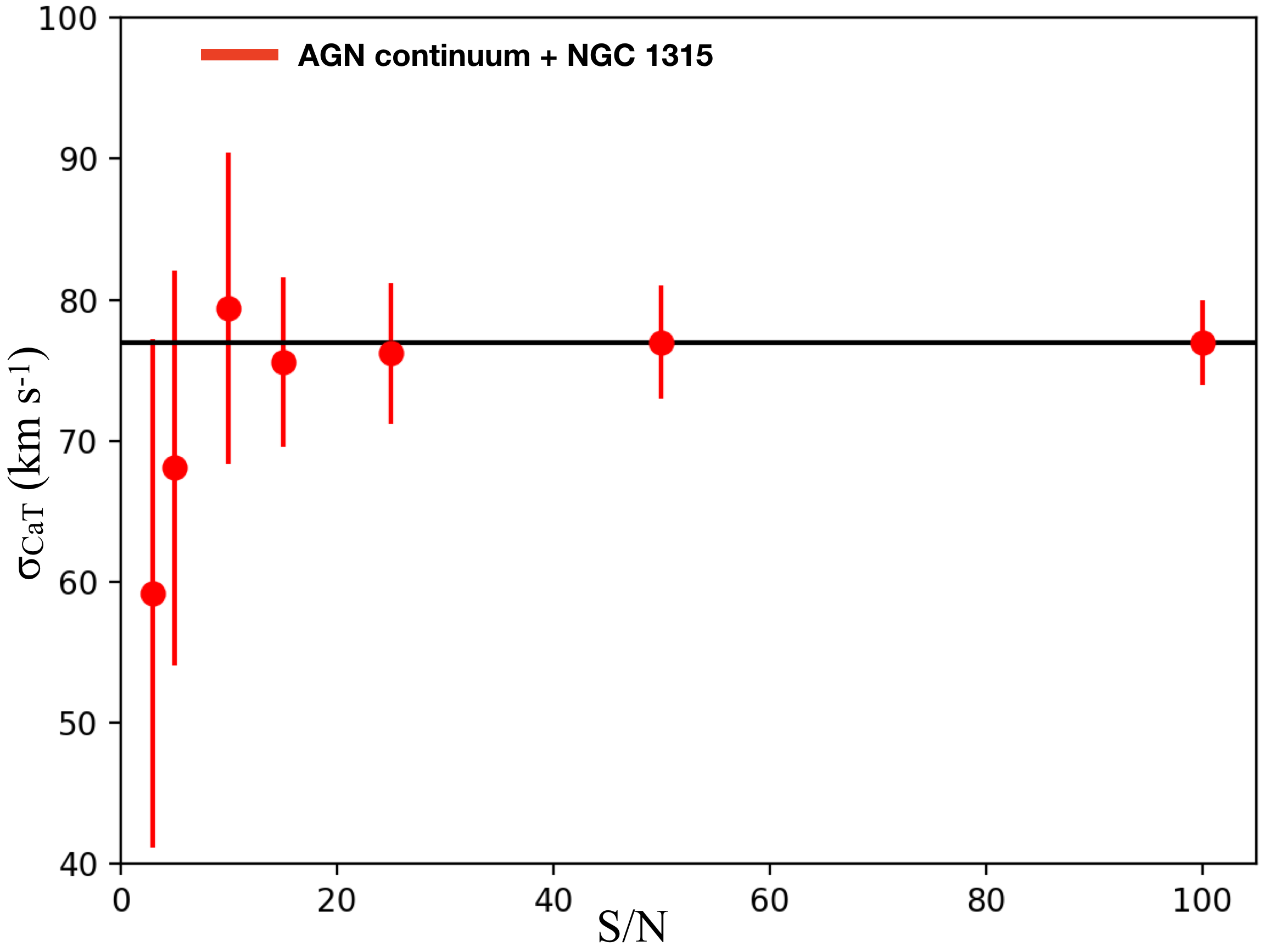}
	\includegraphics[width=8.5cm]{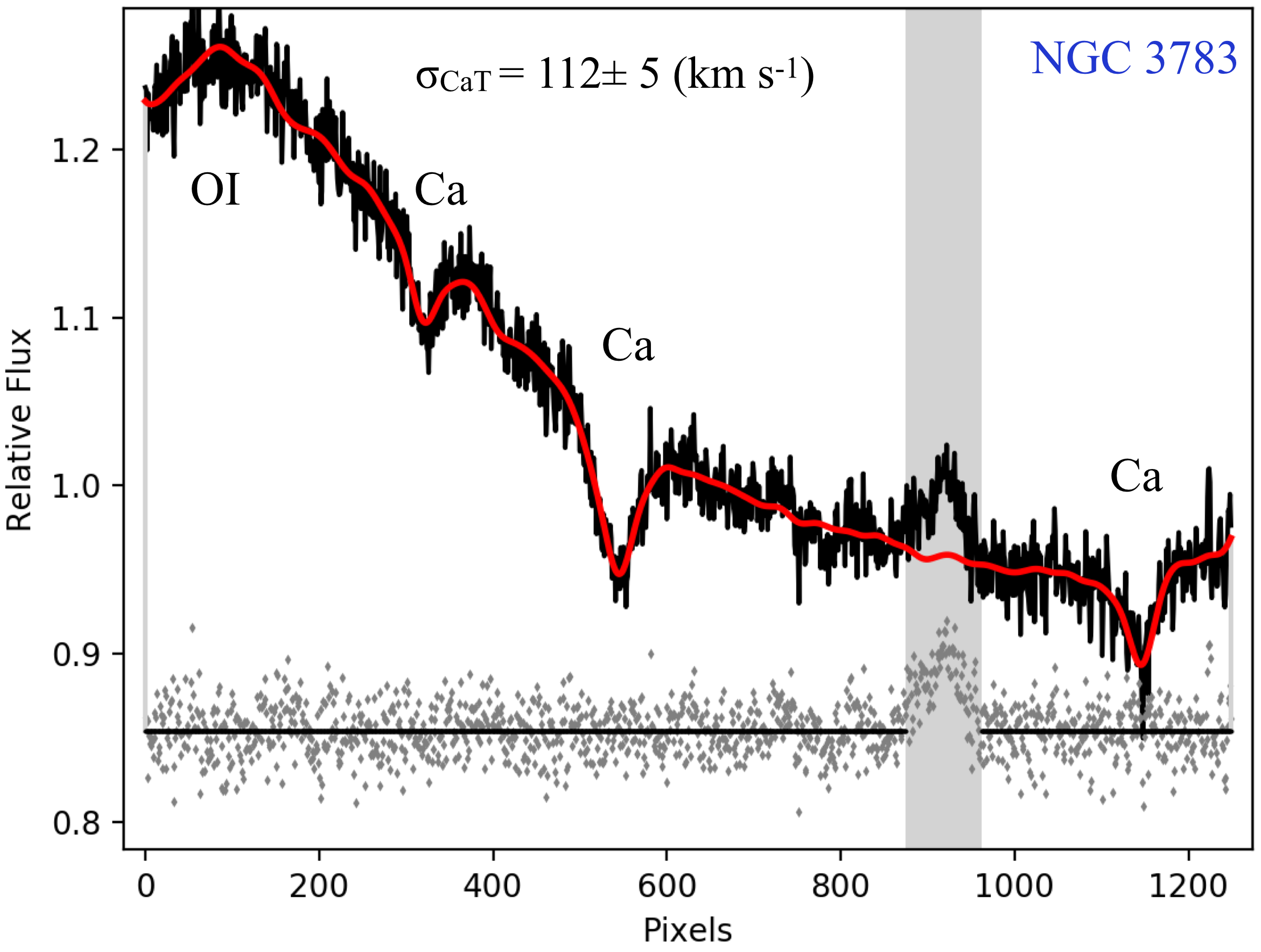}
   \caption{\textbf{Top:} An example of the spectrum from the control galaxy NGC1315, which is combined with the model AGN continuum and the assumed AGN continuum are presented as the red and the blue, respectively.  \textbf{Middle:} The stellar velocity dispersion estimates relative to the signal to noise ratio of the AGN continuum for NGC 1315 (red). The solid black line represents the stellar velocity dispersion estimate from the X-shooter spectrum, which has a S/N $\sim$ 44 per pixel. \textbf{Bottom:}  An example of ppx fit for NGC 3783. Position of the O I emission line and the CaT absorption lines are demonstrated in the plot for visual aid. The gray masked feature represents the Fe II emission line at 8616 \AA. }
              \label{synthetic}%
    \end{figure}

Recent studies report that stellar velocity dispersion estimates can be affected by AGN contamination \citep{GreeneHo2006b,Harris,Woo2013,Batiste2}. Firstly, we address the question of whether the AGN continuum affects the stellar velocity dispersion estimations. In optical bands, the AGN continuum behaves like a power-law function \citep{Oke}, and can be defined as $f_{\lambda}$ $\propto$ $\lambda^{-(\alpha_{v}+2)}$, where $\alpha_{v}$ is the arithmetic mean of the power-law index. We adopt $\alpha_{v}$ = -2.45 \citep{Berk} to model a synthetic AGN continuum. First, we select an inactive control galaxy (NGC 1315) from the LLAMA sample; the stellar velocity dispersion of  this galaxy is estimated as $\sigma_{\star}$ =  77 $\pm{5}$ km s$^{-1}$ using pPXF. Then, the synthetic AGN continuum was combined with the NGC 1315 spectrum. As expected, the AGN continuum has no direct effect on the $\sigma_{\star}$ estimations for any reasonable AGN continuum level (< 70\%), if the continuum is modelled using an adequate number of additive polynomials. In the top panel of Fig. \ref{synthetic}, we present a synthetic AGN spectrum which consists of the spectrum of the inactive galaxy NGC 1315 (shown as red line) and a fairly strong ($\sim$ 70\%) model AGN continuum (blue line). Our active galaxies typically show a much smaller AGN contribution than 70\% at the CaT, which is why this serves as a good test for our fitting accuracy.

On the other hand, the continuum level cannot be estimated accurately, if the spectrum is noisy. To test this, we applied a Monte-Carlo approach to generate noise for every pixel of the synthetic AGN spectra. In this approach, a normal distribution of numbers are allowed to vary within a specified range, and the test was repeated 10$^{4}$ times to obtain the mean distribution of each noise level (S/N: 3, 5, 10, 15, 25, 50, 100). For each S/N level, we fit the data 10$^{2}$ times using pPXF. The stellar velocity dispersion estimates are obtained from the mean of the Gaussian distribution of resulting $\sigma_{\star}$ values for each S/N. In Fig. \ref{synthetic} (middle), we present the comparison between S/N and $\sigma_{\star}$ estimates. By considering this result, one can achieve reliable $\sigma_{\star}$ estimations using data with high S/N ($>$ 15). We confirm that S/N is one of the most important factors, leading to an uncertainty of up to 20\% for a S/N $\lesssim$ 5, which needs to be included into the total uncertainty of $\sigma_{\star}$. We note that our sample of AGNs are observed with S/N $>$ 40; therefore, our calculations are not affected by this issue.

Moreover, AGN emission lines can also affect $\sigma_{\star}$ estimations. The broad O I (8446 $\AA$) emission line, which is detected for some of the AGNs in our sample, is a good example of this (see the bottom Fig. \ref{synthetic}). Correspondingly, we modelled an extremely broad O I 8446 $\AA$ line using a Gaussian model ($\sigma_{OI}$ $\sim$ 2500 km s$^{-1}$), which is added to the synthetic AGN spectrum. By fitting spectra around the CaT regime with different noise levels, we find evidence that the broad O I 8446 $\AA$ emission line can cause inaccurate stellar velocity dispersion estimations of up to 15\%. Since the existence of a broad emission line affects the continuum level determination, such AGNs with broad O I 8446 $\AA$ have been treated specially by masking the part of the spectrum that is affected by the emission line. In a few cases, this can cut off the first CaT line (8498 \AA), but we report that this does not affect the determination of the stellar velocity dispersion.

For disk galaxies, the galaxy rotation makes an important contribution to the measured stellar velocity dispersion from a larger aperture.. The rotational dynamics of spiral galaxies are characterized by galaxy's total luminosity, line-of-sight and maximum rotation velocities and the inclination angle of the disk \citep{TuFi}. Since the LLAMA AGN sample is dominated by spiral galaxies, the galaxy rotation is another effect that may affect the stellar velocity dispersion estimates. By using the velocity-shifted SINFONI data cubes from Shimizu et al. in preparation, we obtained an average inclination-corrected rotational velocity for the LLAMA sample.The contribution from the rotational effects will be further discussed in Section \ref{sec:rotext}.

\subsection{The Robustness of Broad-Line Based $M_{BH}$ estimates}{\label{sec:robo}}

\begin{figure}[hbt!]
   \centering
   	\includegraphics[width=9.0cm]{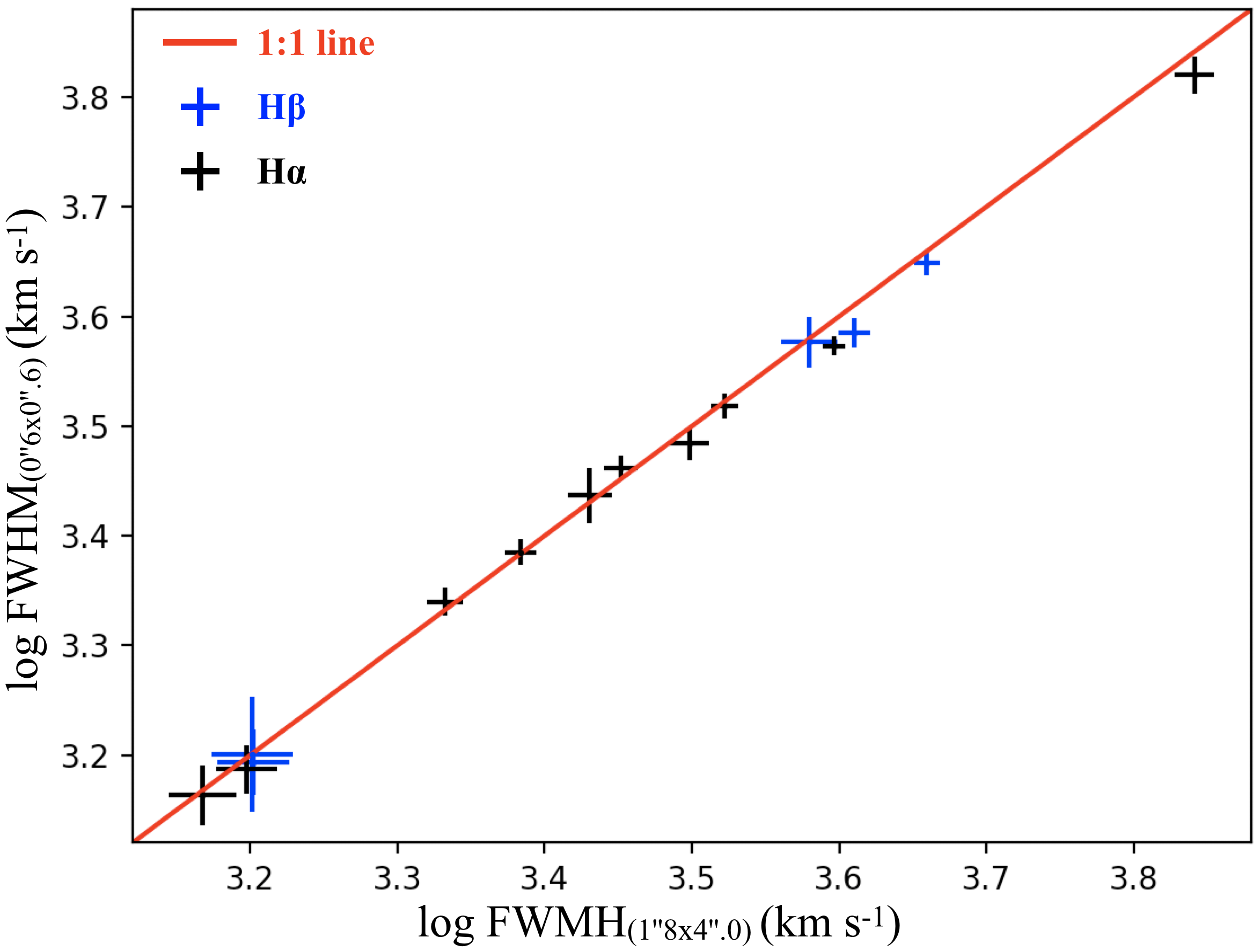}
   \caption{The resulting FWHM comparisons for the small (0$\arcsec$.6$\times$0$\arcsec$.6) and the big aperture (1$\arcsec$.8$\times$4$\arcsec$) for our sample. Note: the black marker represents the resulting estimates from the H$\alpha$, whereas the blue is obtained results from H$\beta$. The black solid line is the 1:1 line. }
              \label{histo}
    \end{figure}

We investigated the broad-line emission of our sample of type 1 AGNs using two different apertures: 0$\arcsec$.6 $\times$ 0$\arcsec$.6 (the central region) and 1$\arcsec$.8 $\times$ 4$\arcsec$ (the FOV of X-SHOOTER data). For each AGN, we fit H$\alpha$ and H$\beta$ emission lines with the same number of the Gaussian curves for each aperture. In Fig. \ref{histo}, we present FWHM comparisons between the central region and the FOV. The broad line FWHM estimates are found to differ up to 5\% due to aperture choice. This difference can be related to the observational seeing or the narrow line contamination. Since we cannot detect the entire BLR gas, this is a systematic error of FWHM estimates and should be added to total uncertainty budget of FWHM estimates.

\begin{figure}[hbt!]
   \centering
	\includegraphics[width=9.0cm]{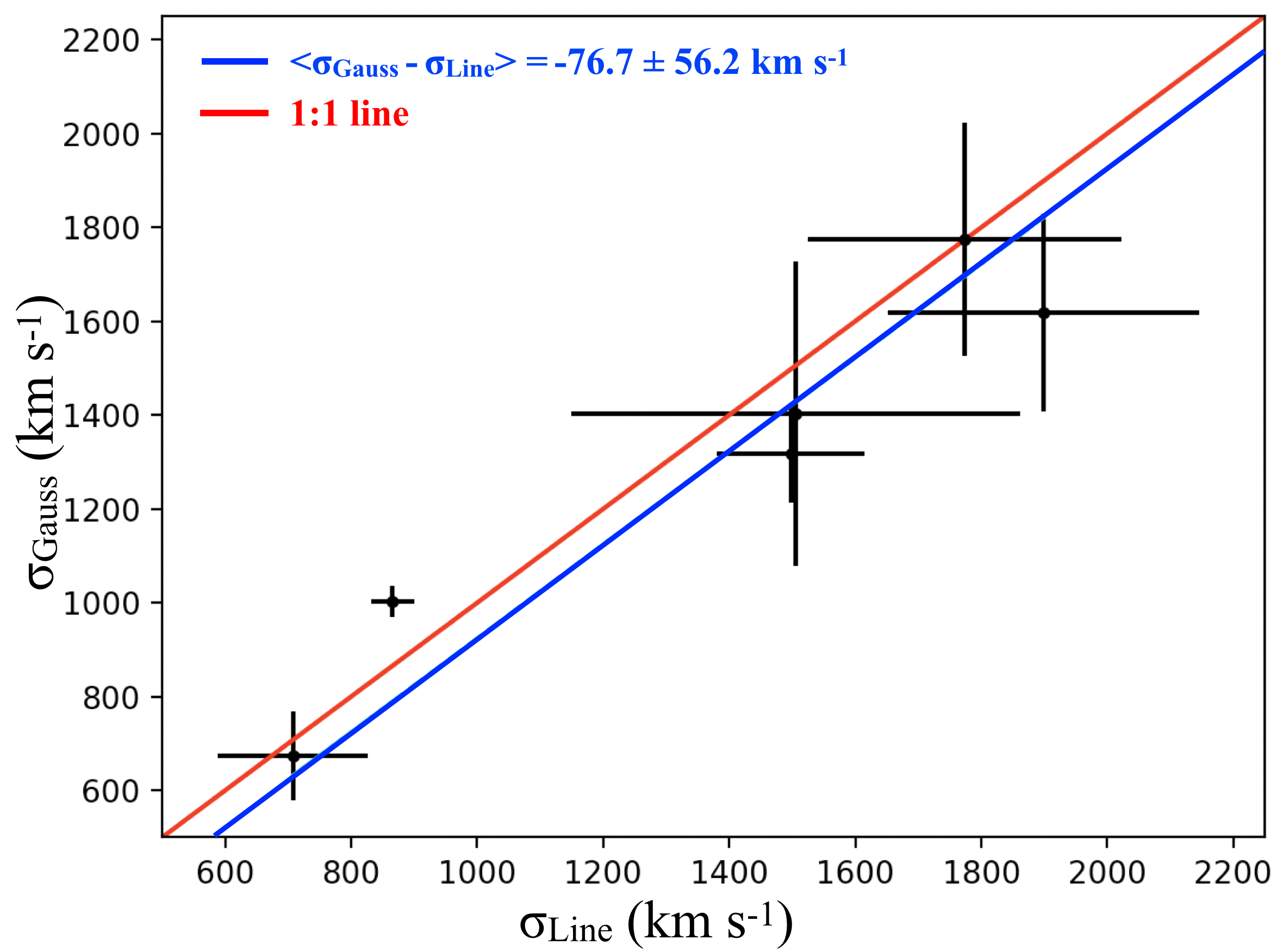}
   \caption{The comparison between $\sigma_{Line}$, which is obtained from the Equation \ref{petersoneq} and $\sigma_{Gauss}$, which is obtained from the line width of Gaussian model. }
              \label{sigmalinevsmodel}%
    \end{figure}

The emission line width of a broad-line can be obtained either from the FWHM or line dispersion ($\sigma_{Line}$). A typical AGN emission line profile can be described by a single Gaussian profile, and FWHM/$\sigma_{Line}$ has a fixed ratio of 2$\sqrt{2 \ln2}$ $\approx$ 2.355 in the Gaussian profile. However, some of AGN emission line widths can only be modelled with multiple Gaussians. In this case, the FWHM needs to be estimated from the combined Gaussian models, and the ratio between FWHM and $\sigma_{Line}$ can vary \citep{Peterson2004,Peterson2011}. \citet{Peterson2018} argue that $\sigma_{Line}$-based $M_{BH}$ estimations are more accurate than FWHM-based ones for H$\beta$, if an AGN emission has an irregular line profile. For the multiple-peaked emission line profiles, the irregular kurtosis can be either positive or negative, and it  can affect the accuracy of emission line estimations. These authors also note that $\sigma_{Line}$-based estimations are less sensitive to the contribution from extended line wings. The $\sigma_{Line}$ can be estimated from the second moment of the emission line profile P ($\lambda$):  

 \begin{equation}
\sigma_{Line} = \left[\int_{}^{}  \frac{(\lambda - \lambda_{0})^{2} P(\lambda) d\lambda}{\int_{}^{} P (\lambda) d\lambda}\right]^{1/2} ,
\label{petersoneq}
\end{equation}
where $\lambda_{0}$ is the center of emission line profile. In Fig. \ref{sigmalinevsmodel}, we compare the $\sigma_{Line}$ obtained from the Equation \ref{petersoneq} and $\sigma_{Model}$ obtained from its ratio with the FWHM (FWHM/$\sigma_{Model}$ $\approx$ 2.355) for the Gaussian profile. We find a slight difference (an offset of 76.7 $\pm{56.2}$ km s$^{-1}$) between the two estimates for our H$\beta$-based investigations. We note that this difference affects our $M_{BH}$ estimates by $\sim$ 0.1 dex. This result is consistent with \citet{Peterson2018}, therefore, we also suggest using $\sigma_{Line}$ in H$\beta$-based $M_{BH}$ investigations. 

\subsection{Black hole masses and the systematical uncertainties}

The $M_{BH}$ values for our sample of type 1 AGNs are presented in Table~\ref{Results}. They are in the range of 6.34 $\leq$ $\log{M_{BH}}$ $\leq$ 7.75 M$_{\odot}$ for H$\alpha$. We note that the average black hole mass of inactive galaxies in the relation by \citet{KoHo} is substantially higher, possibly indicating that our sample of AGNs did not yet go through a major merger phase \citep{Wandel1999}.

Black hole mass uncertainties are determined from the bootstrapping analysis. In this approach, we used all uncertainties from parameters we used, such as uncertainties from single-epoch calibration parameters, $f$ factor, FWHM and luminosity. First, we generated 10$^{8}$ random numbers from a normal distribution for each parameter. Then, these numbers are added to all parameters of $M_{BH}$ estimations. Finally, using the Gaussian distribution of obtained 10$^{8}$ $M_{BH}$ values, we measured black hole mass uncertainties within the 1 $\sigma$ confidence level.

However. single-epoch based $M_{BH}$ estimations have been reported to have a systematical uncertainty, which is reported as a lower limit of 0.40 dex  by \citet[e.g.,][]{Pancoast}. The uncertainty of $f$ factor introduces an uncertainty of 0.12 dex \citep{Woo2015}, which is obtained from the comparison of the $M_{BH}$ - $\sigma_{\star}$ relation between the RM AGNs and inactive galaxies. Second uncertainty is the intrinsic scatter of BLR radius-luminosity relation, which is reported as 0.13 dex for reliable estimates \citep{Bentz13}. Third, we variability in luminosity and line-width bring a 0.1 dex uncertainty \citep{Parkb}. Last, we adopt an uncertainty of 0.15 dex, which is assumed to come from redshift-independent distance measurements. Third, the uncertainty in distance measurement also plays a big role  Correspondingly, the total uncertainty of $M_{BH}$ estimates can be 0.3 - 0.4 dex.

\begin{table*}
\begin{center}
\caption{The spectral results of the LLAMA AGN sample. Columns are from left to right as follows: 1) Object names,  2) The extinction in the BLR are taken from \citet{Allan}. $\bullet$: the extinction in the BLR are estimated in this study using the same method provided by \citet{Allan}, 3) FWHMs of H$\alpha$ emission line, 4) FWHMs of H$\beta$ (or Pa$\beta$) emission line, 5) Black hole masses estimated from the following methods (for the different sections of the table) from top to bottom: 5A) the H$\alpha$ - FWHM (extinction-corrected), 5B) Megamaser disk, 5C) the LLAMA $M_{BH}$ - $\sigma_{\star}$, 6) The extinction-corrected black hole masses estimated from the H$\beta$ - $\sigma_{Line}$ (or Pa$\beta$ - FWHM), 7) Accretion rates, 8) Eddington ratios, 9) The offset from the $M_{BH}$ - $\sigma_{\star}$ relation of KH13 for given $\sigma_{\star}$.
The first section of the table lists LLAMA Seyfert (Sy) 1-1.5 AGNs, the second section lists the three LLAMA Seyfert 1.8 and 1.9 AGNs, the third section lists the two LLAMA Seyfert 2 galaxies for which megamaser observations are available and the fourth section lists the rest LLAMA Seyfert 2 galaxies for which $M_{BH}$ are estimated from the LLAMA $M_{BH}$ - $\sigma_{\star}$ relation.} 
\begin{tabular}{ccccccccc} 
\hline
\hline
1 & 2 & 3& 4& 5A & 6 & 7 & 8 & 9 \\ 
\hline
Object &	$A_{V}$ (BLR)   & 	FWHM (H$\alpha$)      & FWHM (H$\beta$)  &  $M_{BH}$ (FWHM)  & $M_{BH}$ ($\sigma_{Line}$)  &	$\dot{M}$	&	$\lambda_{Edd}$  & $\Delta{M}$ 	\\

	&	 &1$\arcsec$.8$\times$4$\arcsec$	&	 1$\arcsec$.8$\times$4$\arcsec$	& (H$\alpha$) 	& (H$\beta$) 		& 	&  &\\
	
	&   mag    &    km s$^{-1}$        &	 km s$^{-1}$   &   10$^{6}$    M$_{\odot}$     	 &    10$^{6}$    M$_{\odot}$    	&	10$^{-2}$ M$_{\odot}$ year$^{-1}$	&	 & dex	\\

\hline
MCG-05-14-12 &	0.0$\pm{0.2}$	& 1836.0$\pm{119}$	&	2019.1$\pm{167}$	&	2.29$\pm{0.68}$	&	2.30$\pm{1.38}$ & 6.12  &	0.120 &  0.23   \\

MCG-06-30-15 &	2.8$\pm{0.4}$	&	1456.8$\pm{122}$	&	1588.4$\pm{198}$	&	7.38$\pm{1.98}$   &	5.97$\pm{1.92}$ 	&	12.0	&	0.073  &  -0.06 \\

NGC 3783 &	0.1$\pm{0.2}$	&	3002.3$\pm{196}$	&	3102.3$\pm{312}$	&	11.2$\pm{3.61}$ 	&	10.1$\pm{4.72}$	&	67.3	&	0.272 &     -0.27 \\ 

NGC 4235 &	1.5$\pm{0.5}$	&	6611.1$\pm{461}$	&	-	&	55.8$\pm{15.9}$ 	&	- 	& 	5.96	&	0.005 &    0.27	\\

NGC 4593 &	0.0$\pm{0.1}$	&	3741.8$\pm{213}$	&	4179.4$\pm{294}$	&	12.4$\pm{3.91}$   &	10.0$\pm{4.38}$	&	25.3	&	0.091   &  -0.50  \\

NGC 6814 &	0.4$\pm{0.4}$	&	3299.3$\pm{191}$	&	3771.0$\pm{279}$	&	11.6$\pm{3.67}$   &	13.4$\pm{4.12}$	& 	7.92	&	0.031 &   -0.16 \\

NGC 7213 &	0.0$\pm{0.3}$ ($\bullet$)	&	2732.8$\pm{264}$	&	3302.0$\pm{701}$	&	6.46$\pm{2.01}$   &	6.45$\pm{2.47}$  &		4.05	&	 0.028  & -1.46 \\
\hline
Object &	$A_{V}$ (BLR)   & 	FWHM (H$\alpha$)      & FWHM (Pa$\beta$)  &  $M_{BH}$ (FWHM)  & $M_{BH}$ (FWHM)  &	$\dot{M}$	&	$\lambda_{Edd}$  & $\Delta{M}$ 	\\

	&	 &1$\arcsec$.8$\times$4$\arcsec$	&	 1$\arcsec$.8$\times$4$\arcsec$	& (H$\alpha$) 	& (Pa$\beta$) 		& 	&  &\\
	
	&   mag    &    km s$^{-1}$        &	 km s$^{-1}$   &   10$^{6}$    M$_{\odot}$     	 &    10$^{6}$    M$_{\odot}$    	&	10$^{-2}$ M$_{\odot}$ year$^{-1}$	&	 & dex	\\
\hline
MCG-05-23-16 &	4.2$\pm{0.9}$	&	2186.1$\pm{166}$	&	1935.4$\pm{196}$	&	27.1$\pm{8.74}$  &	25.3$\pm{8.84}$    &		54.8	&	0.091  &	0.17 	 \\

NGC 1365 &	4.4$\pm{0.9}$	&	2406.1$\pm{180}$	&	1872.0$\pm{352}$	&	19.7$\pm{5.77}$   &	13.8$\pm{5.96}$ 	&	5.38	&	0.012   &	-0.48 \\

NGC 2992 &	4.5$\pm{0.8}$	&	2085.5$\pm{189}$	&	2180.9$\pm{260}$	&	22.8$\pm{6.74}$  &	26.4$\pm{9.02}$ 	&	4.38	&	0.004   &	-0.10 \\

\hline
1 & 2 & 3& 4& 5B & 6 & 7 & 8 & 9 \\ 

\hline
Object	    & 		 &   & 	   & $M_{BH}$  	&	 & $\dot{M}$		 &	$\lambda_{Edd}$	&  $\Delta{M}$ 	\\

	&          &       &	   &      (Megamaser)  	 &  	&    	\\
	&	&	&		&	10$^{6}$    M$_{\odot}$	&		& 10$^{-2}$ M$_{\odot}$ year$^{-1}$	&  & dex \\
\hline

NGC 4388	&	-	&	-	&	-	&	8.40$\pm{0.2}^{\bullet}$ &	-	&	91.8	&	0.489	&  -0.24	\\
NGC 5728	&	-	&	-	&	-	&	23.0$\pm{2.3}^{\star}$ &		-	&	38.2	& 	0.074 	&  -0.42	\\

\hline
1 & 2 & 3& 4& 5C & 6 & 7 & 8 & 9 \\ 
\hline
Object	    & 		&    & 	   & $M_{BH}$  	&	 & $\dot{M}$		 &	$\lambda_{Edd}$		 & $\Delta{M}$  \\

	&           &      &	   &      ($M_{BH}$ - $\sigma_{\star}$)  	 &  	&    	&	\\
	&	&	&		&	10$^{6}$    M$_{\odot}$	&		& 10$^{-2}$ M$_{\odot}$ year$^{-1}$	&	& dex\\
\hline
ESO 137-G034 &	-	&	-	&	-	&	21.5$\pm{15.8}$ &	-	&	8.12	&	0.017	&	0.20	 \\
ESO 021-G004 &	-	&	-	&	-	&	52.1$\pm{38.4}$ &	-	&	6.96	&	0.006	&	0.08	 \\
NGC 2110 &	-	&	-	&	-	&	150$\pm{110}$ &	-	&	76.6	&	0.023	 &	-0.05	\\
NGC 3081 &	-	&	-	&	-	&	36.6$\pm{26.9}$ &	-	&	31.9	&	0.039	 &	0.13	\\
NGC 5128 &	-	&	-	&	-	&	66.3$\pm{48.9}$	&	-	&	 15.9   &   0.011    &  0.05 \\
NGC 5506 & -    &   -   &   -   &   22.4$\pm{17.2}$ &   -   &  32.7    &   0.065    &  0.19 \\
NGC 7172 &	-	&	-	&	-	&	53.4$\pm{39.3}$ &	-	&	34.4	&	0.029	 &	0.08	\\
NGC 7582 &	-	&	-	&	-	&	30.5$\pm{22.4}$ &	-	&	31.9	&	0.047	 &	0.15 \\
\hline

\end{tabular}

 {\raggedright Notes: NGC 5128 has also black hole mass estimates from the other methods: $M_{BH}$ 4.5$^{+1.7}_{-1.0}$ 10$^{7}$ M$_{\odot}$ from H$_{2}$ gas kinematics by \citet{Neumayer}, $M_{BH}$ = 5.5$\pm{3.0}$ 10$^{7}$ M$_{\odot}$ from stellar kinematics by \citet{Cap2009}. We emphasize that our $M_{BH}$ estimate for NGC 5128 is consistent with these results. The $\sigma_{CO (2-0)}$ is used to obtain $M_{BH}$ for NGC 5506 due to the absence of $\sigma_{CaT}$. The $M_{BH}$ estimates from $\star$: \citep{Braatz}, $\bullet$: \citep{Greene2016}. We note that we adopted 10\% uncertainty for the $M_{BH}$ of NGC 5728 due to absence of uncertainty in the related study. We adopted the $A_{V}$ (BLR) estimates obtained from the He II line ratios for MCG-05-23-16, NGC 1365 and NGC 2992 reported by \citet{Allan}, since this method gives better results for Sy 1.8 and Sy 1.9 galaxies.}

\label{Results}
\end{center}
\end{table*}

   \begin{figure*}
   \centering
   \includegraphics[width=6.05cm]{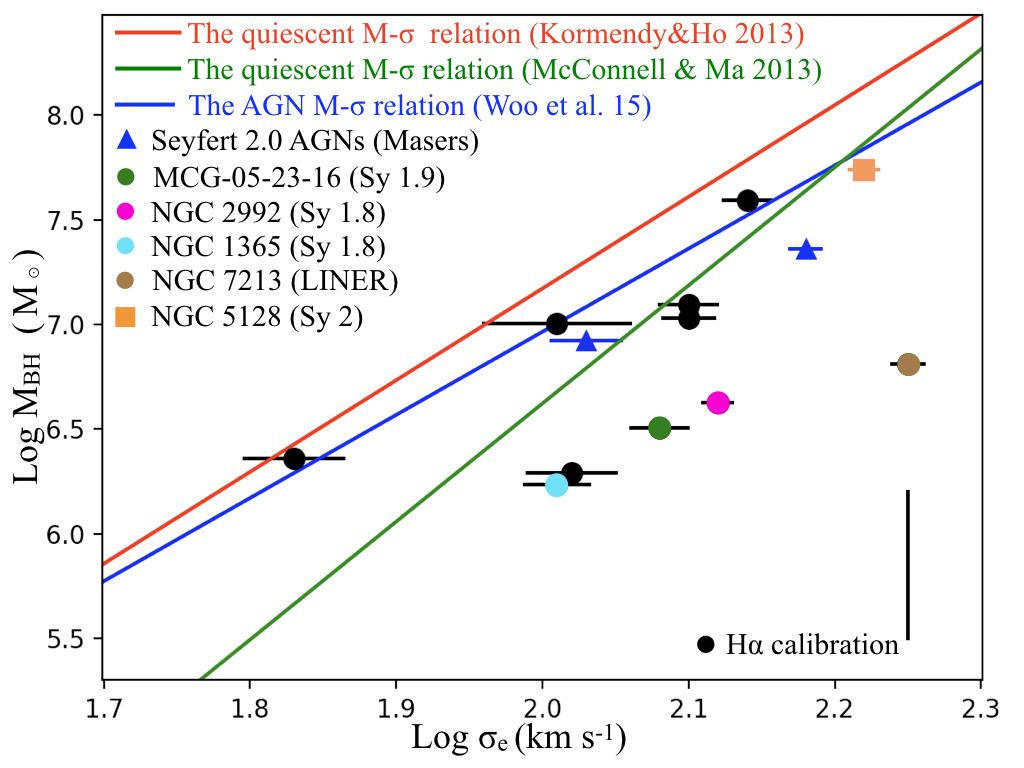}
   \includegraphics[width=6.05cm]{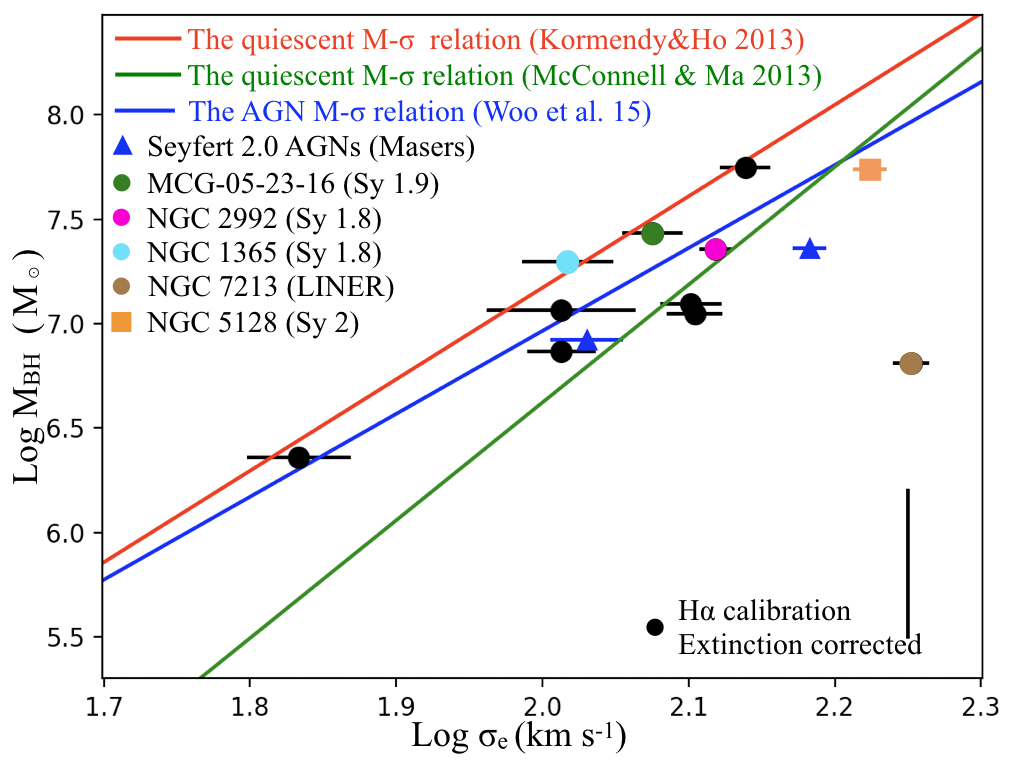}
   \includegraphics[width=6.05cm]{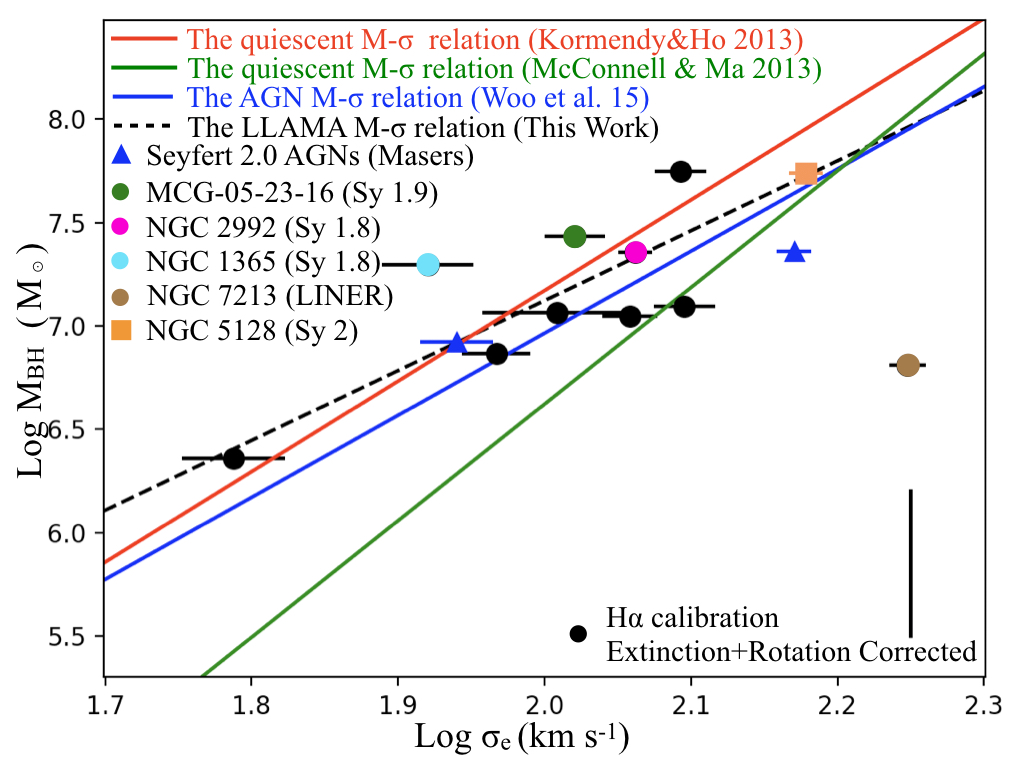}
      \caption{$\textbf{Left:}$ The $M_{BH}$ - $\sigma_{\star}$ relation of our sample of galaxies, where $M_{BH}$ values are estimated using the H$\alpha$ based calibration. The $M_{BH}$ - $\sigma_{\star}$ relation of KH13, MM13 and W15 are presented as red, green and blue solid lines, respectively. Sy 1.8, Sy 1.9, Sy 2 and LINER galaxies are presented in different color for visual aid. We additionally present the location of the two LLAMA Seyfert 2 (NGC 4388 and NGC 5728) galaxies that have Megamasers $M_{BH}$ estimates as blue triangles. In addition, we present the $M_{BH}$ estimates of NGC 5128 obtained from stellar kinematics as an orange box \citep{Cap2009}. Finally, the average uncertainties on the black hole mass estimates of the LLAMA AGNs ($\sim$ 0.40 dex) are presented as a vertical black line in the legend, in order to avoid confusion of data points. $\textbf{Middle:}$ The $M_{BH}$ - $\sigma_{\star}$ relation of our sample of galaxies, where $M_{BH}$ values are estimated using the extinction corrected fluxes and the H$\alpha$ calibration. $\textbf{Right:}$ $M_{BH}$ - $\sigma_{\star}$ relation of our sample of galaxies, where the H$\alpha$ $M_{BH}$ values are presented as the extinction and rotation corrected. The LLAMA $M_{BH}$ - $\sigma_{\star}$ relation is presented as a black dashed line. \label{M_sigma}}
   \end{figure*}

\subsection{Accretion Rates}{\label{sec:accretion}}

Many properties of the AGN \citep[e.g. the torus phenomenology:][]{Wada} are expected to depend on the Eddington ratio of the ``central engine''. One of the main drivers of our study is to provide the Eddington ratio for the whole LLAMA sample.

We compute the accretion rates following Eqs.~\ref{eq:lbol} and \ref{eq:macc} and find them in the range 0.04 $<$ $\dot{M}$ $<$ 0.92 M$_{\odot}$ yr$^{-1}$ assuming an accretion efficiency of 10\% (see Table \ref{Results}). Using our estimated BH masses, we further calculate the Eddington ratio $\lambda$ following Eq.~\ref{eq:ledd} for all of our AGNs. They are in the regime 0.004 $\leq$ $\lambda$ $\leq$ 0.49. These results indicate that the most LLAMA AGNs are growing at a rate that is well below Eddington, though likely in the radiatively efficient regime via a geometrically thin, optically thick disk \citep{ShSu}.

\subsection{The $M_{BH}$ - $\sigma_{\star}$ Relation of the LLAMA Sample} {\label{sec:llama_m_sigma}}
In Fig. \ref{M_sigma}, we present the $M_{BH}$ - $\sigma_{\star}$ relation for the LLAMA AGN sample adopting the broad-line based single-epoch black hole masses derived using the H$\alpha$ emission line profiles. Using the high signal to noise data, we report 38 stellar velocity dispersion estimates (20 AGNs and 18 inactive galaxies) in total ( Table \ref{catvscotable}), which are derived from the CaT and/or CO (2-0) absorption features. We provide $M_{BH}$ of 10 type 1 AGNs in the LLAMA sample (see Table \ref{Results}). In addition, we adopt a stellar kinematic $M_{BH}$ estimate of NGC 5128 \citep{Cap2009} and two megamaser $M_{BH}$ estimates of  NGC 4388 \citep{Greene2016} and  NGC 5728 \citep{Braatz}. Therefore, we constructed an $M_{BH}$ - $\sigma_{\star}$ relation for thirteen AGNs in the LLAMA sample.

We then performed a linear regression where we allowed both the intercept and the slope to vary. For this fit, we used FITEXY and the extinction-corrected $M_{BH}$ and the rotation-corrected $\sigma_{\star}$ estimates for our sample. We exclude NGC 7213 from this fit since it shows LINER-like properties; also the H $\beta$ fit for this galaxy fails.

The resulting $M_{BH}$ - $\sigma_{\star}$ relation for the LLAMA AGNs is then:

\begin{equation}
\label{eq:llama_m_sigma}
\log{(M_{BH}/M_{\odot})} = 8.14 (\pm 0.20) + 3.38 (\pm 0.65) \log{\left( \frac{\sigma_{\star}}{ 200 \: \: km \: s^{-1}} \right)},
\end{equation}

and the intrinsic scatter of this relation is $\epsilon$ = 0.32$\pm{0.06}$. We note that our slope (3.38$\pm{0.65}$) is smaller than the slope reported by \citet{Woo2015} ($3.97 \pm 0.56$) who included narrow-line Seyfert AGNs in order to extend to lower black hole masses, and consistent with \citet{Woo2013} who found a slope of $3.46 \pm 0.61$. Within the uncertainties of our small sample, our slope is consistent with both of these AGN relations, but not consistent with the slope reported by \citet{KoHo} for more massive, inactive galaxies. This result still shows that the LLAMA sample of AGNs, which is a volume complete sample of the most luminous local AGNs, is representative for the larger AGN population sampled with reverberation mapping in terms of its location and slope on the M-sigma relation.

For reference for future publication, and using the LLAMA $M_{BH}$ - $\sigma_{\star}$ relation (Eq.~\ref{eq:llama_m_sigma}), we estimate $M_{BH}$ values also for our Type 2 AGNs (Table~\ref{Results}).   

\subsection{The LLAMA $M_{BH}$ - $\sigma_{\star}$ Relation versus spheroidal $M_{BH}$ - $\sigma_{\star}$ Relation}{\label{sec:rotext}}

In the left panel of Fig. \ref{M_sigma}, we present the $M_{BH}$ values without extinction-correction and $\sigma_{\star}$ parameter without rotation-correction. We compare the LLAMA $M_{BH}$ - $\sigma_{\star}$ relation with the $M_{BH}$ - $\sigma_{\star}$ relation of KH13, MM13 and the AGN $M_{BH}$ - $\sigma_{\star}$ relation by W15. First, we found a high offset (0.75 dex) from the KH13 relation using these parameters. 

In previous works, some authors concentrated on correcting the broad Balmer fluxes and/or the monochromatic accretion luminosities in various wavelengths (i.e., 1350, 3000, 5100 $\AA$), which are used in single-epoch $M_{BH}$ estimations, using galactic extinction maps \citep[e.g.,][]{VesPet,Denney2009,ShLi,Bentz2016,Kozlow}. In our study, we additionally corrected H$\alpha$ and the continuum fluxes, which are used for deriving black hole masses, using the estimated BLR extinction of LLAMA sample by \citet{Allan}. In the middle panel of Fig. \ref{M_sigma}, we present the $M_{BH}$ - $\sigma_{\star}$ relation obtained using extinction-corrected black hole masses. The extinction correction increased the estimated $M_{BH}$ by a factor of 0.02 - 0.93 dex for our sample, and reduced the average offset from the KH13 relation to 0.38 dex. This result indicates that the extinction in BLR can cause significantly under-estimation of $M_{BH}$, unless it is taken into account.

In an upcoming LLAMA study, Shimizu et al. (in preparation) fit for the spatially resolved stellar kinematics within the SINFONI cubes. The stellar velocity fields were then modelled as an exponential disk. Using the model velocity field, we then shifted the spectra within the original SINFONI cubes such that the stellar velocity is removed. In this way, we can measure a rotation corrected stellar velocity dispersion for the whole SINFONI FOV and compare it to the original one to produce a rotation correction that can be applied to our X-SHOOTER based velocity dispersion. Correspondingly, we obtained a rotation correction factor for our AGNs (see Table \ref{catvscotable}). Therefore, we reduced the obtained stellar velocity dispersion using this rotation correction factor. However, We are still missing SINFONI observations for the following galaxies: MCG-05-14-12, NGC 4235, NGC 5128, and the spatially resolved stellar kinematics for NGC 3783, MCG-06-30-15. For these galaxies, the obtained stellar velocity dispersion estimates are reduced 10\% the average galaxy rotation contribution to $\sigma_{\star}$ for the LLAMA sample (Shimizu et al. in preparation). After the $\sigma_{\star}$ estimates are corrected for galaxy rotation, the LLAMA galaxies are found to agree with $M_{BH}$ - $\sigma_{\star}$ relation of \citet{KoHo}. The average intrinsic scatter of LLAMA sample obtained adopting the slope and intercept of \citet{KoHo} relation is found to be is 0.30 dex, which is consistent with the intrinsic scatter of \citet{KoHo} $M_{BH}$ - $\sigma_{\star}$ relation (see Fig. $\ref{M_sigma}$). This result shows that the rotation can make a significant contribution to stellar velocity dispersion (up to 20 \%), which is consistent with previous investigations \citep[e.g.,][]{Kang,Batiste,Eun}.

We additionally compared our results with the $M_{BH}$ - $\sigma_{\star}$ relation reported by MM13. By adopting a slope of 5.64 reported by MM13, we find an average offset of 0.46 dex for our sample relative to the relation of MM13. However, the majority of our sample (8 out of 10) are found to be above the relation reported by MM13. There are two possible explanations for the discrepancy between our results and MM13; in the MM13 sample, the disk galaxies are not corrected for their rotation component, and their sample includes brightest cluster galaxies, which are located in a different environment than the LLAMA sample.

Even though a few studies in the literature report that pseudo-bulges do not follow the $M_{BH}$ - $\sigma_{\star}$ relation \citep{Greene2010,Korm2011,KoHo}, the pseudo-bulge dominated LLAMA sample follow the $M_{BH}$ - $\sigma_{\star}$ relation of elliptical and spheroidal bulge-dominated galaxies after applying the extinction-correction to our $M_{BH}$ and the rotation-correction to  our $\sigma_{\star}$ estimates. Therefore, we argue that, in order to reduce the offset from the elliptical-dominated $M_{BH}$ - $\sigma_{\star}$ relation, a correction to $M_{BH}$ for the dust extinction (derived via the H $\alpha$ or continuum flux) and a correction of $\sigma_{\star}$ for a rotational component of the disk/bulge must be applied to spiral-dominated local Seyfert AGNs.

\section{CONCLUSIONS} {\label{sec:conclusion}}
In a volume limited complete sample of the most luminous, X-ray selected, local Sy 1 AGNs, comprising the LLAMA sample, we examine the spatially resolved stellar kinematics and the properties of the broad emission lines using medium spectral resolution (R $\sim$ 8000) X-SHOOTER data. We additionally compare our results with SINFONI data which extend our analysis to the H+K bands. We itemize our main results below: 
\begin{itemize}
\item The stellar velocity dispersions obtained via the CaT at $\sim$ 8500 $\AA$ is in the range 73 $\leq$ $\sigma_{\star CaT}$ $\leq$ 227 km s$^{-1}$.  We also estimate the stellar velocity dispersions from the near-infrared stellar CO (2-0) absorption feature for a sub-set of galaxies using SINFONI data and find them to be in the range of 101 $\leq$ $\sigma_{\star CO (2-0)}$ $\leq$ 231 km s$^{-1}$. For the galaxies for which we have both observations, the two stellar velocity dispersion measurements are in good agreement. On average, the stellar velocity dispersion derived from the near-IR CO feature is higher by $\sim$ 3.69 $\pm{0.93}$ km s$^{-1}$ than the value derived from the CaT.

\item We apply Monte-Carlo-like simulations to test the robustness of stellar velocity dispersion estimations for bright AGNs in which we test the effects of signal/noise and of the AGN continuum and emission lines. We conclude that stellar velocity dispersions can be obtained accurately for AGNs if the data have a S/N $>$ 15.

\item The SMBH masses of the LLAMA sample of Seyfert~1 AGNs are derived from single-epoch broad-line based black hole mass estimates, which result in 6.34 $\leq$ $\log{M_{BH}}$ $\leq$ 7.75 M$_{\odot}$ using the H$\alpha$ line width and flux as a tracer of black hole mass. We additionally estimate H$\beta$ emission line black hole masses for our sample of AGNs. When the H$\beta$ was not available, we used the Pa$\beta$ emission line instead (see Table $\ref{Results}$).

\item The Eddington ratio and accretion rates of the LLAMA sample are found to be within 0.004 $\leq$ $\lambda$ $\leq$ 0.49 and 0.04 $<$ $\dot{M}$ $<$ 0.92 M$_{\odot}$ yr$^{-1}$, respectively. The median for Type 1 and Type 2 is $\sim$ 0.08 less than expected of Seyfert galaxies (10\%), but perhaps consistent with the selection method (hard X-ray).

\item The best fitting parameters for the LLAMA $M_{BH}$ - $\sigma_{\star}$ relation are $\alpha$ = 8.14 $\pm{0.20}$, $\beta$ = 3.38 $\pm{0.65}$, $\epsilon$ = 0.32 $\pm{0.06}$. Within our uncertainties, the LLAMA AGN sample is consistent with the $M_{BH}$ - $\sigma_{\star}$ relations reported by \citet{Woo2013,Woo2015} in terms of slope. The average intrinsic scatter of LLAMA sample around the \citet{KoHo} $M_{BH}$ - $\sigma_{\star}$ relation is found to be 0.30 dex. This intrinsic scatter is consistent with with the intrinsic scatter of \citet{KoHo} $M_{BH}$ - $\sigma_{\star}$ relation. Correspondingly, we report that the pseudo-bulge dominated LLAMA AGNs are now on the $M_{BH}$ - $\sigma_{\star}$ relation reported by \citet{KoHo}. (see the right panel of Fig. $\ref{M_sigma}$).

\item Using the $M_{BH}$ - $\sigma_{\star}$ relation of the LLAMA AGNs with single-epoch RM or maser black hole masses, we infer black hole masses for the other LLAMA Seyfert 2 AGNs as well as the inactive galaxies in the sample.

\item We argue that, in order to reduce the offset from the elliptical-dominated $M_{BH}$ - $\sigma_{\star}$ relation, a correction to $M_{BH}$ for the dust extinction (derived via the H $\alpha$ or continuum flux) and a correction of $\sigma_{\star}$ for a rotational component of the disk/bulge must be applied to spiral-dominated local Seyfert AGNs.

\item Our main finding implies that the $M_{BH}$ - $\sigma_{\star}$ relation could be same for both elliptical and pseudo-bulge hosting galaxies. Correspondingly, we encourage further investigations with a larger sample.

\end{itemize} 

\begin{acknowledgements} We would like to thank the anonymous referee for the comments and suggestions. TC would like to thank Hojin Cho, Stefano Marchesi, Federica Ricci, Julián Mejía-Restrepo, Murillo Marinello Assis de Oliveira, Swayamtrupta Panda, Nathen Nguyen, Kimberly Emig, Kirsty Butler, Fraser Evans, Dirk van Dam and Walter Jaffe for very useful discussions. T.C. is partly supported by a DFG grant within the SPP 1573 "Physics of the interstellar medium". R.A.R thanks partial support from Conselho Nacional de Desenvolvimento Cient\'ifico e Tecnol\'ogico and  Funda\c c\~ao de Amparo \`a pesquisa do Estado do RS. S.V. acknowledges support from a Raymond and Beverley Sackler Distinguished Visitor Fellowship and thanks the host institute, the Institute of Astronomy, where this work was concluded. S.V. also acknowledges support by the Science and Technology Facilities Council (STFC) and by the Kavli Institute for Cosmology, Cambridge. S.V. acknowledges support from a Raymond and Beverley Sackler Distinguished Visitor Fellowship and thanks the host institute, the Institute of Astronomy, where this work was concluded. S.V. also acknowledges support by the Science and Technology Facilities Council (STFC) and by the Kavli Institute for Cosmology, Cambridge. VNB gratefully acknowledge assistance from a National Science Foundation (NSF) Research at Undergraduate Institutions (RUI) grant AST-1909297. Note that findings and conclusions do not necessarily represent views of the NSF.

\end{acknowledgements}


\begin{appendix}

\section{Broad-Line Fittings}

\begin{figure*}
\centering
\includegraphics[width=9.0cm]{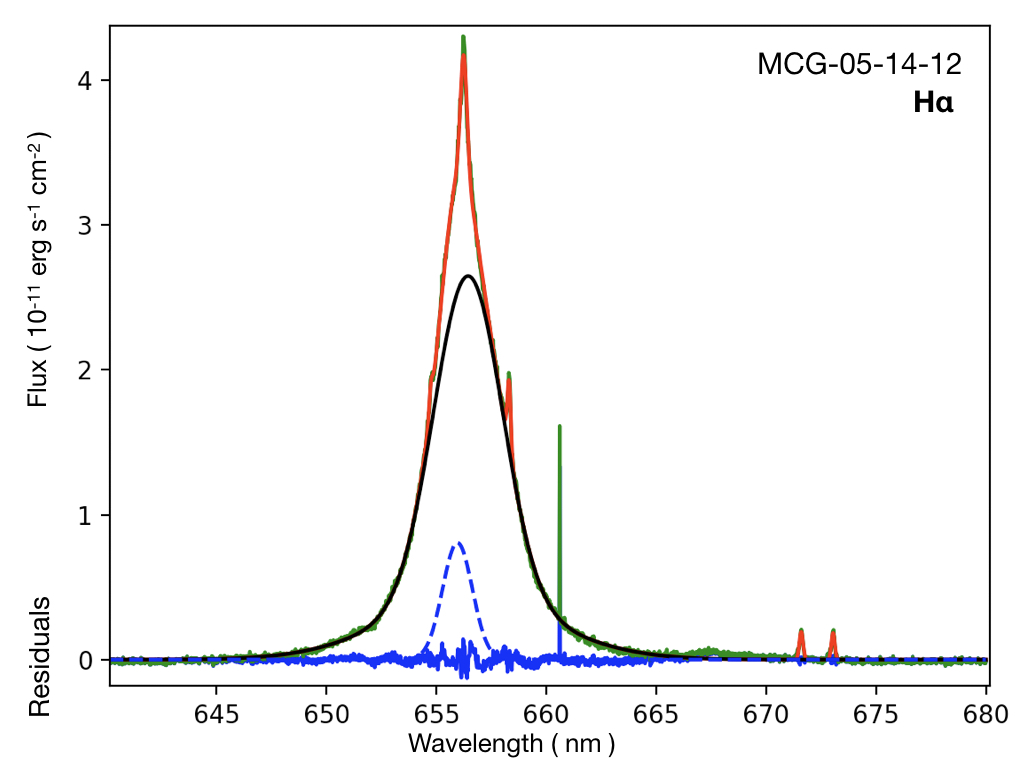}
\includegraphics[width=9.0cm]{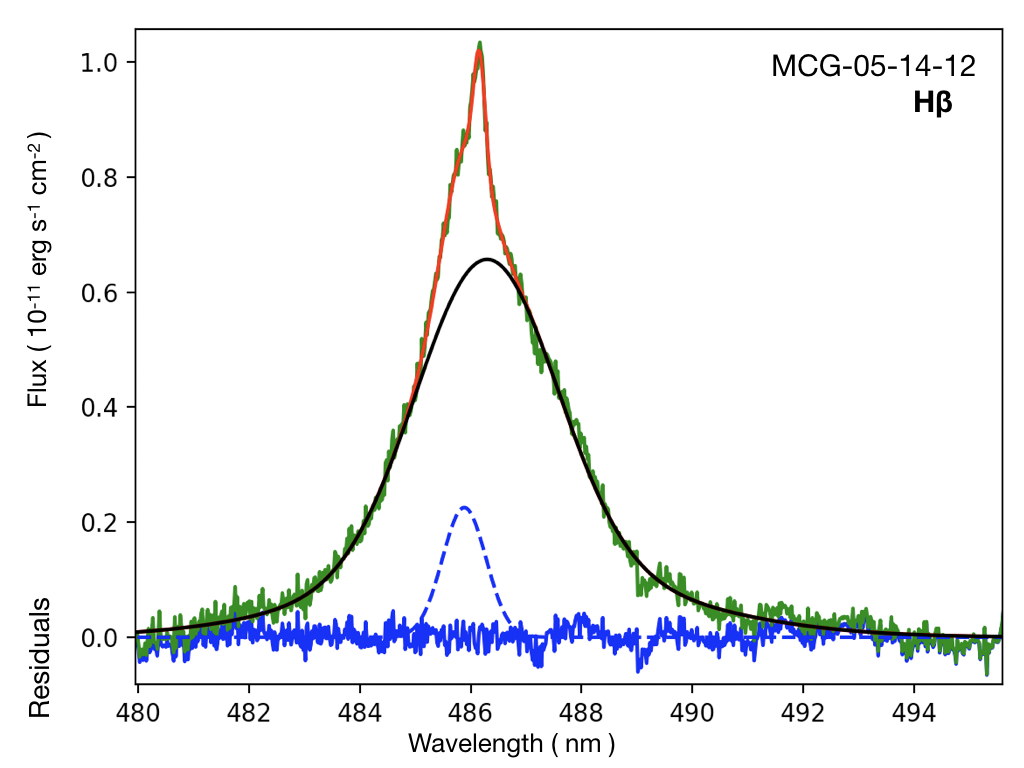}
\includegraphics[width=9.0cm]{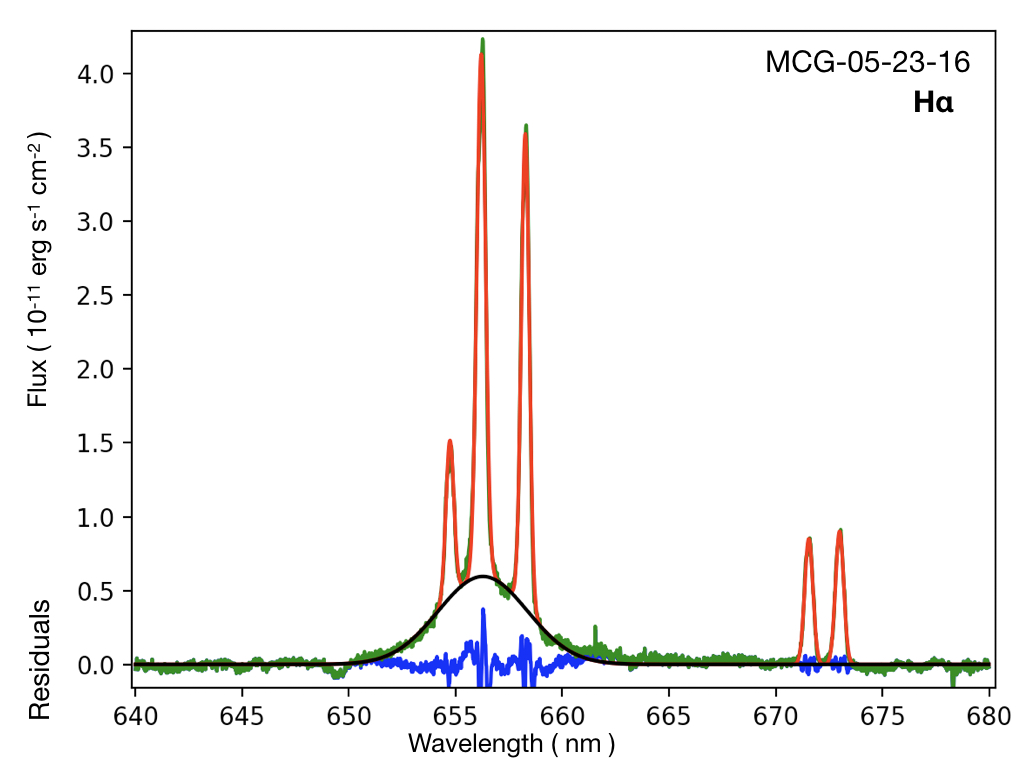}
\includegraphics[width=9.0cm]{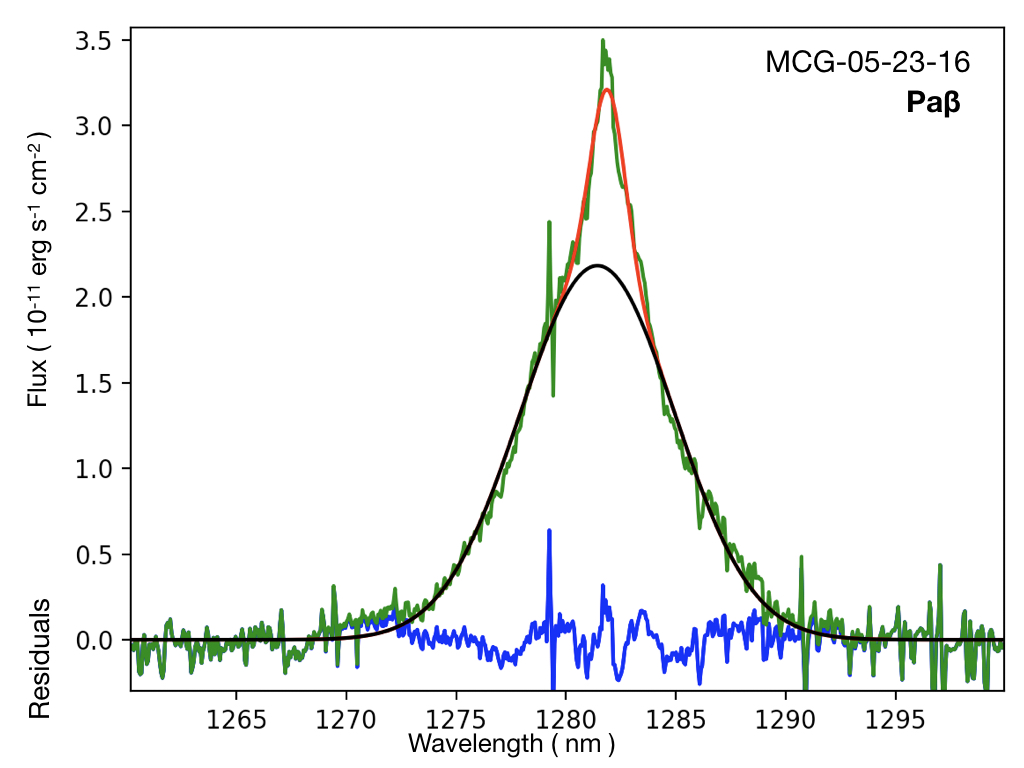}
\includegraphics[width=9.0cm]{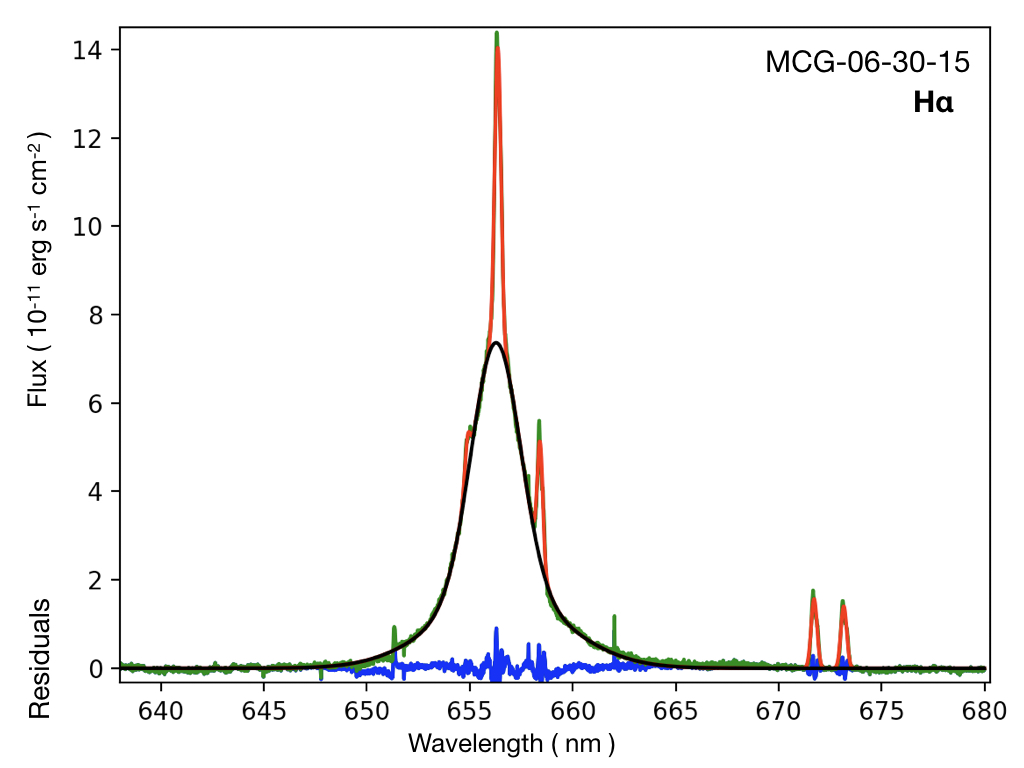}
\includegraphics[width=9.0cm]{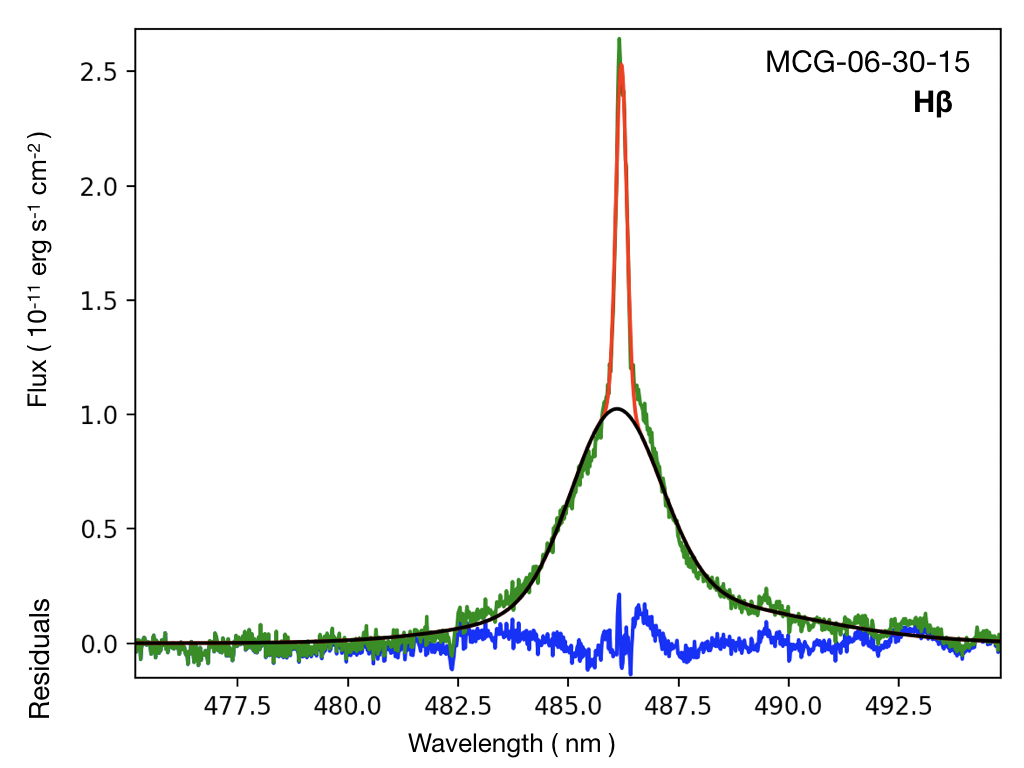}

\caption{The BLR emission fittings of our sample. The black solid line represents the broad-line emission line width, whereas the red solid line represents the best fit. Residuals are presented as blue for visual aids. The unidentified blue-shifted broad emission lines of NGC 1365, NGC 2992 and MCG-05-14-12 are presented as blue dashed line.  }
\label{appendix1}%
\end{figure*}

\begin{figure*}
\centering
\includegraphics[width=9.0cm]{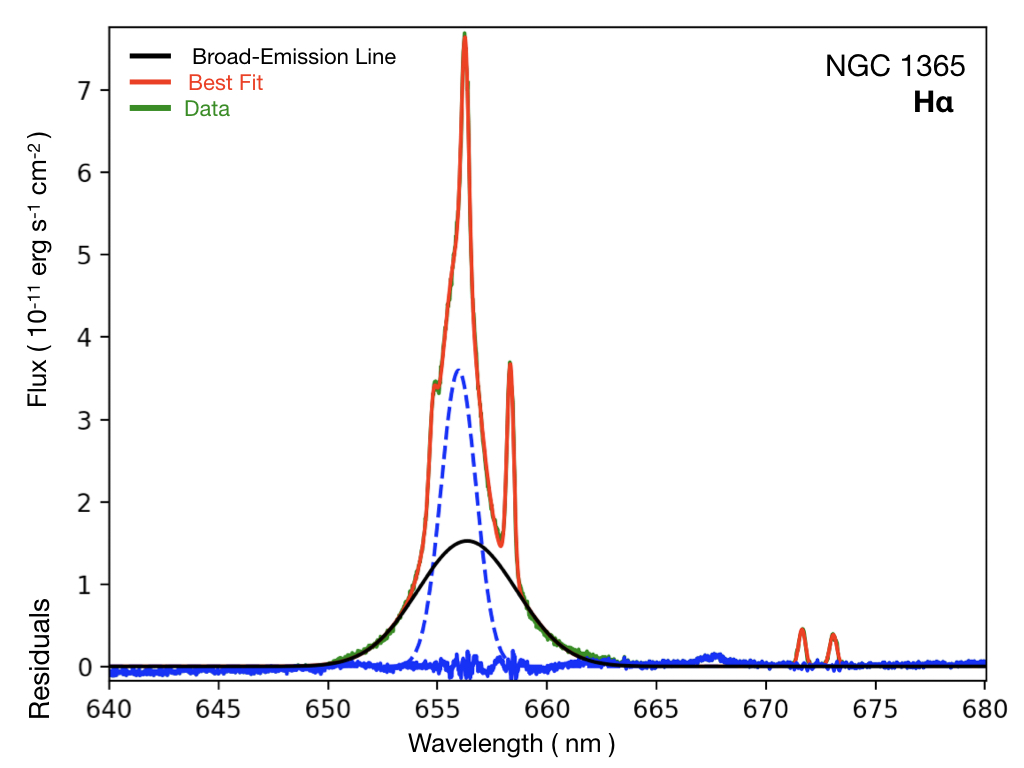}
\includegraphics[width=9.0cm]{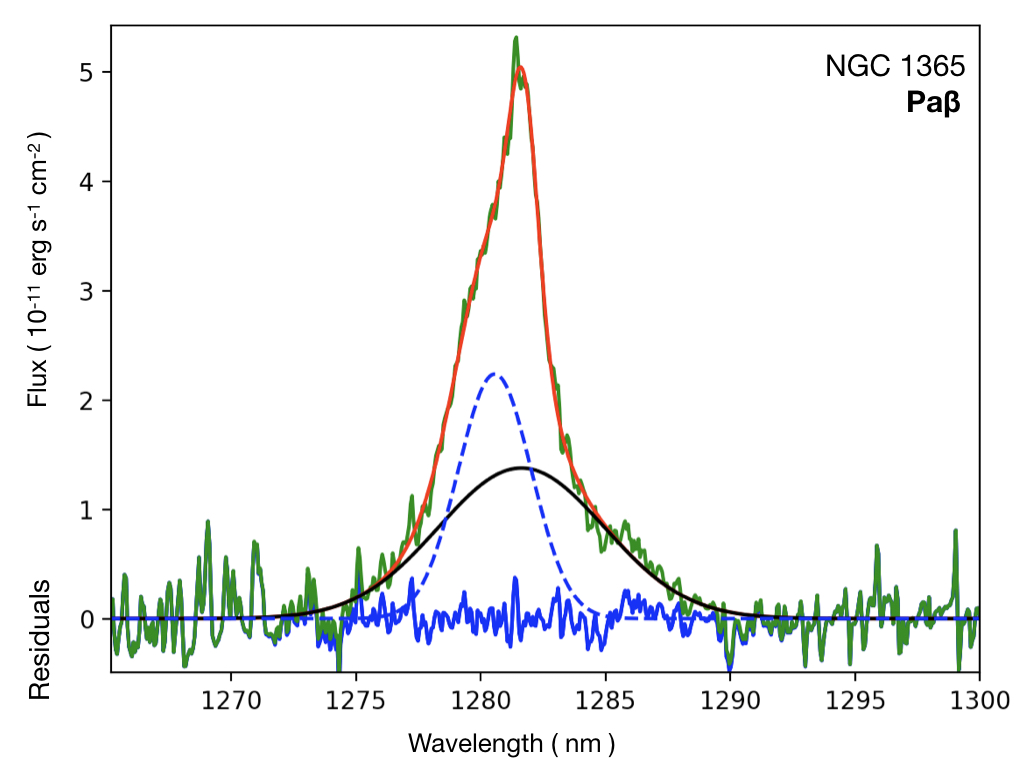}
\includegraphics[width=9.0cm]{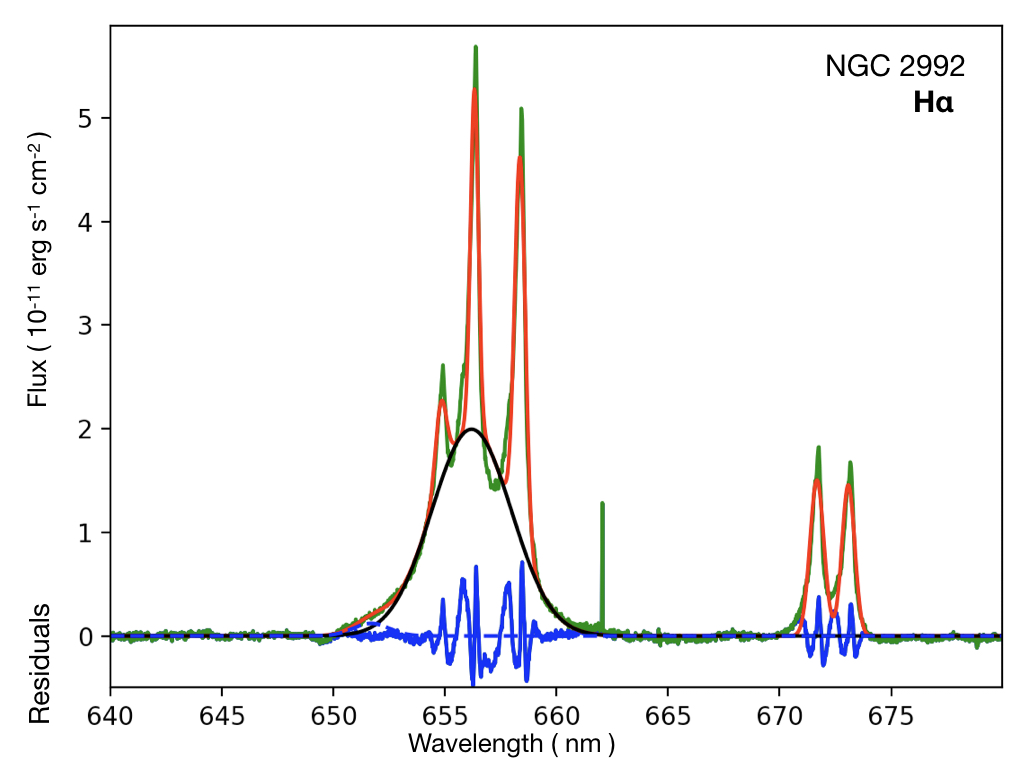}
\includegraphics[width=9.0cm]{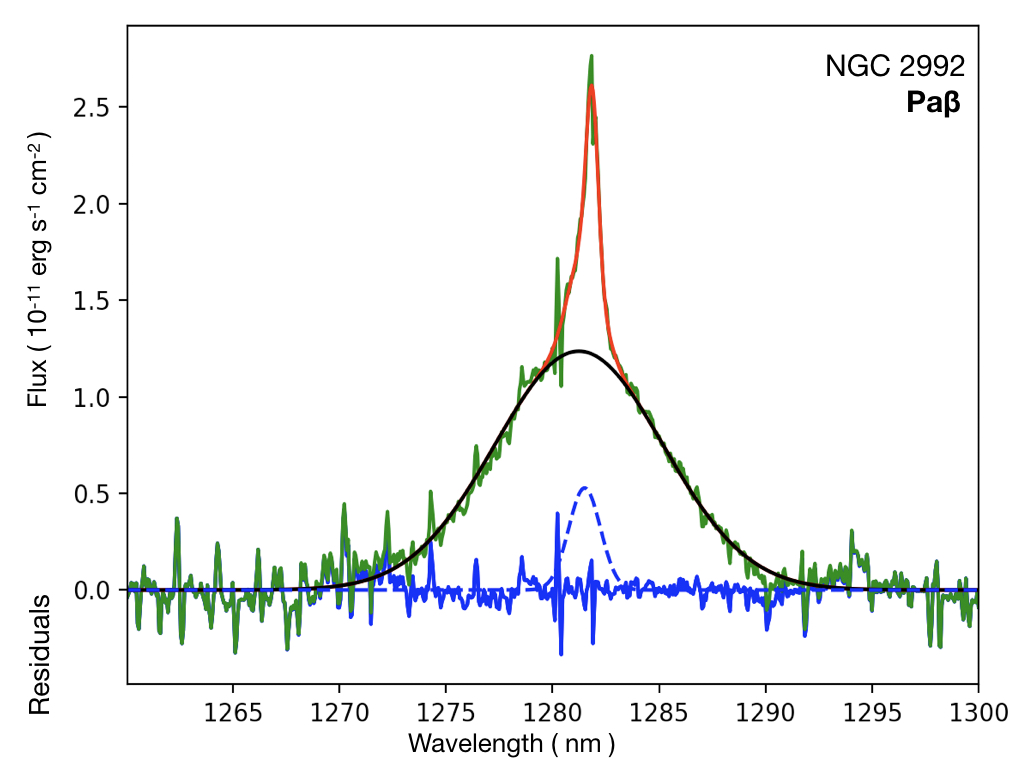}
\includegraphics[width=9.0cm]{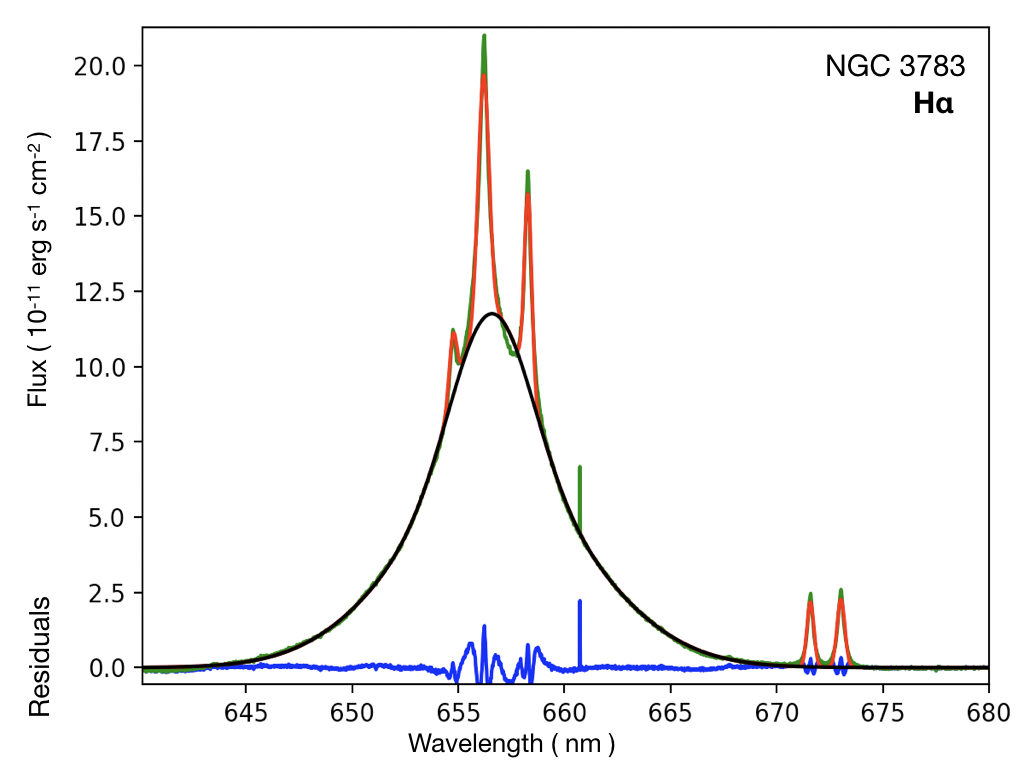}
\includegraphics[width=9.0cm]{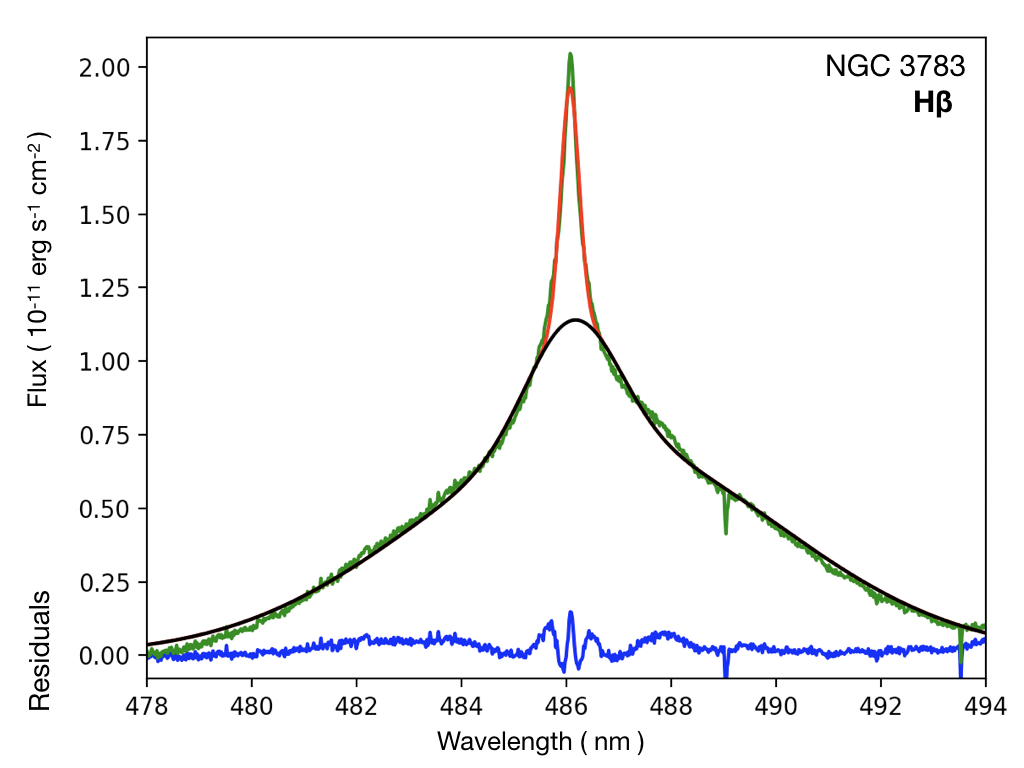}
\caption{Continued}
\label{appendix2}%
\end{figure*}

\begin{figure*}
\centering
\includegraphics[width=9.0cm]{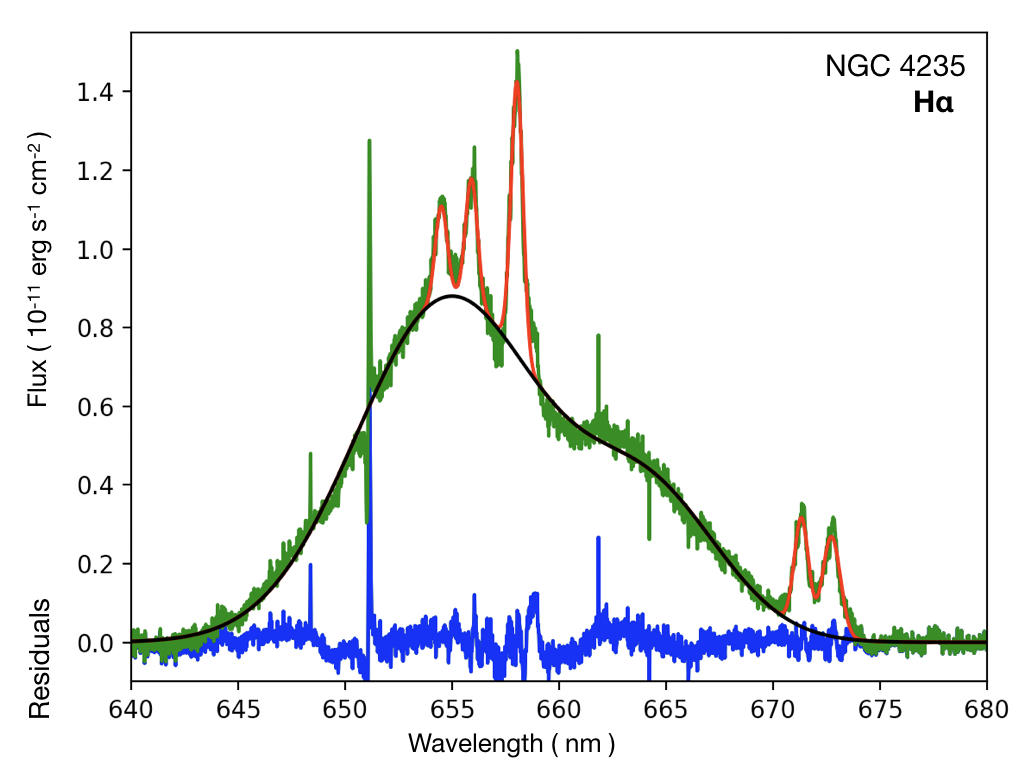}
\includegraphics[width=9.0cm]{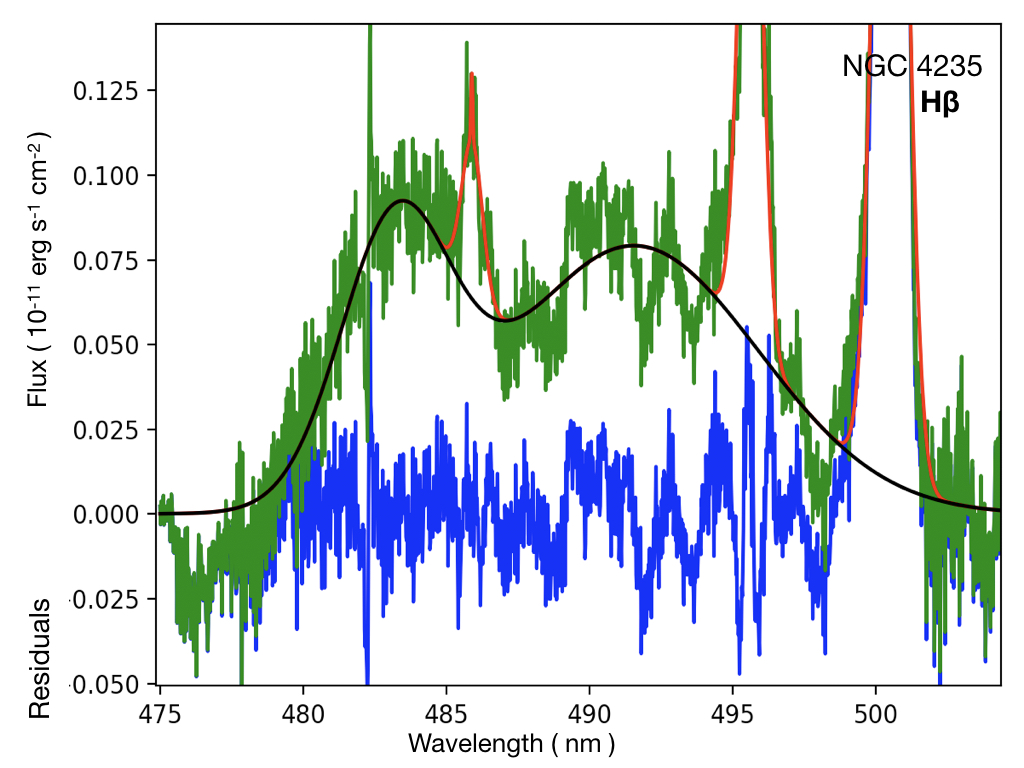}
\includegraphics[width=9.0cm]{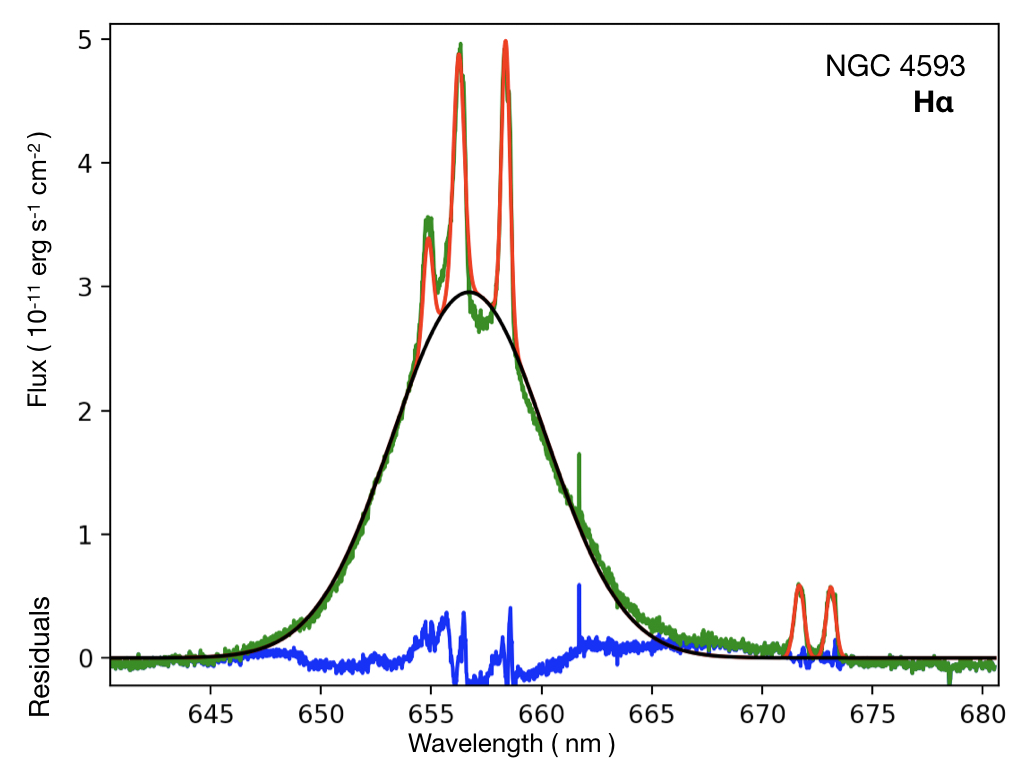}
\includegraphics[width=9.0cm]{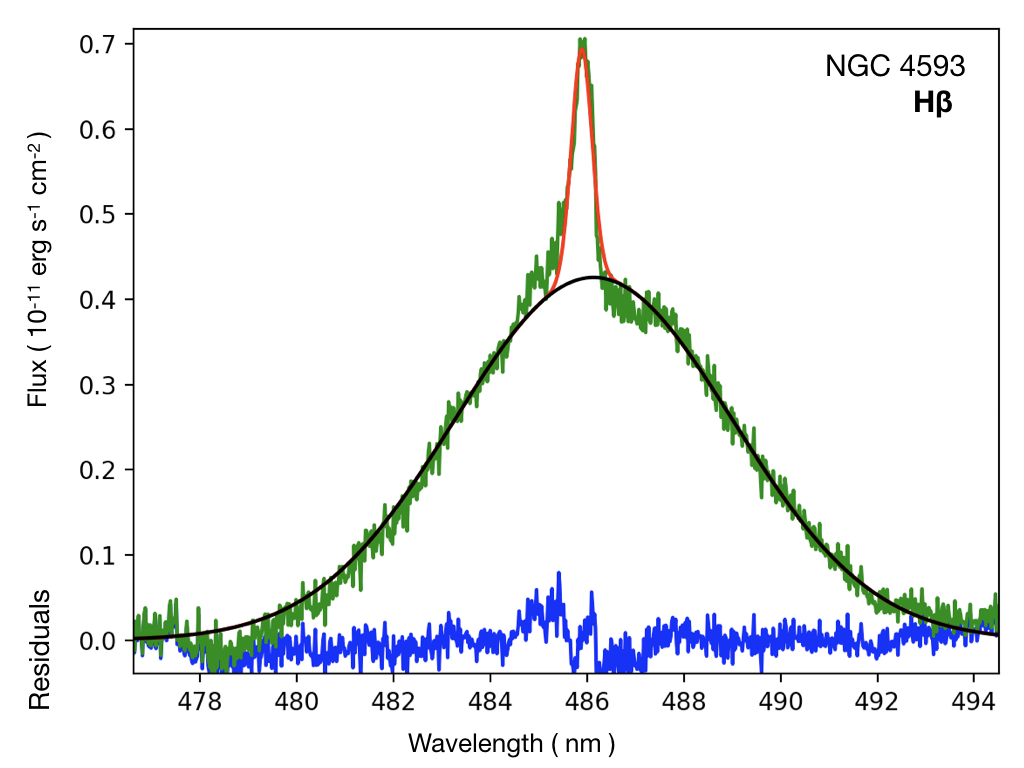}
\includegraphics[width=9.0cm]{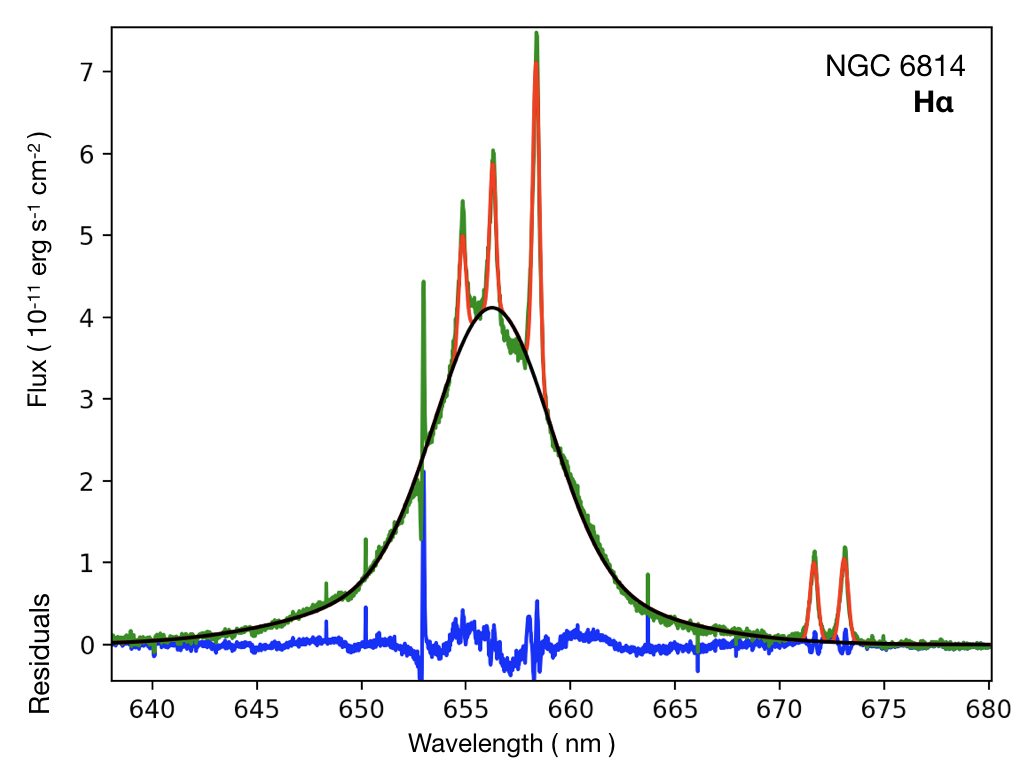}
\includegraphics[width=9.0cm]{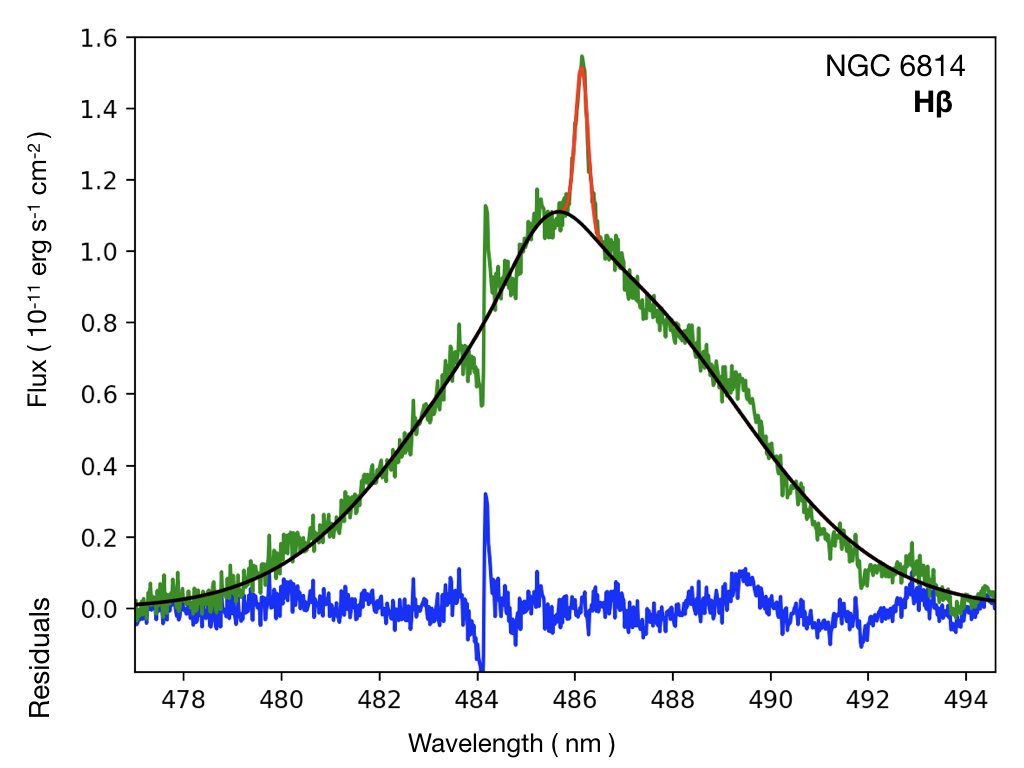}
\caption{Continued}
\label{appendix3}%
\end{figure*}

\begin{figure*}
\centering
\includegraphics[width=9.0cm]{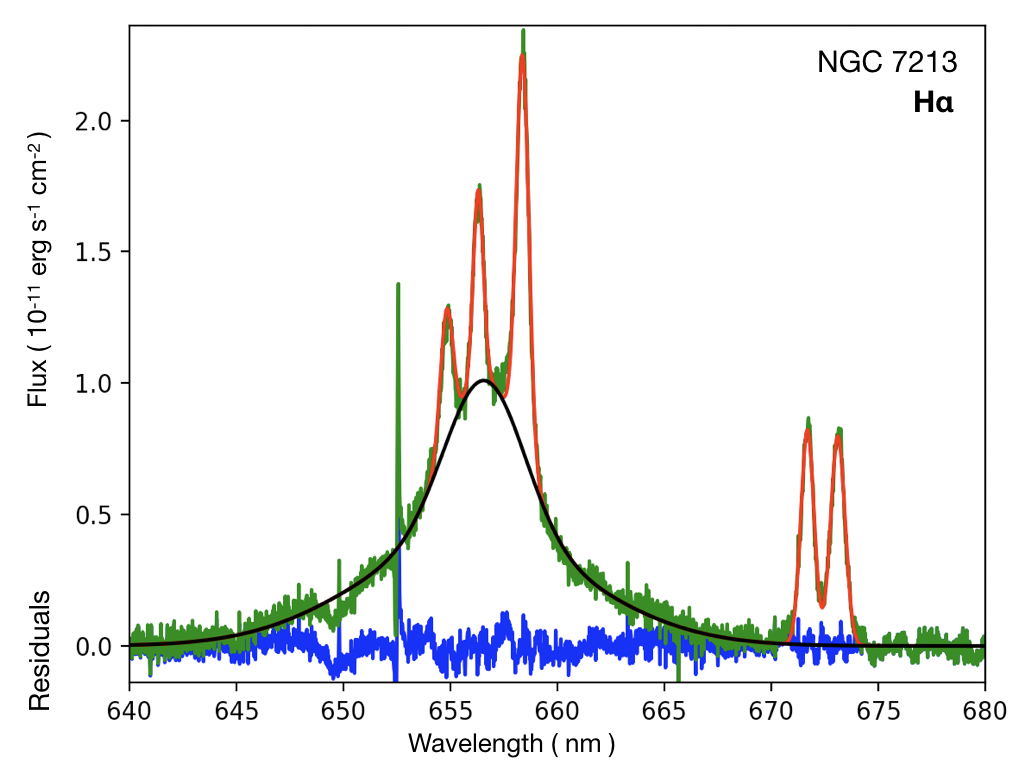}
\includegraphics[width=9.0cm]{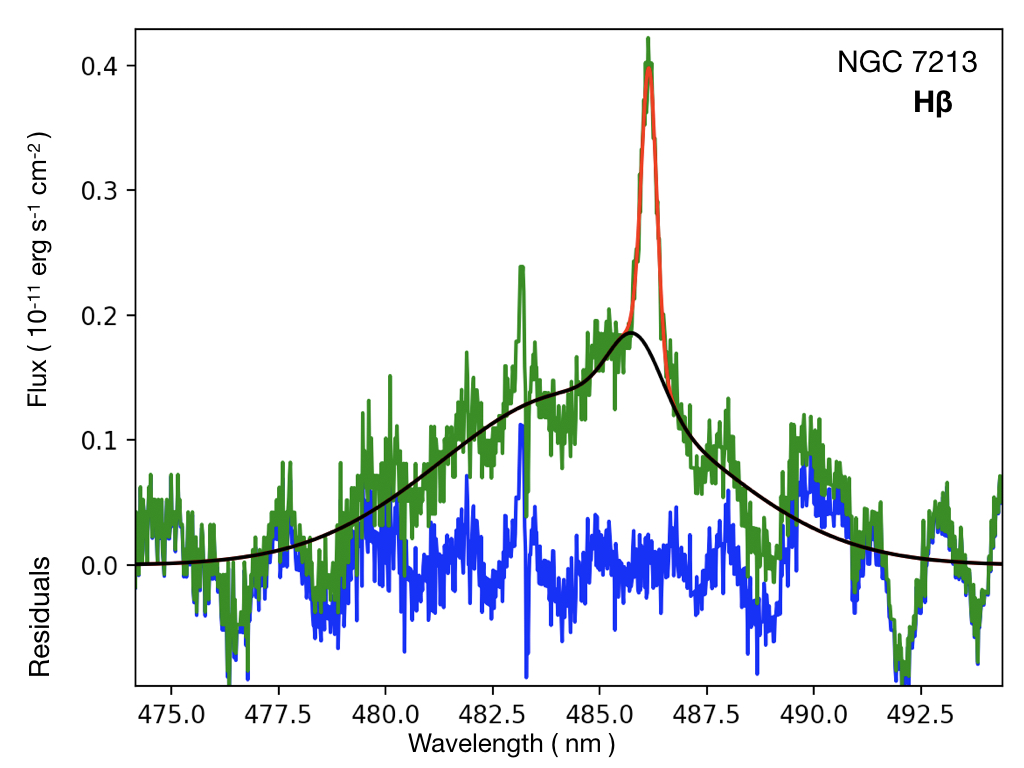}
\caption{Continued}
\label{appendix4}%
\end{figure*}

\section{The pPXF Fittings}
We present the pPXF stellar velocity dispersion fitting results from CaT absorption lines (from Fig. \ref{appendixCAT1} to \ref{appendixcat6}), whereas CO (2-0) fitting results are presented from Fig. \ref{appendixco1} to  \ref{appendixco5}.

\begin{figure*}
\centering
\includegraphics[width=9.0cm]{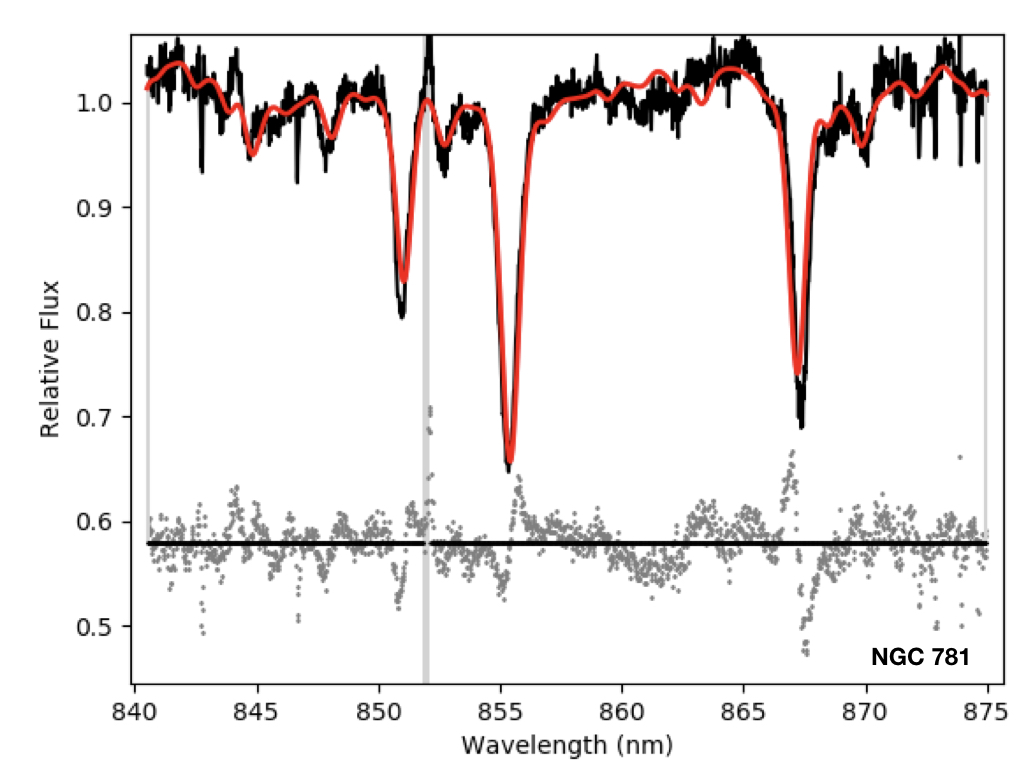}
\includegraphics[width=9.0cm]{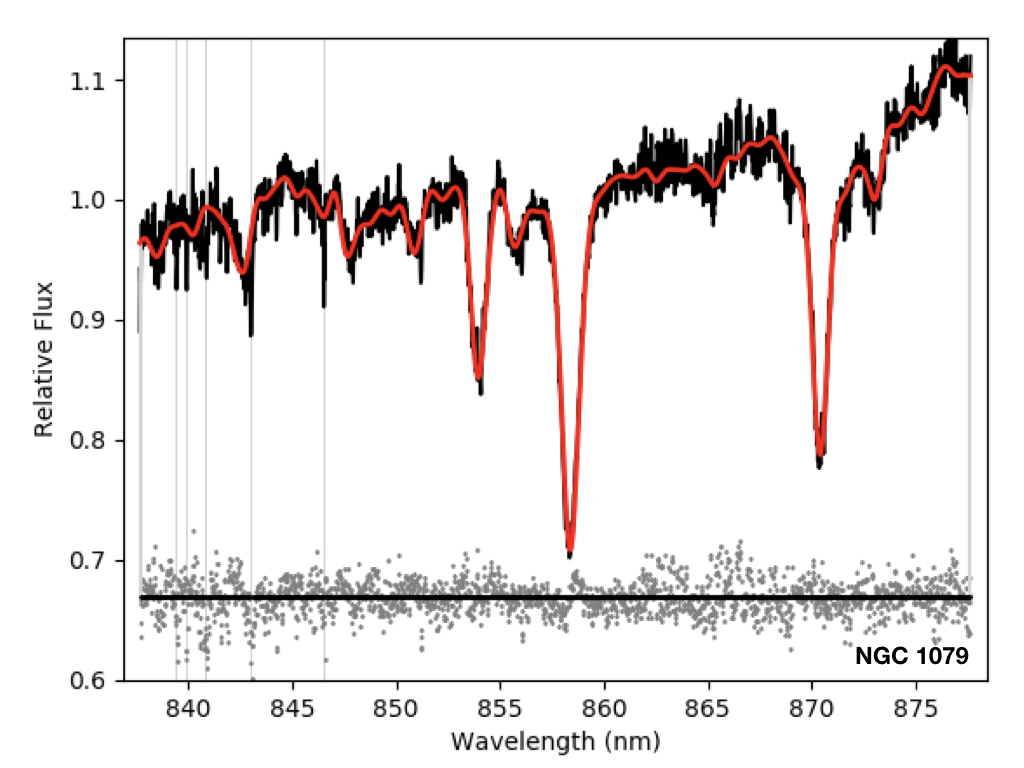}
\includegraphics[width=9.0cm]{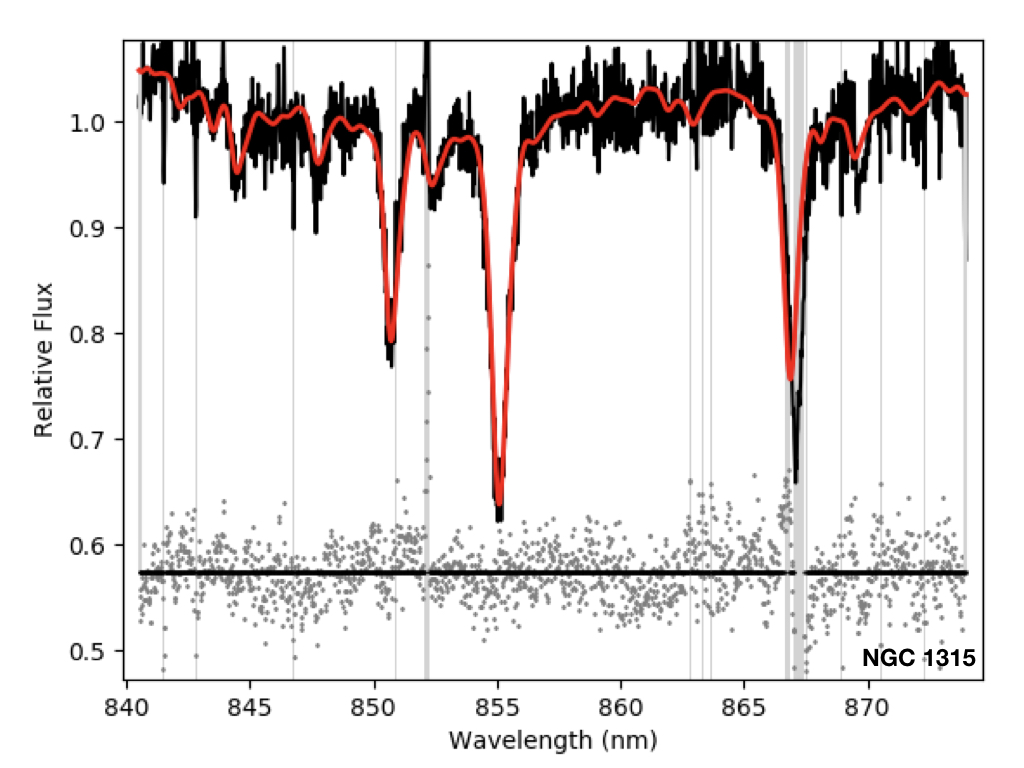}
\includegraphics[width=9.0cm]{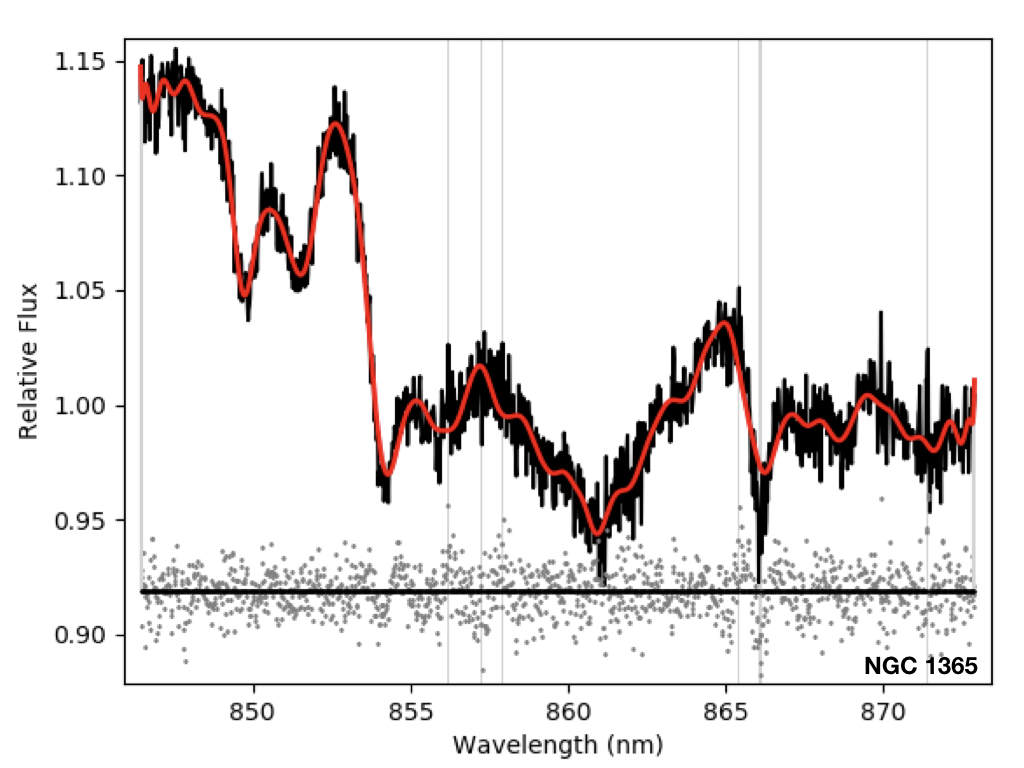}
\includegraphics[width=9.0cm]{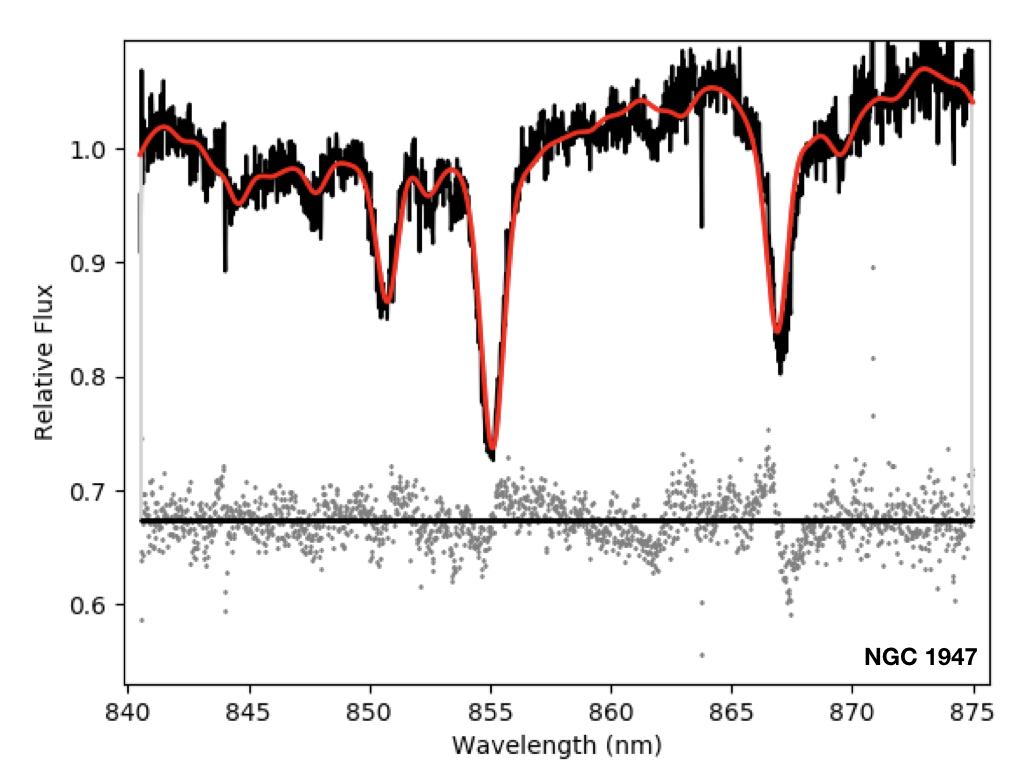}
\includegraphics[width=9.0cm]{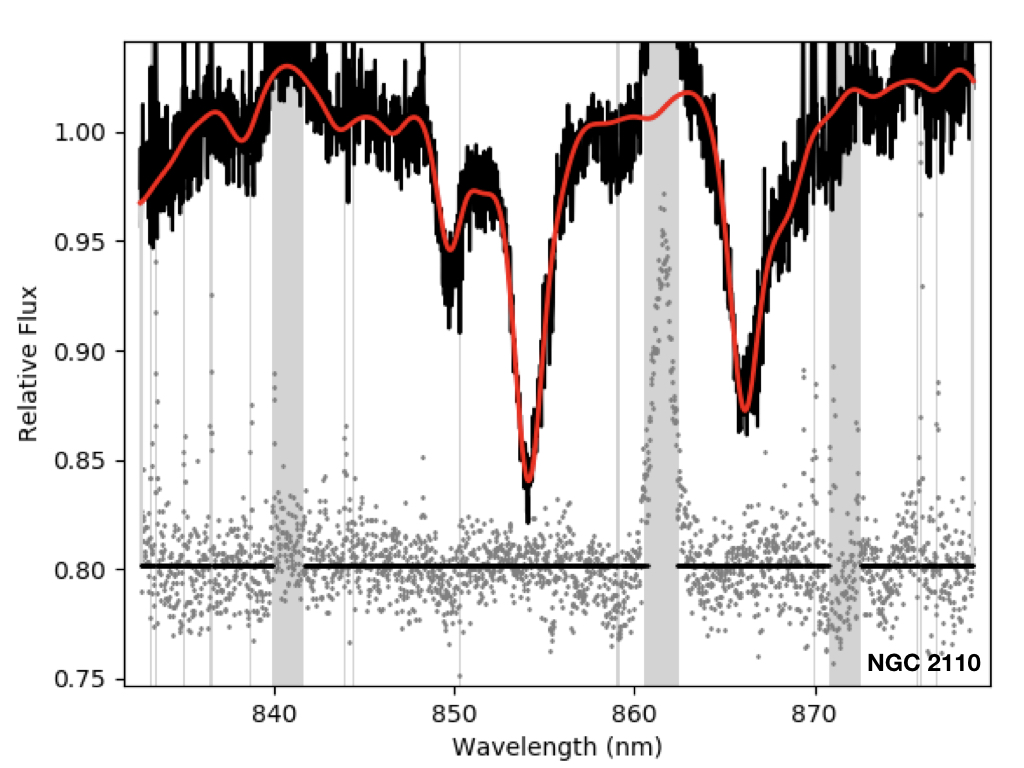}
\caption{The pPXF fitting plots for CaT. The red solid line represents the best-fit, whereas the residuals are shown as gray. The vertical gray lines represents masked features.}
\label{appendixCAT1}%
\end{figure*}

\begin{figure*}
\centering
\includegraphics[width=9.0cm]{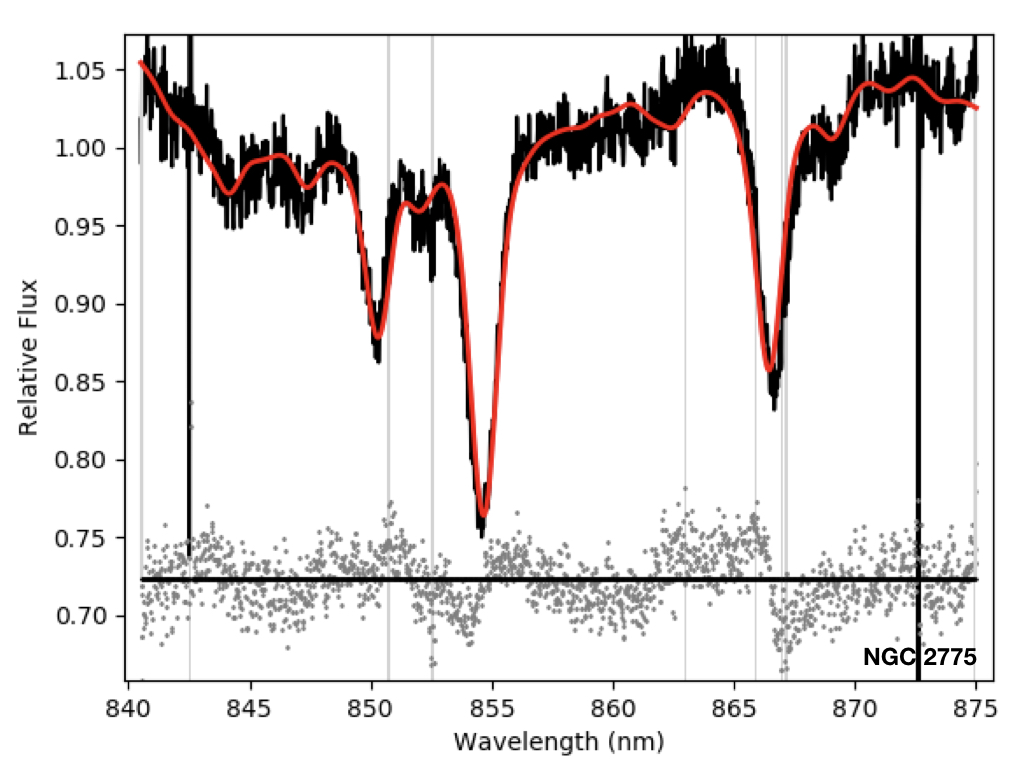}
\includegraphics[width=9.0cm]{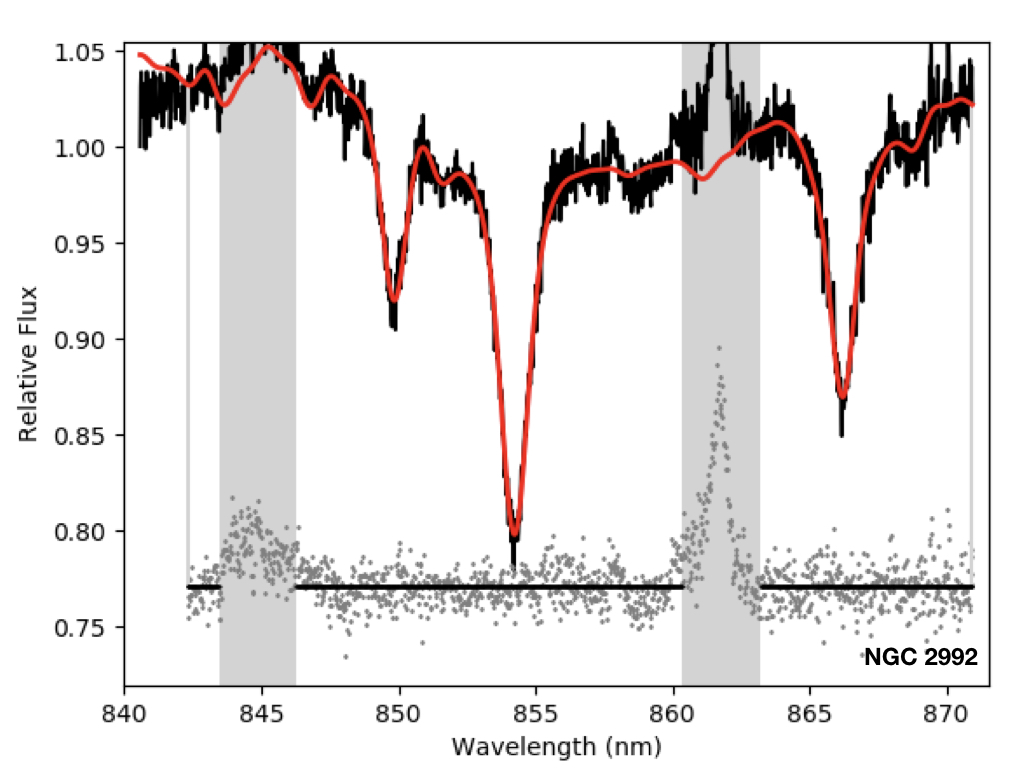}
\includegraphics[width=9.0cm]{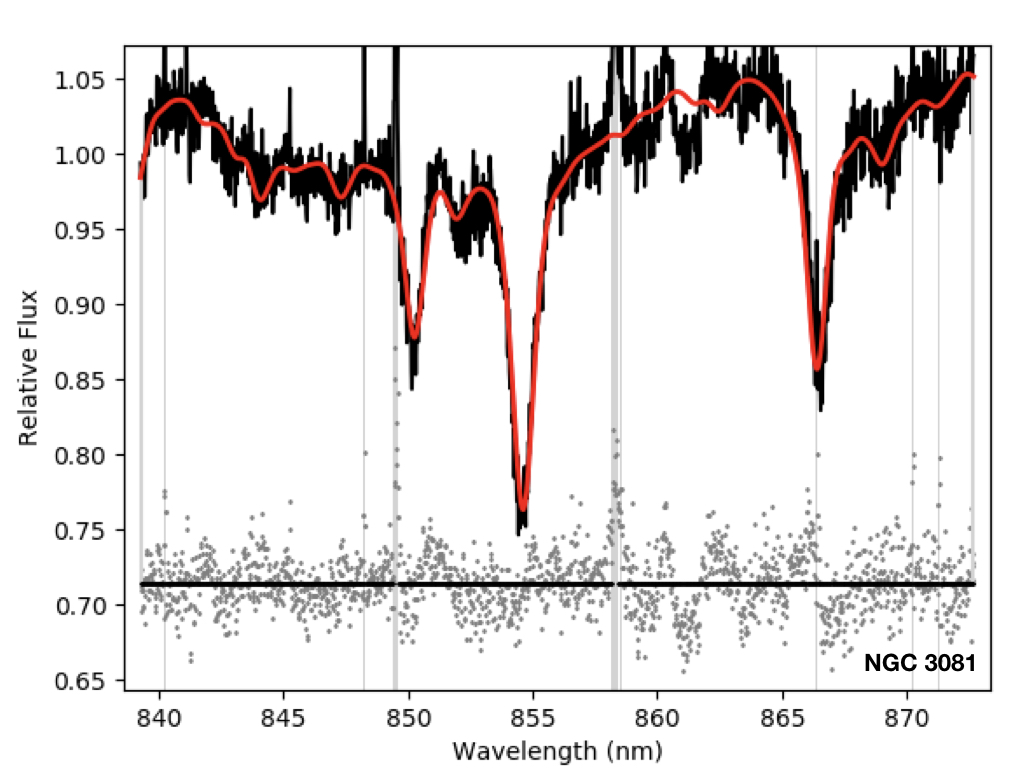}
\includegraphics[width=9.0cm]{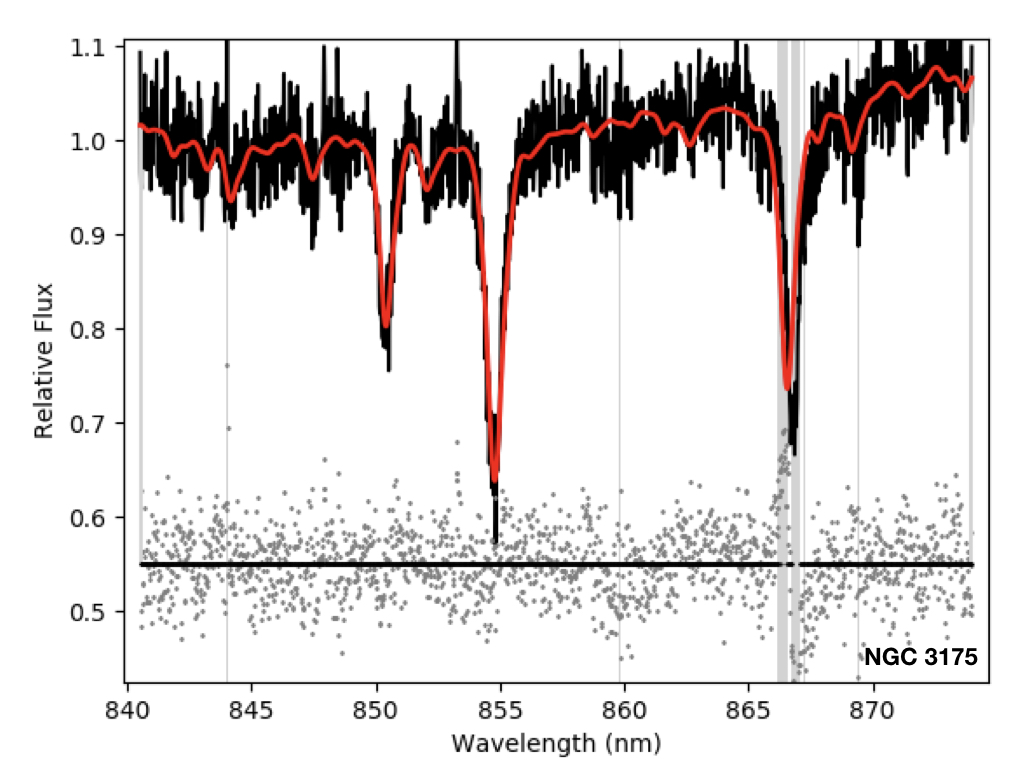}
\includegraphics[width=9.0cm]{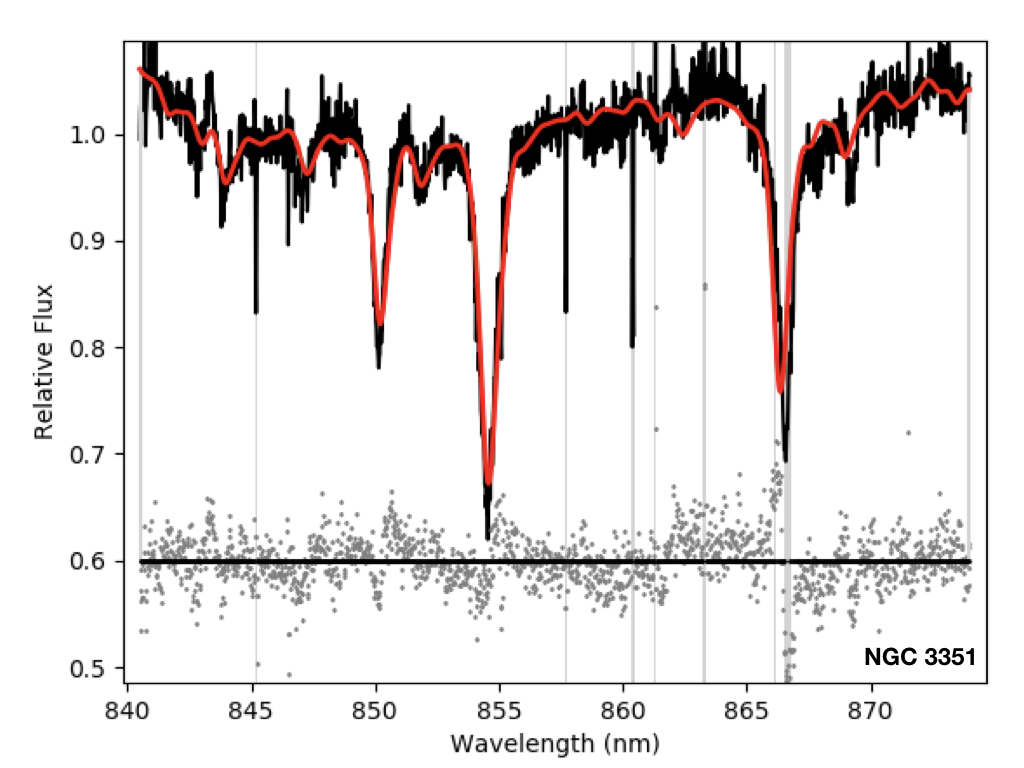}
\includegraphics[width=9.0cm]{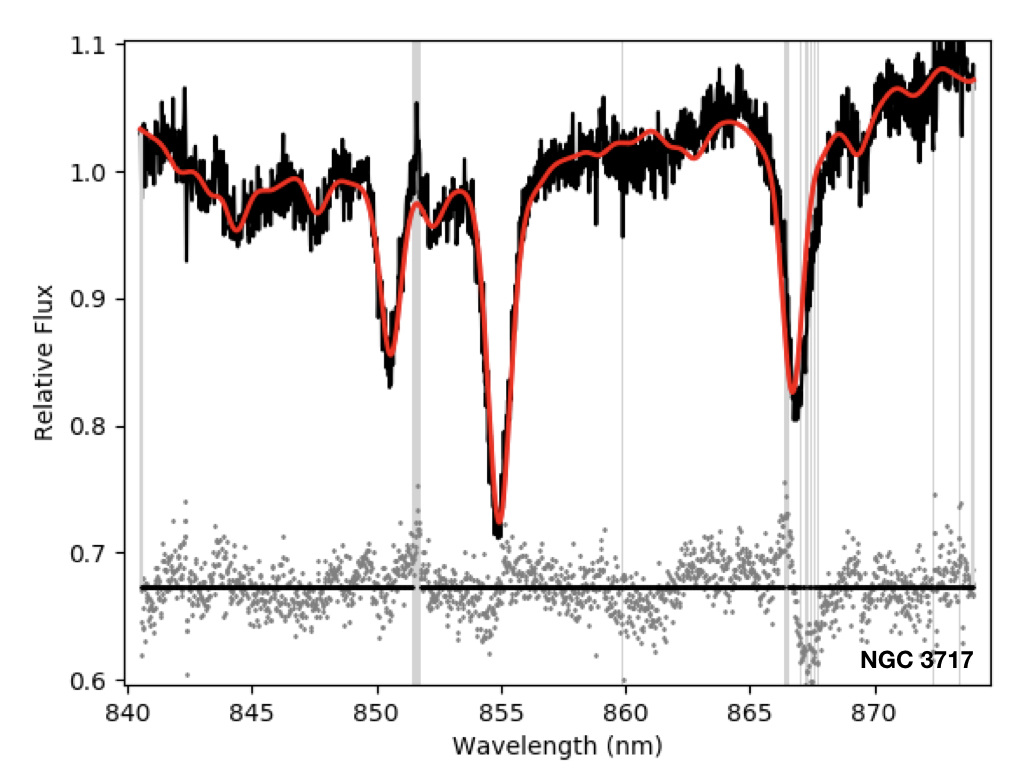}
\caption{Continued }
\label{appendixCAT2}%
\end{figure*}

\begin{figure*}
\centering
\includegraphics[width=9.0cm]{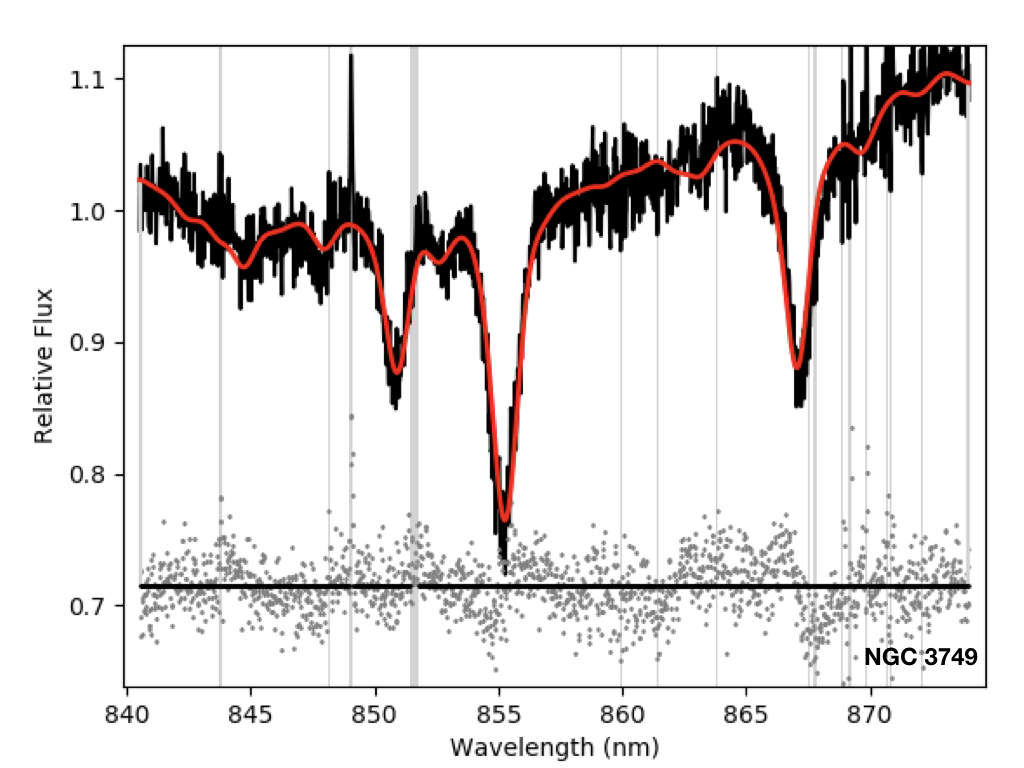}
\includegraphics[width=9.0cm]{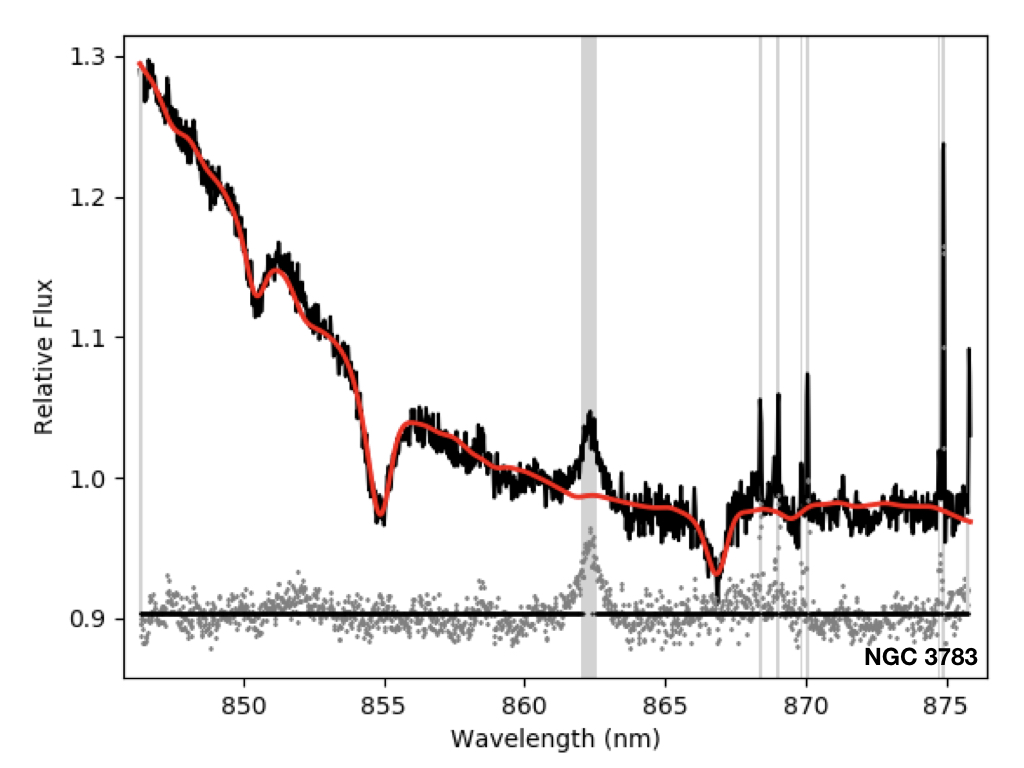}
\includegraphics[width=9.0cm]{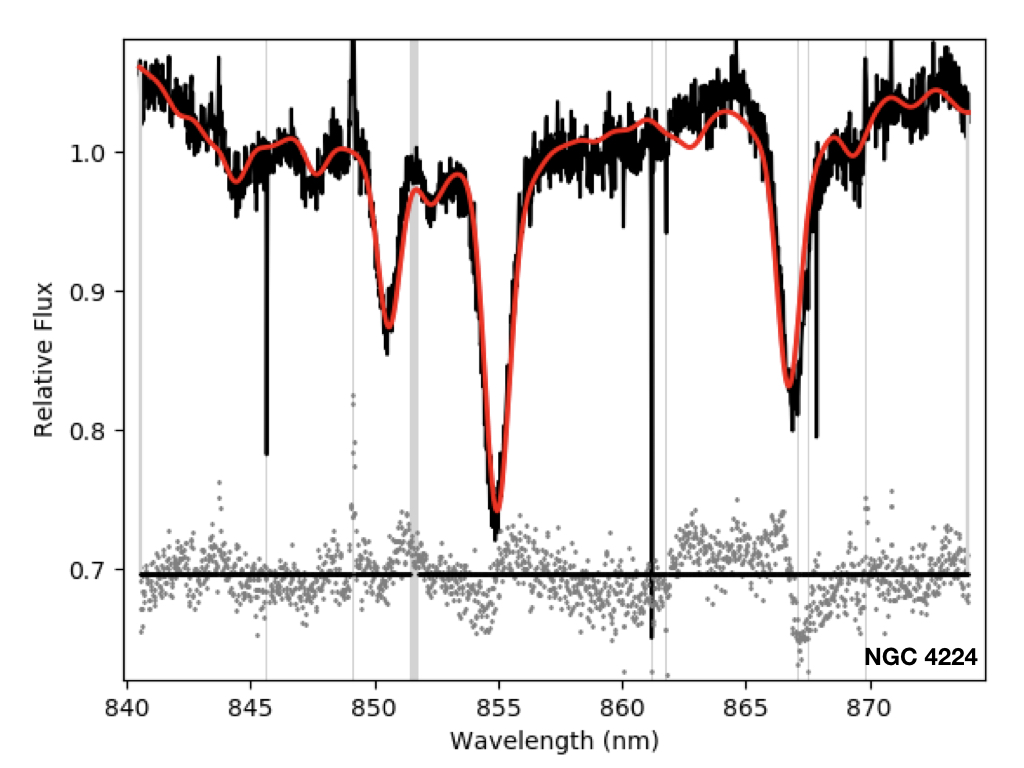}
\includegraphics[width=9.0cm]{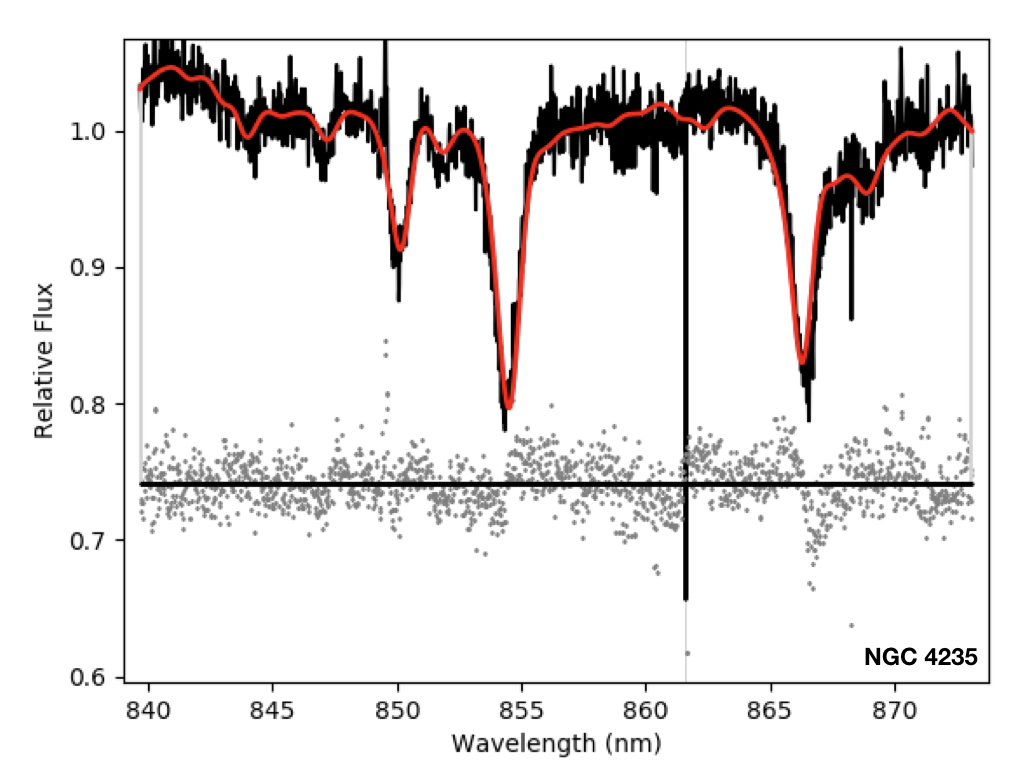}
\includegraphics[width=9.0cm]{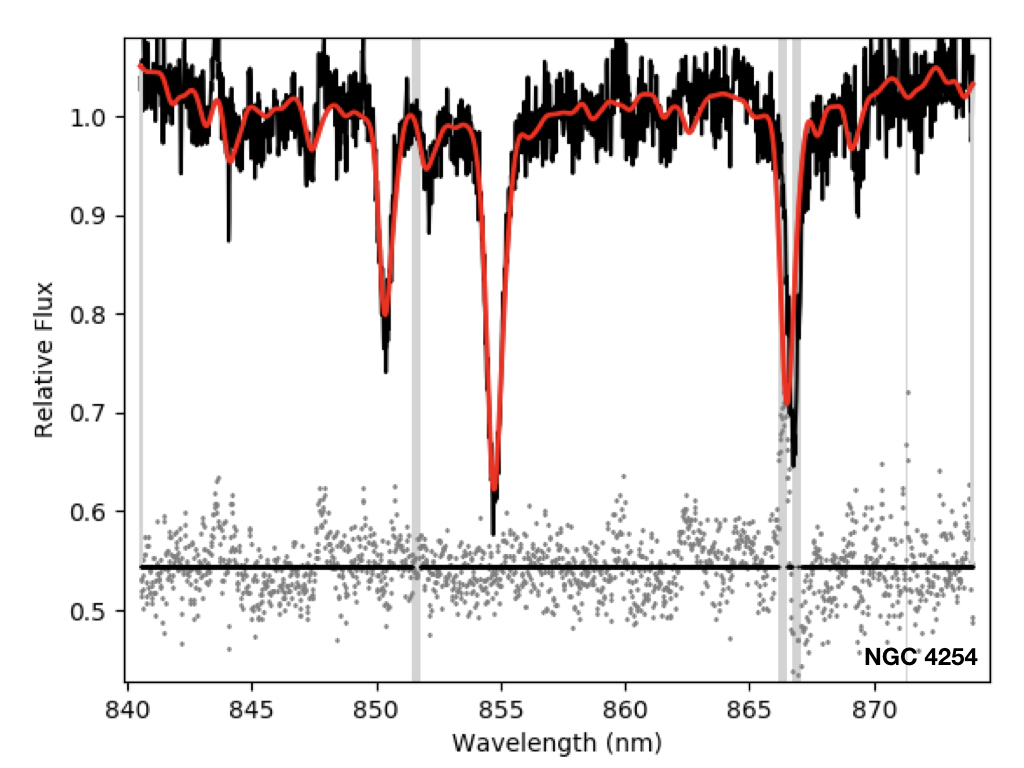}
\includegraphics[width=9.0cm]{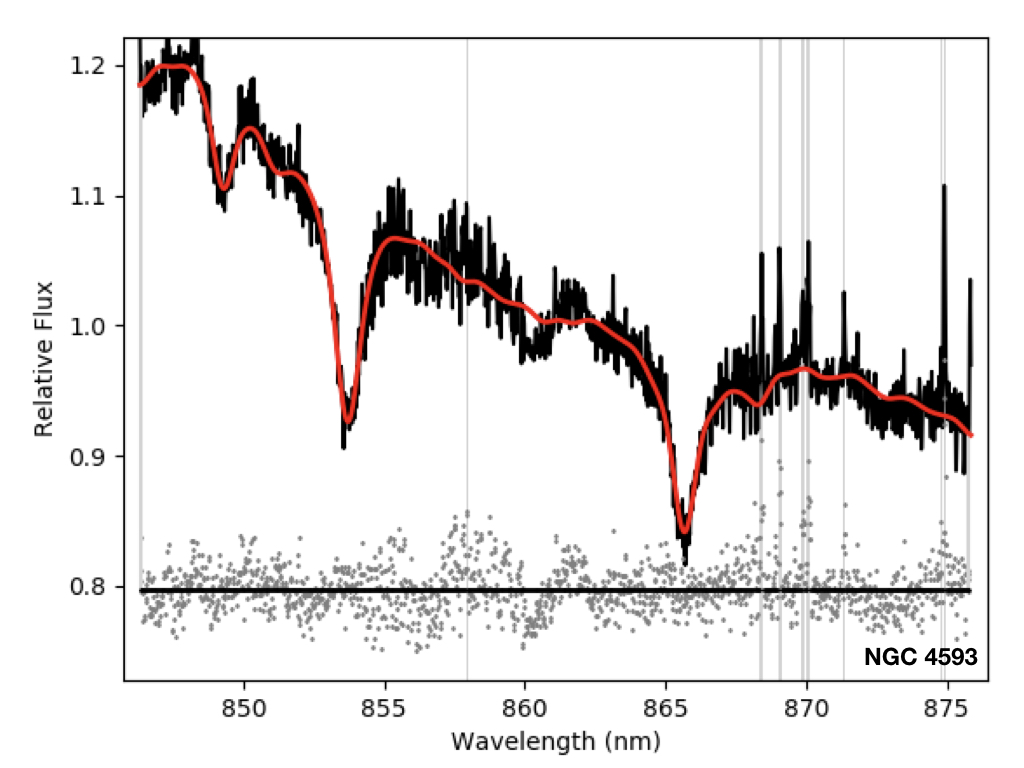}
\caption{Continued }
\label{appendixCAT3}%
\end{figure*}

\begin{figure*}
\centering
\includegraphics[width=9.0cm]{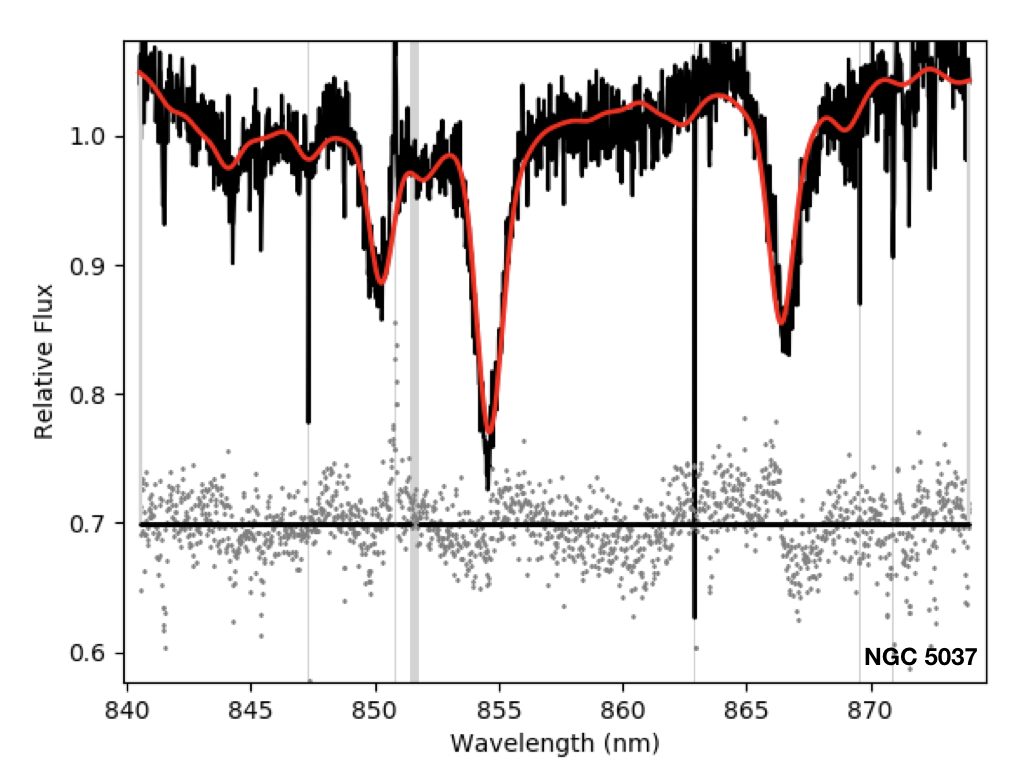}
\includegraphics[width=9.0cm]{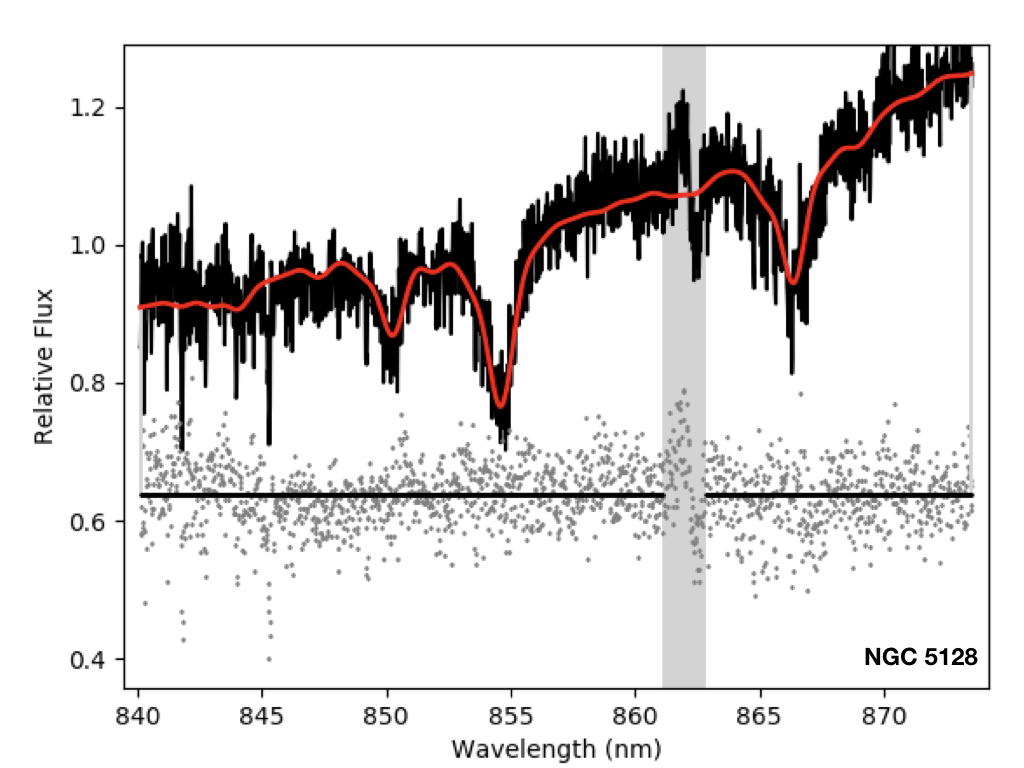}
\includegraphics[width=9.0cm]{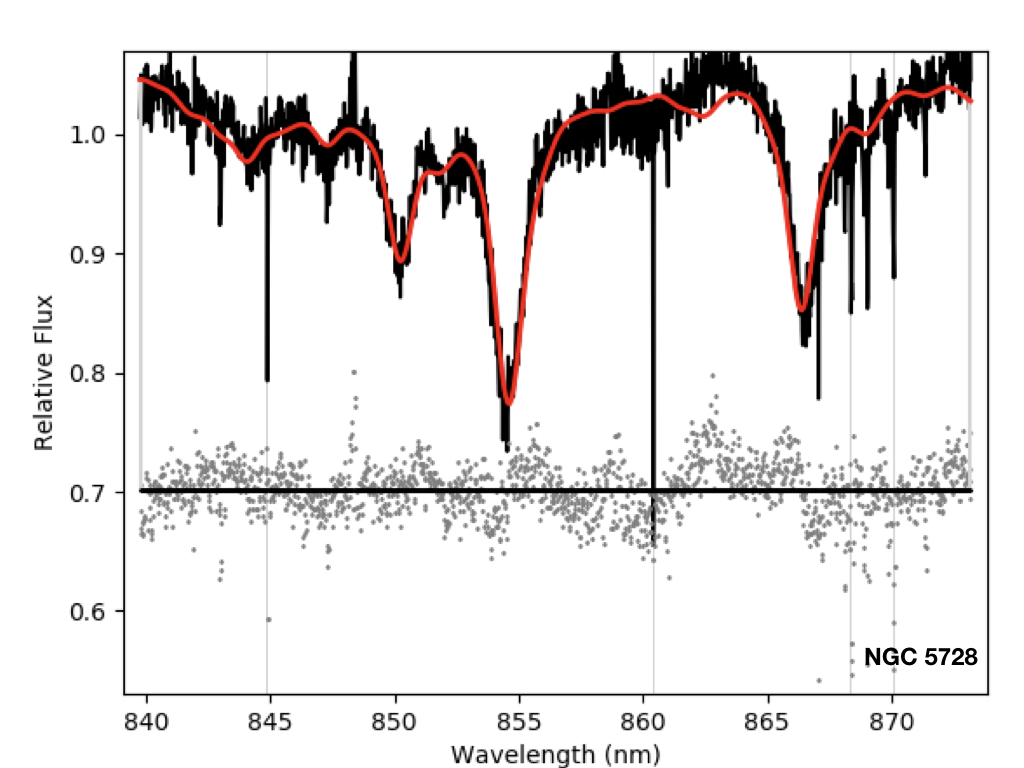}
\includegraphics[width=9.0cm]{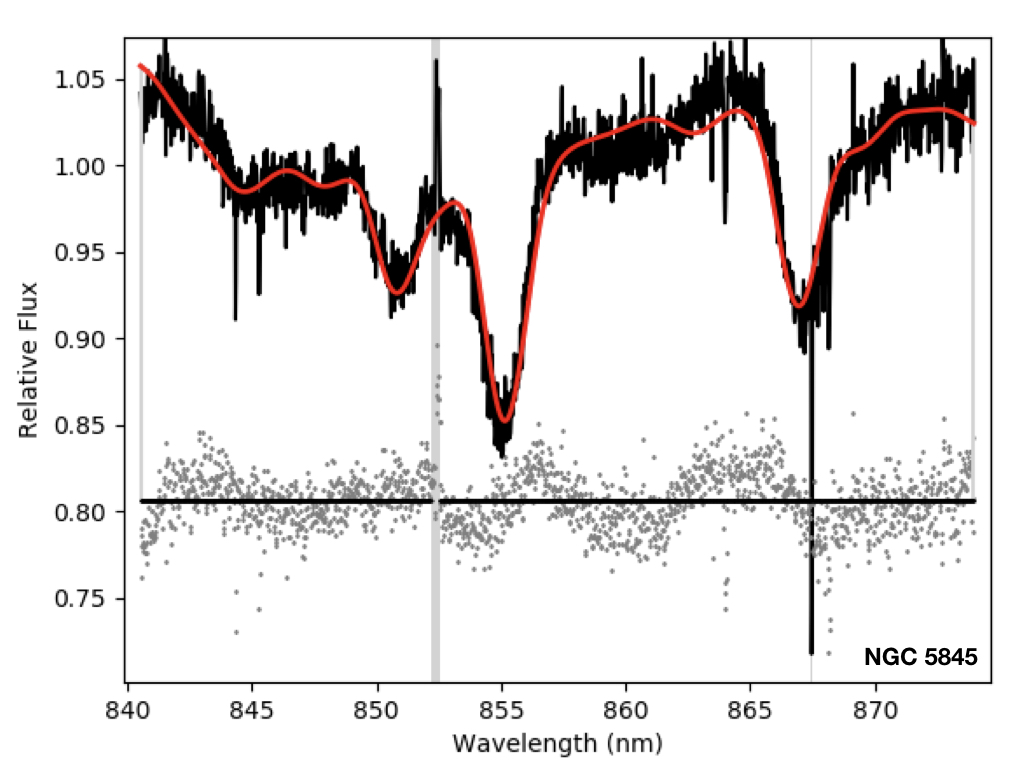}
\includegraphics[width=9.0cm]{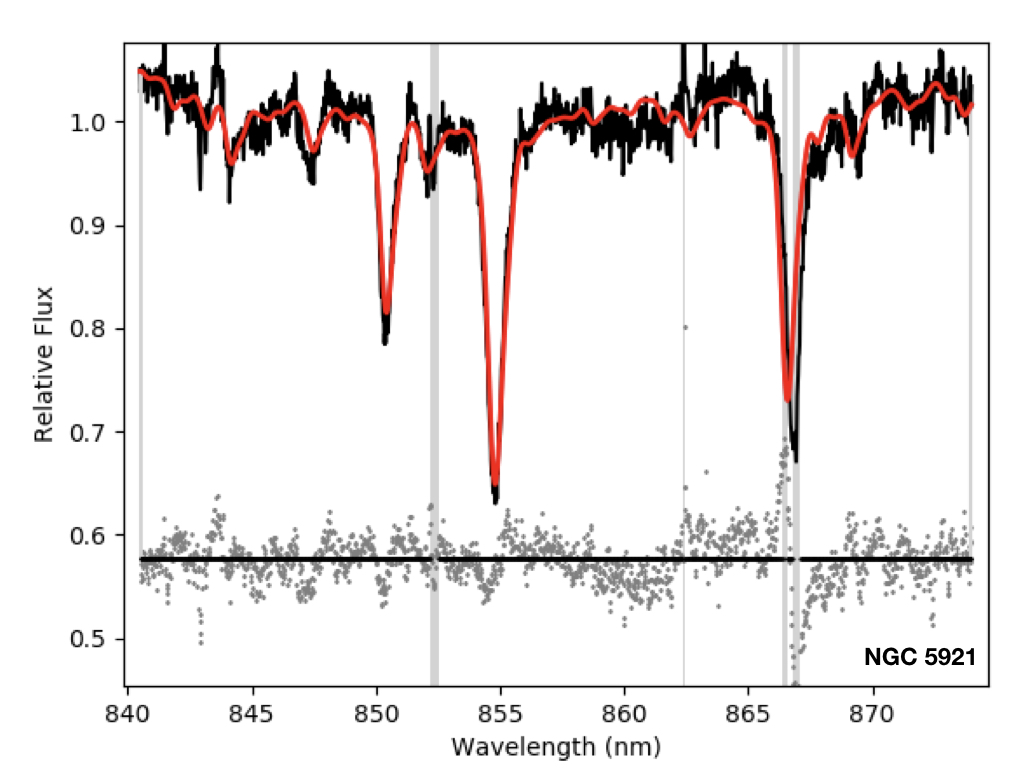}
\includegraphics[width=9.0cm]{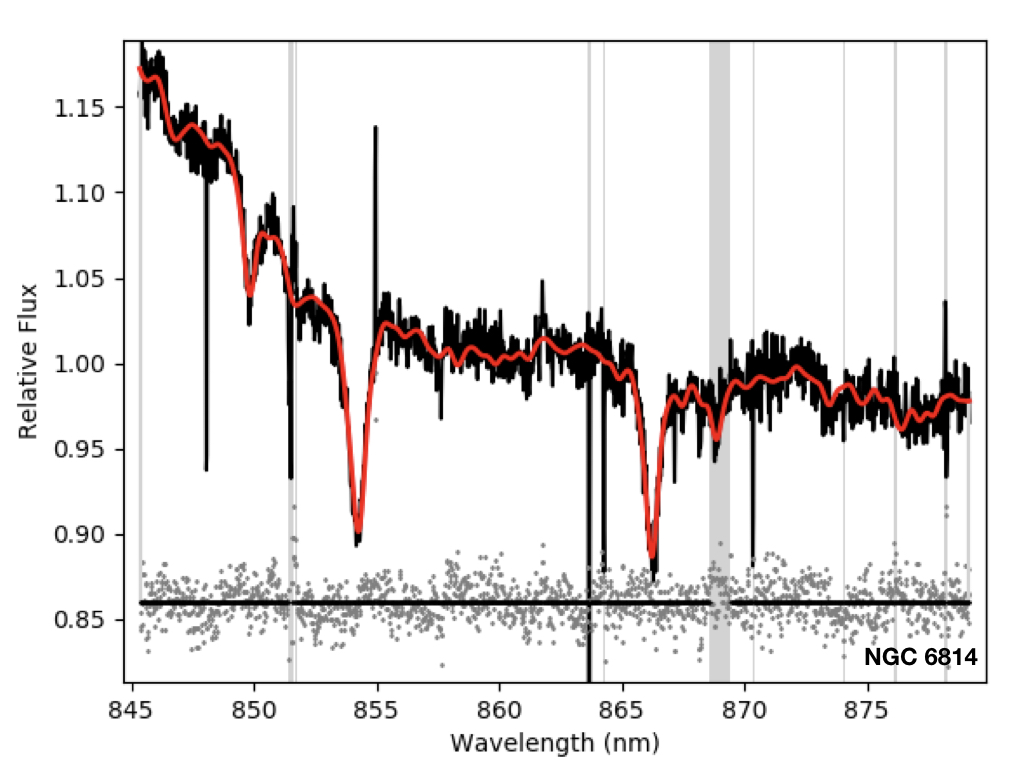}
\caption{Continued }
\label{appendixCAT4}%
\end{figure*}

\begin{figure*}
\centering
\includegraphics[width=9.0cm]{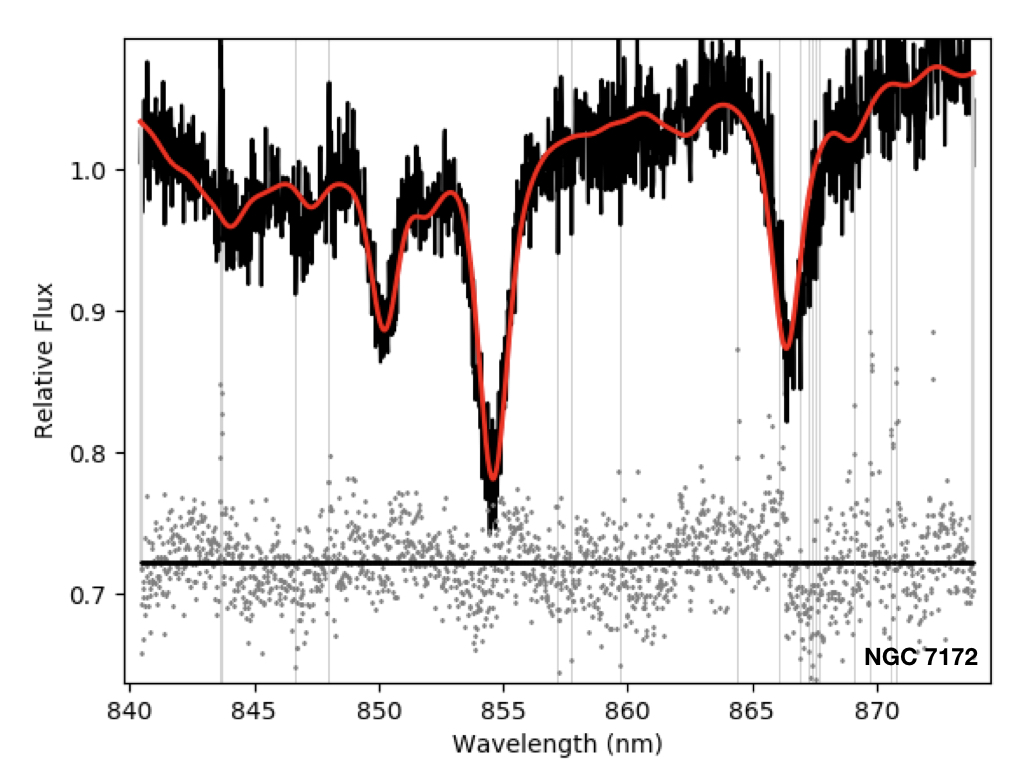}
\includegraphics[width=9.0cm]{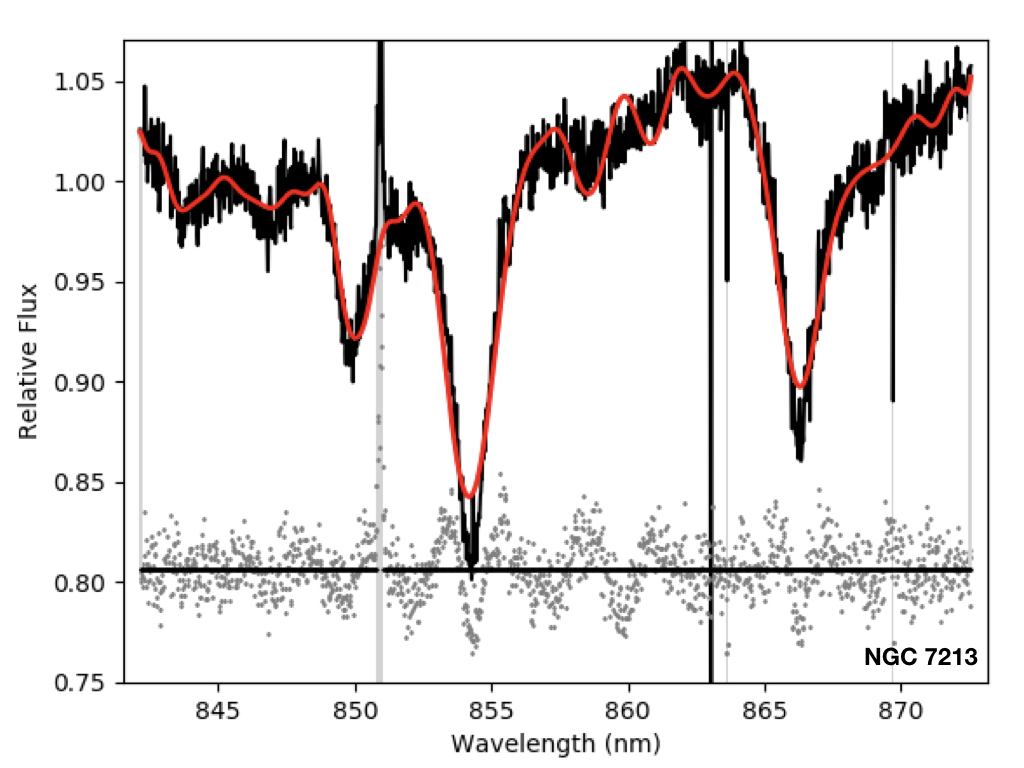}
\includegraphics[width=9.0cm]{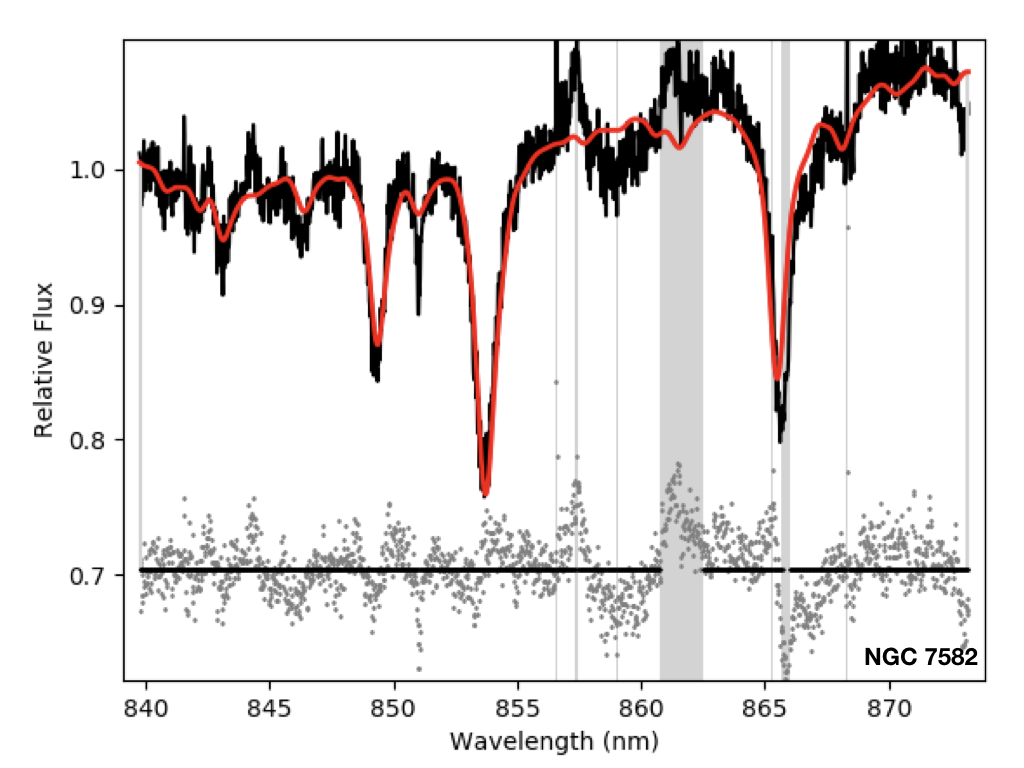}
\includegraphics[width=9.0cm]{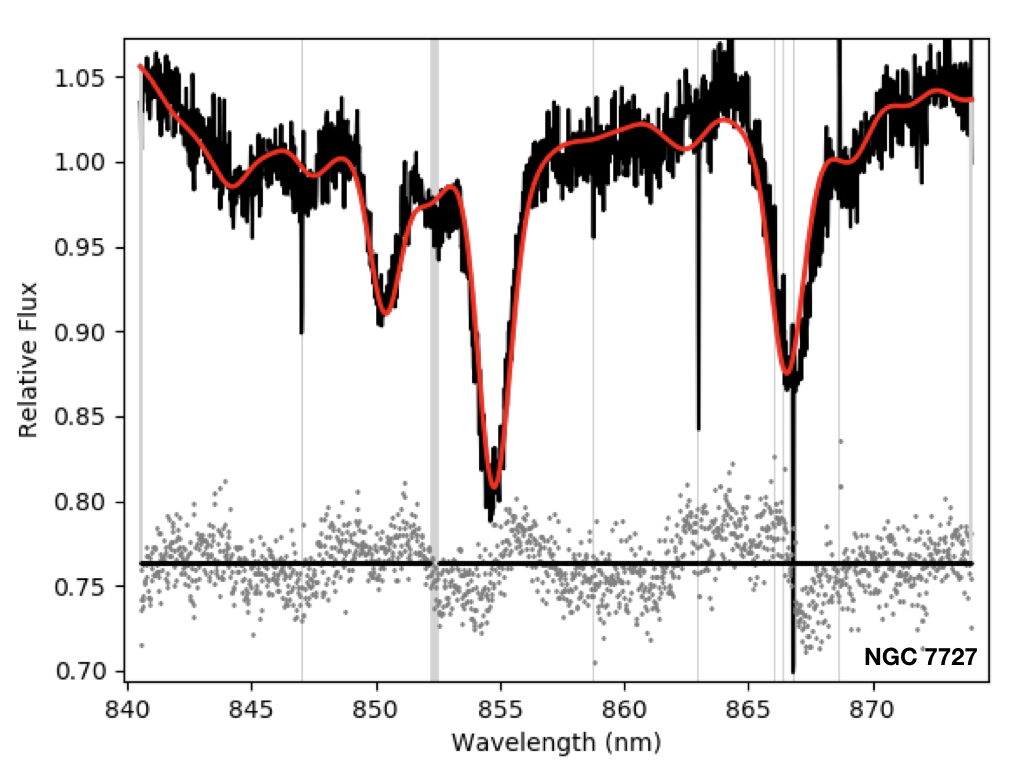}
\includegraphics[width=9.0cm]{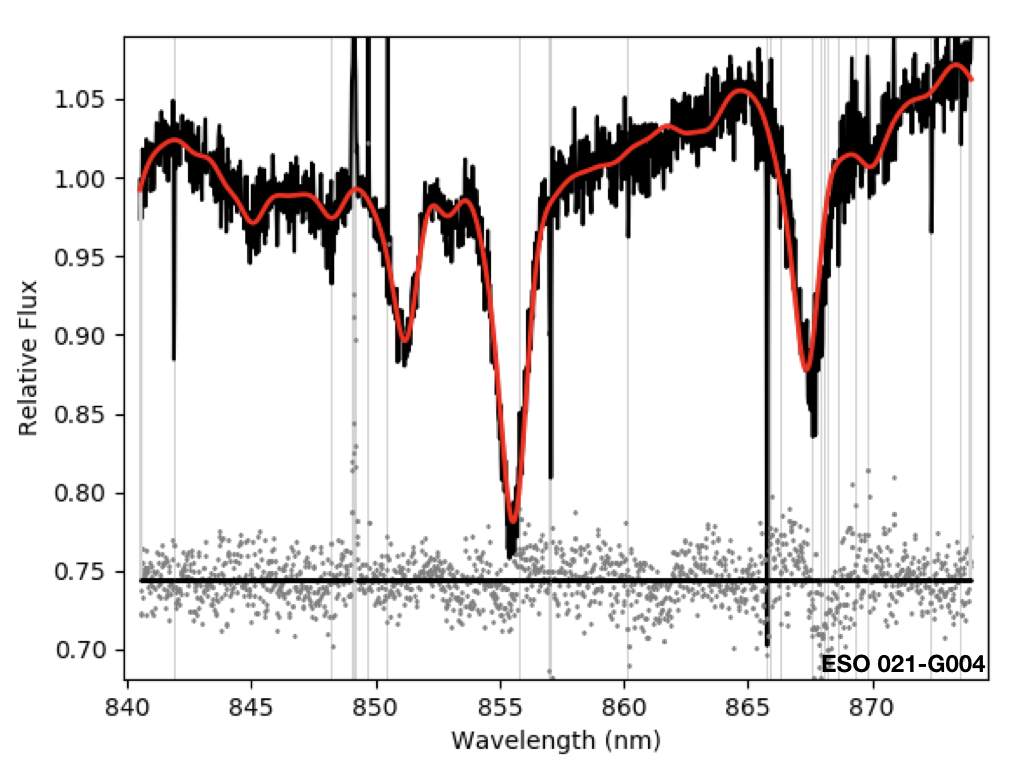}
\includegraphics[width=9.0cm]{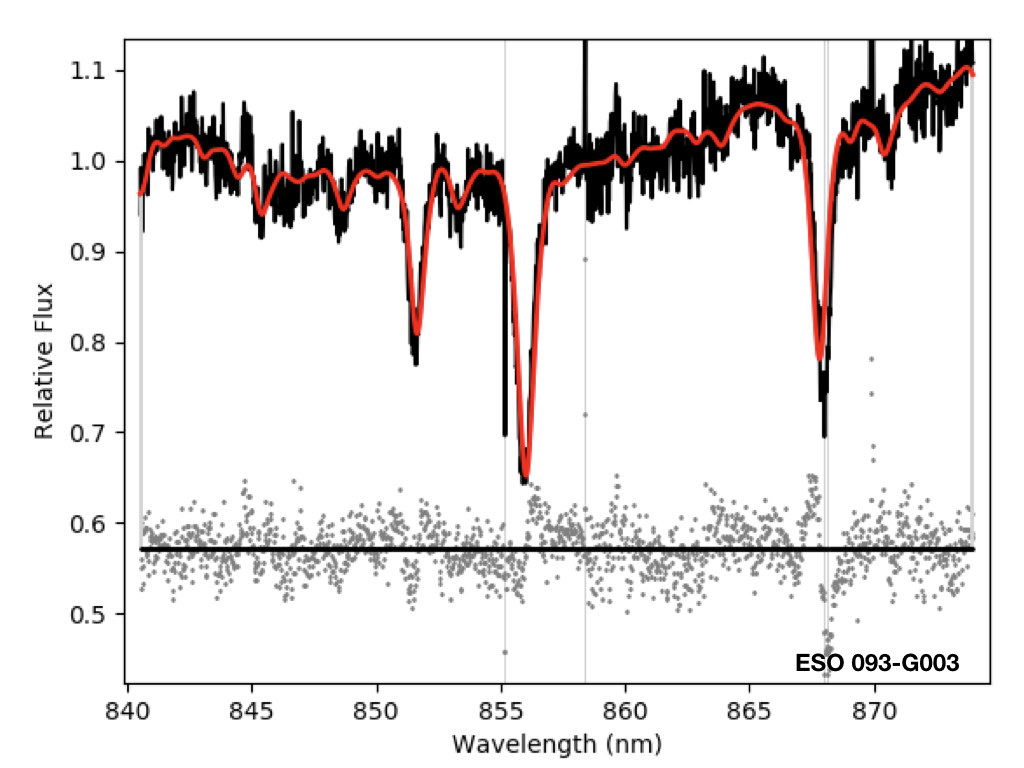}
\caption{Continued }
\label{appendixCAT5}%
\end{figure*}

\begin{figure*}
\centering
\includegraphics[width=9.0cm]{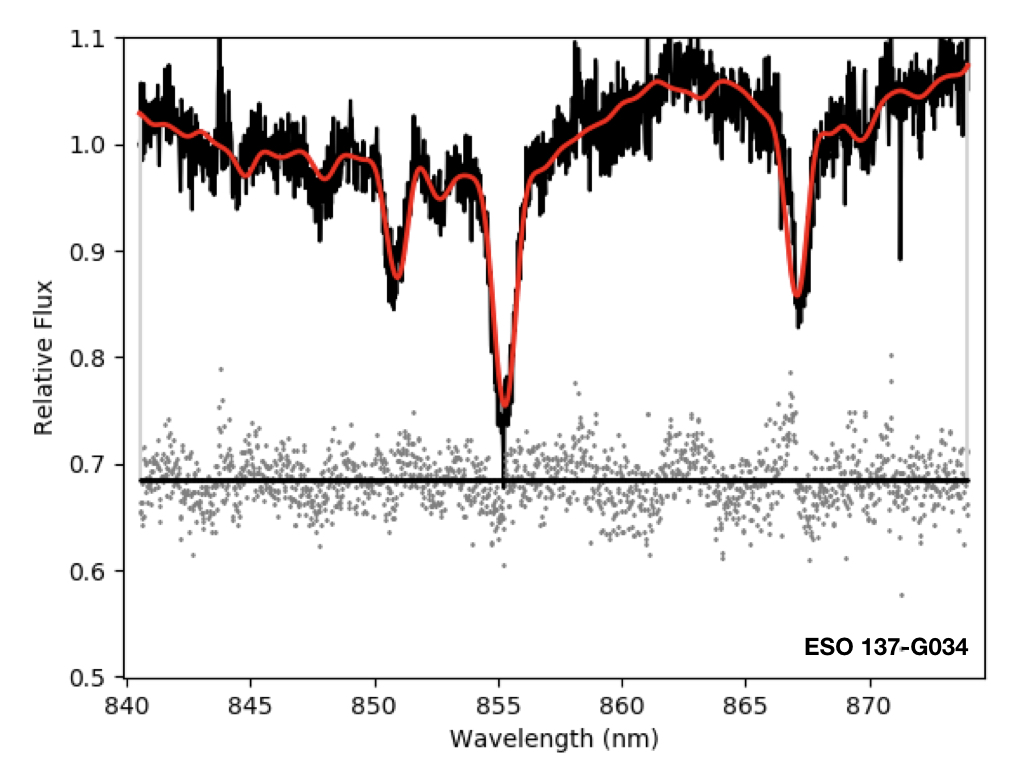}
\includegraphics[width=9.0cm]{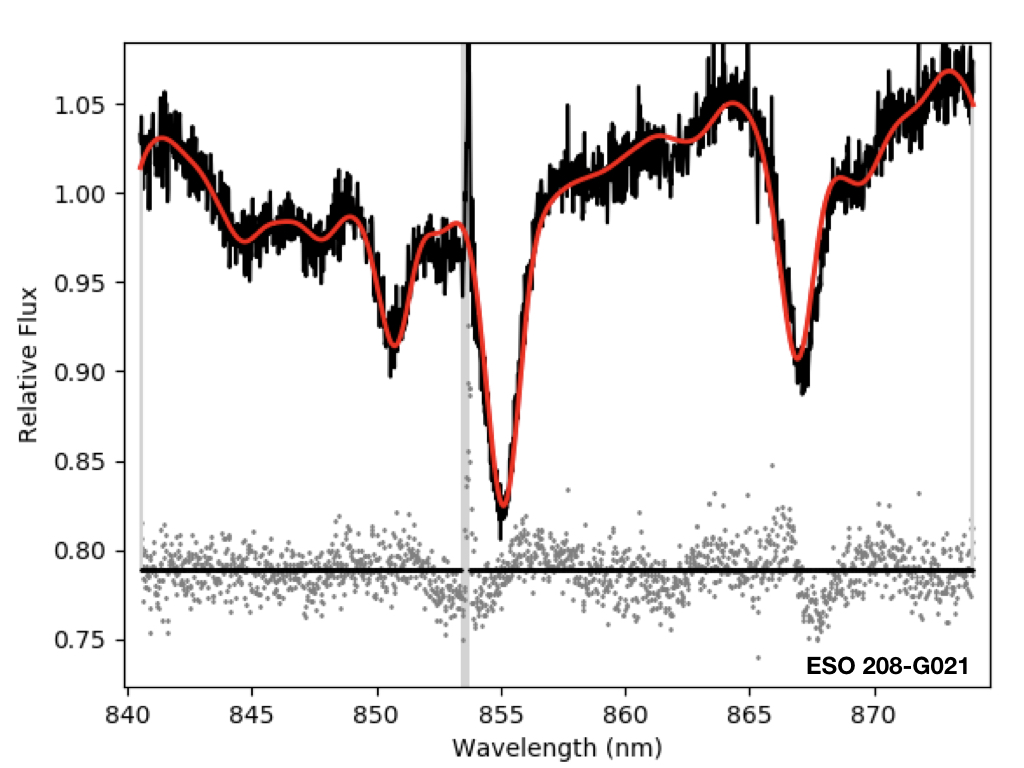}
\includegraphics[width=9.0cm]{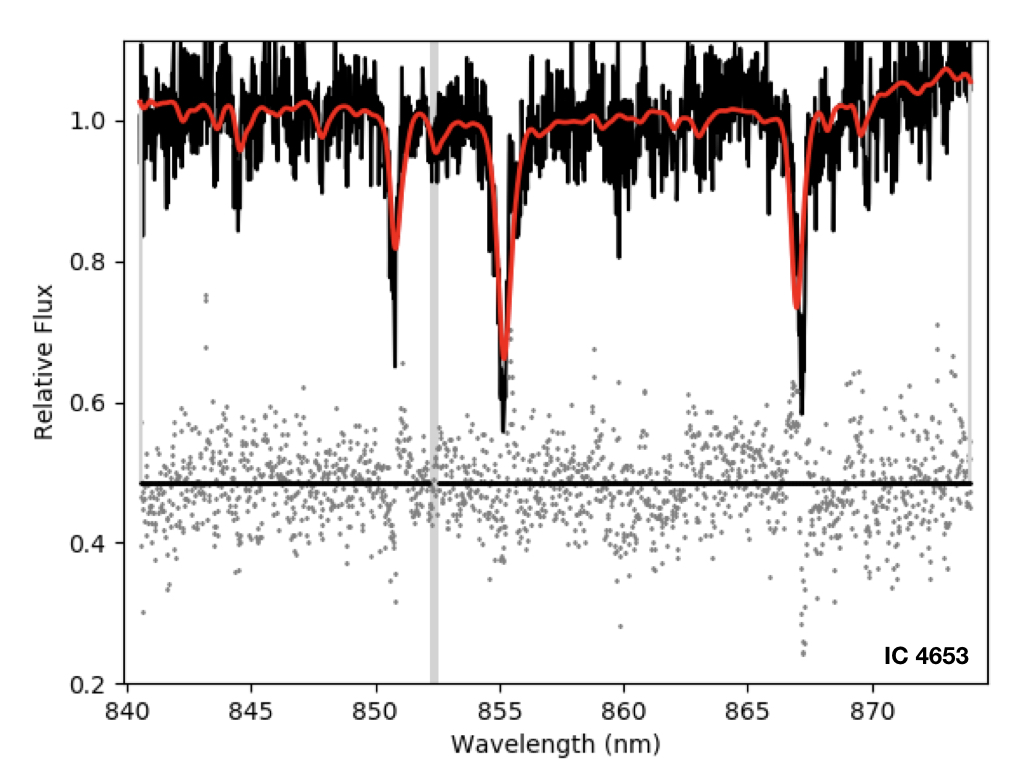}
\includegraphics[width=9.0cm]{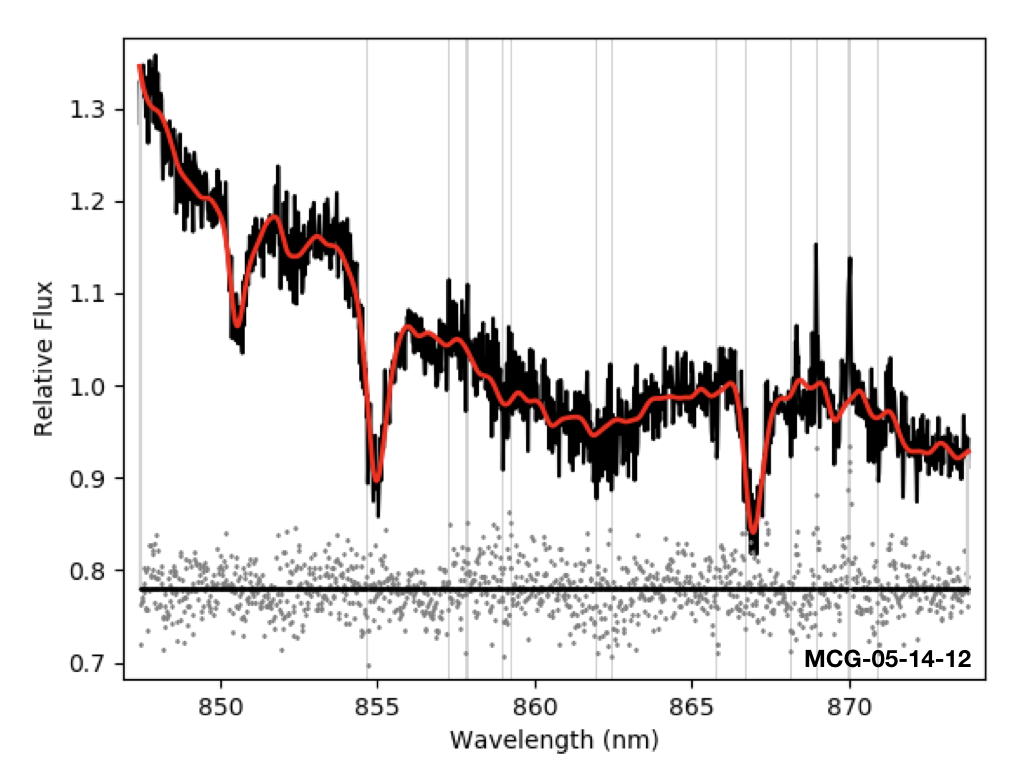}
\includegraphics[width=9.0cm]{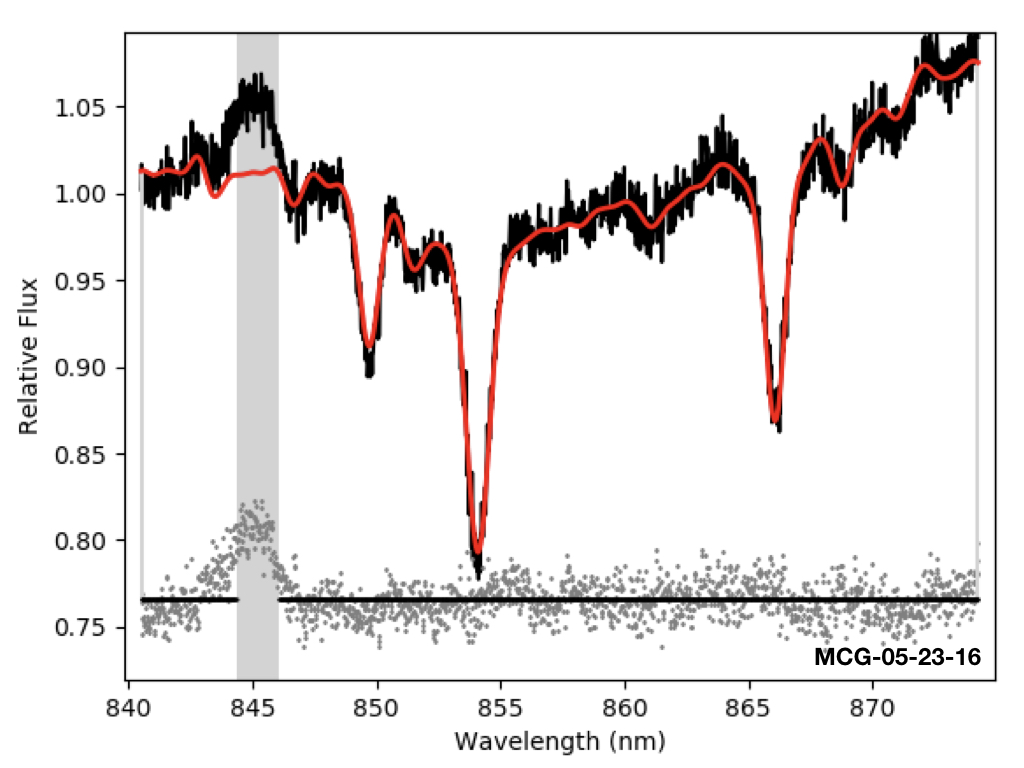}
\includegraphics[width=9.0cm]{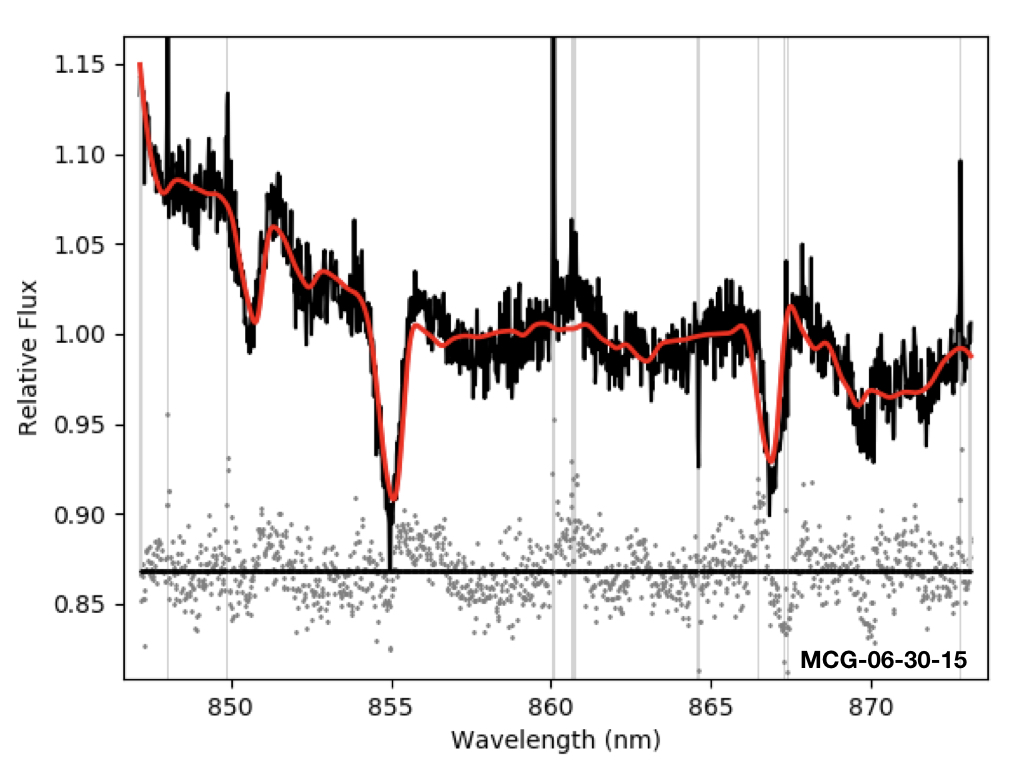}
\caption{Continued }
\label{appendixcat6}%
\end{figure*}

\begin{figure*}
\centering
\includegraphics[width=9.0cm]{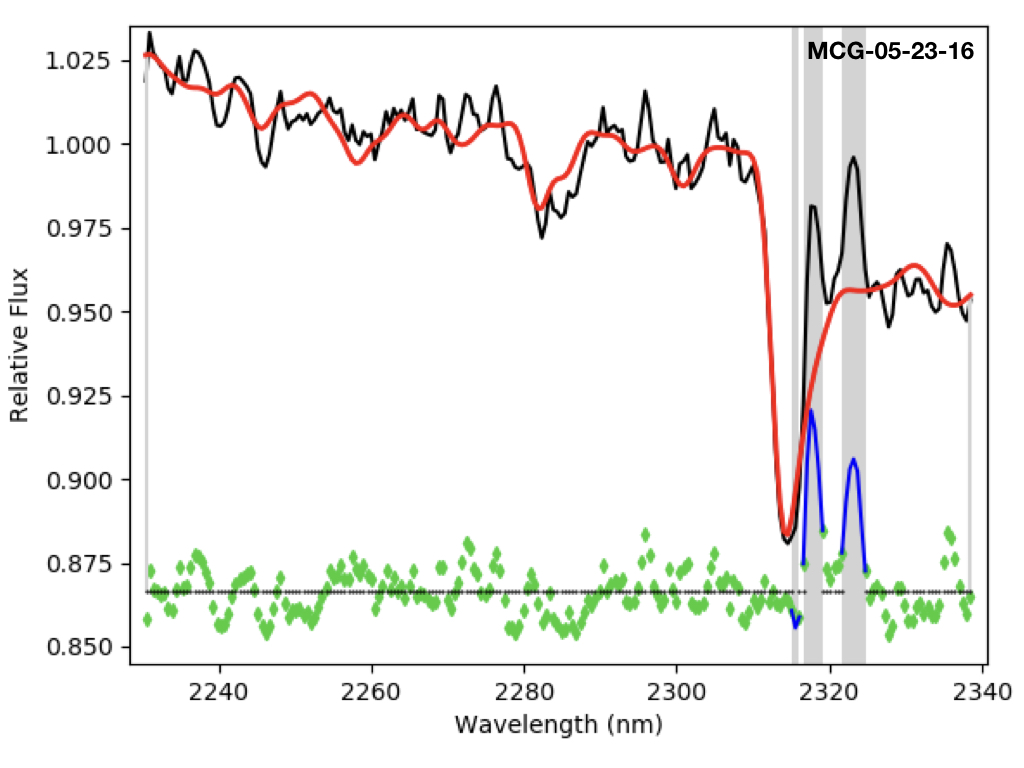}
\includegraphics[width=9.0cm]{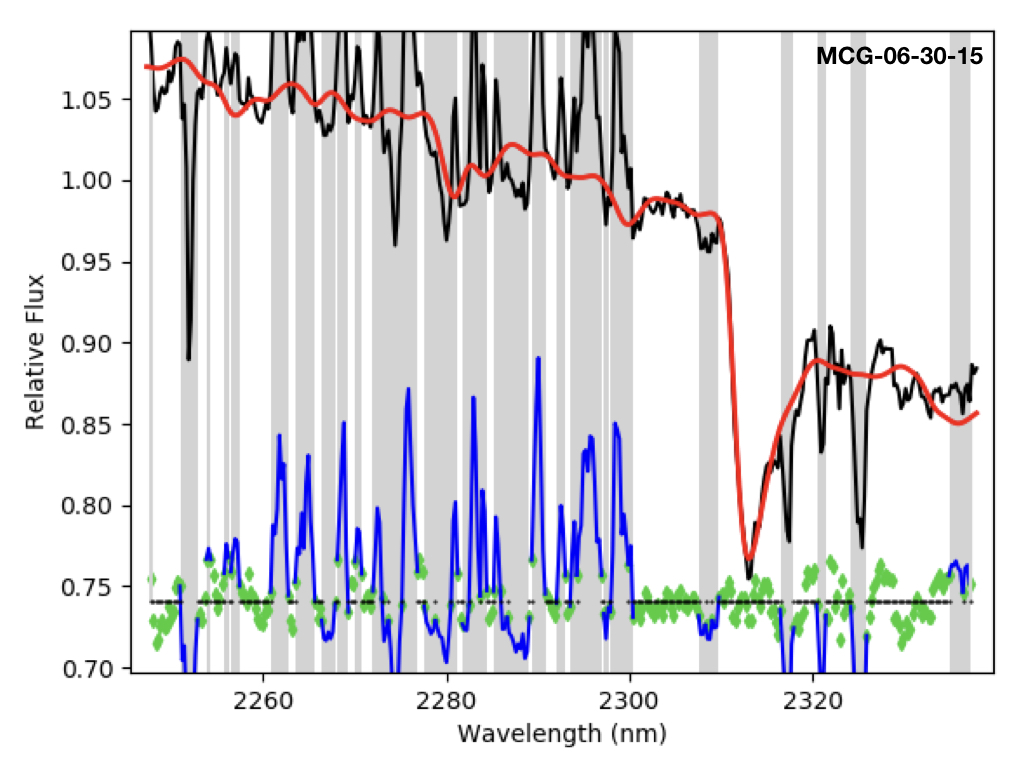}
\includegraphics[width=9.0cm]{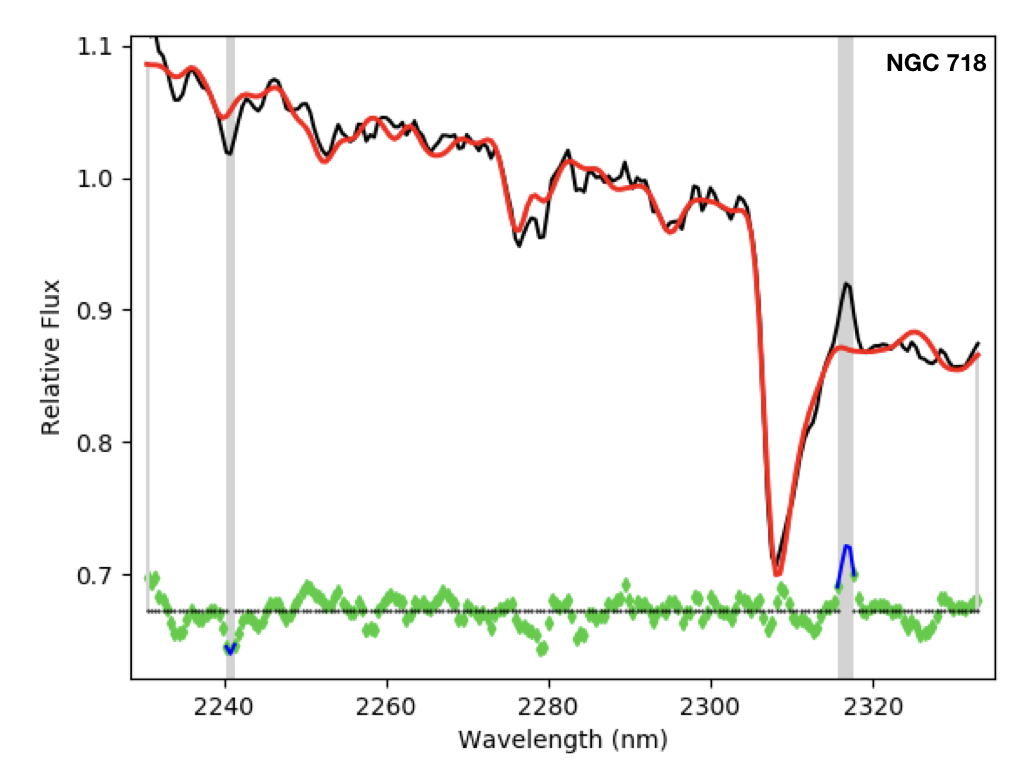}
\includegraphics[width=9.0cm]{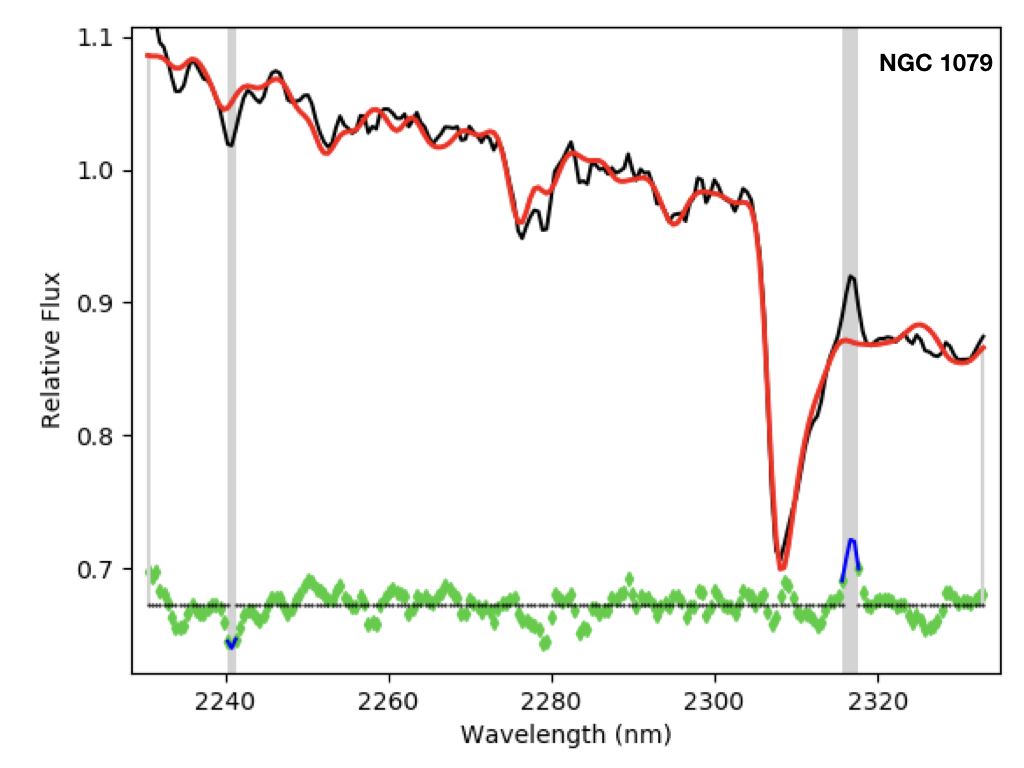}
\includegraphics[width=9.0cm]{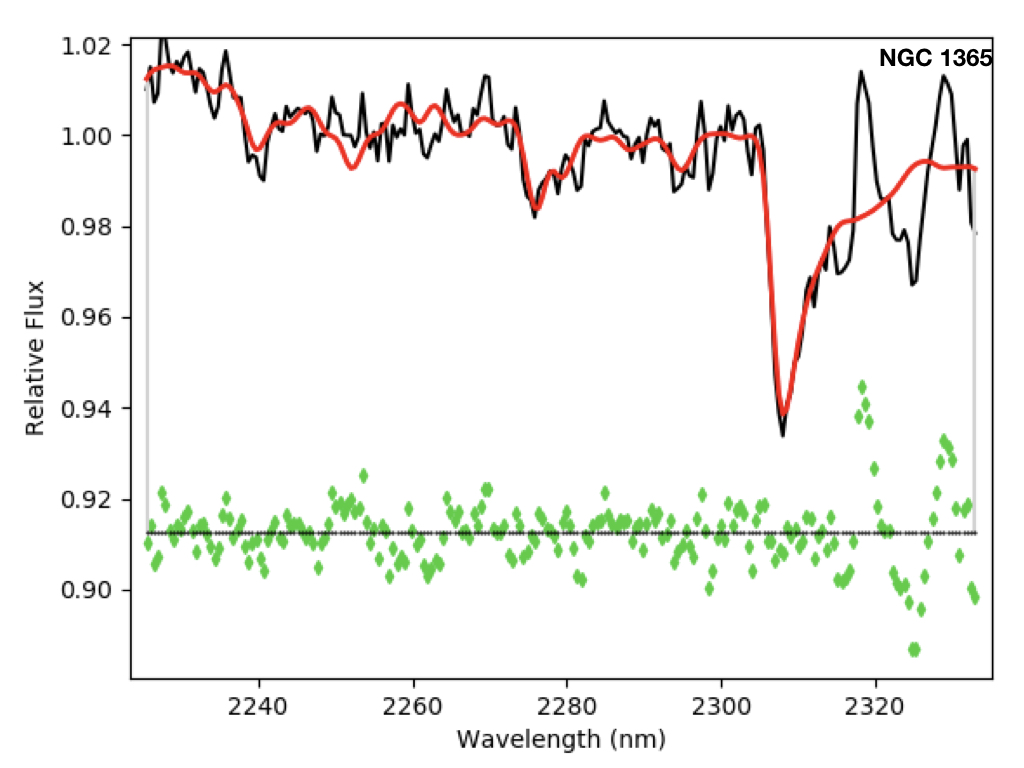}
\includegraphics[width=9.0cm]{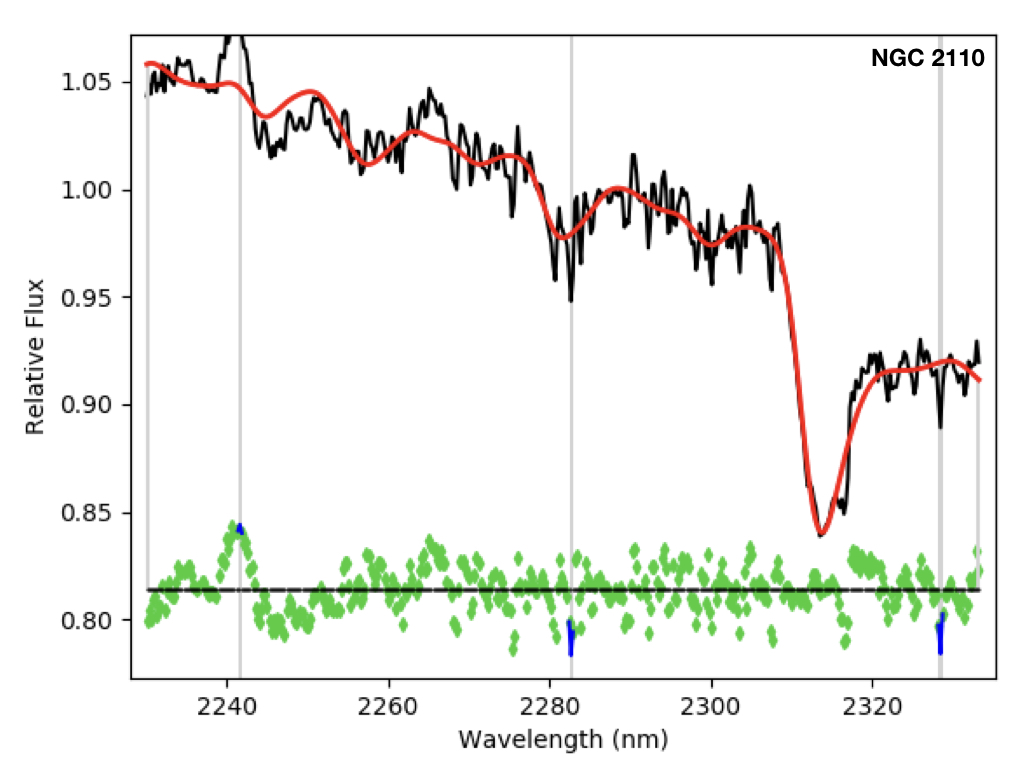}
\caption{The pPXF fitting plots for CO (2-0). The red solid line represents the best-fit, whereas the residuals are shown as green. The vertical gray lines represents masked features.}
\label{appendixco1}%
\end{figure*}

\begin{figure*}
\centering
\includegraphics[width=9.0cm]{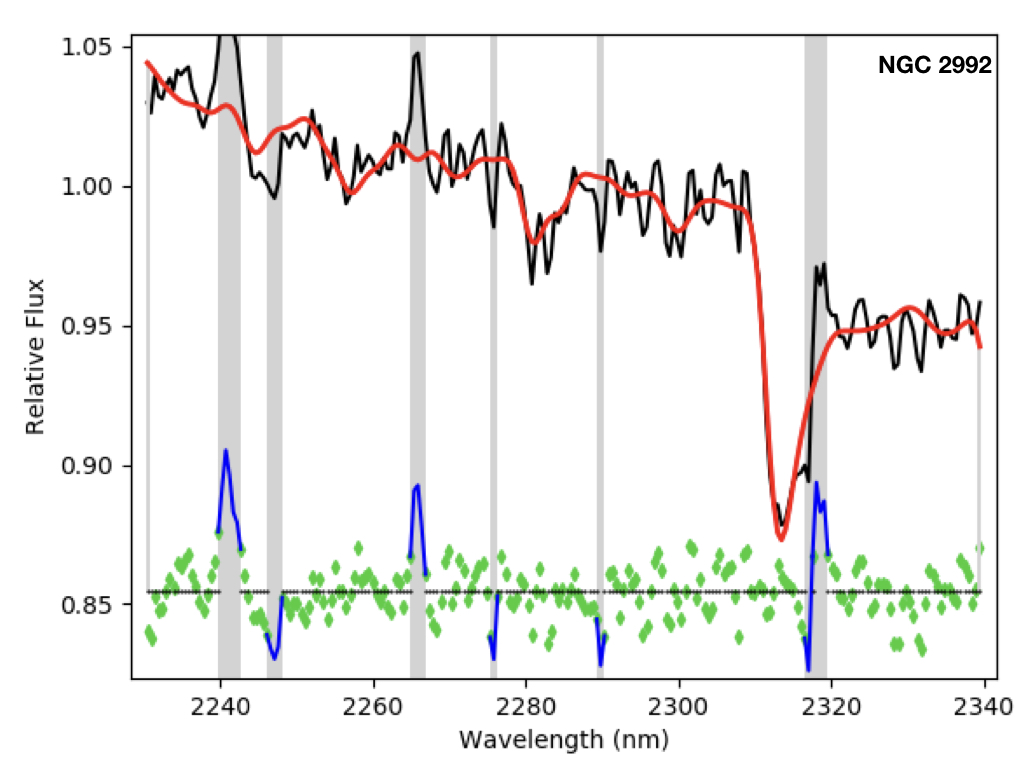}
\includegraphics[width=9.0cm]{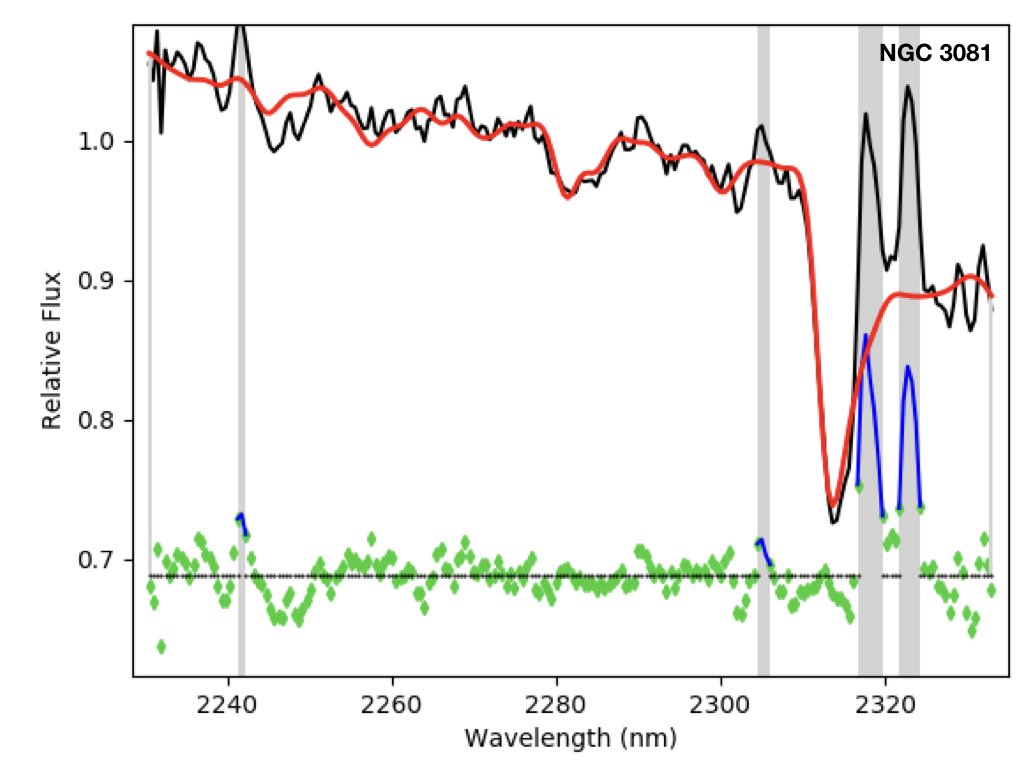}
\includegraphics[width=9.0cm]{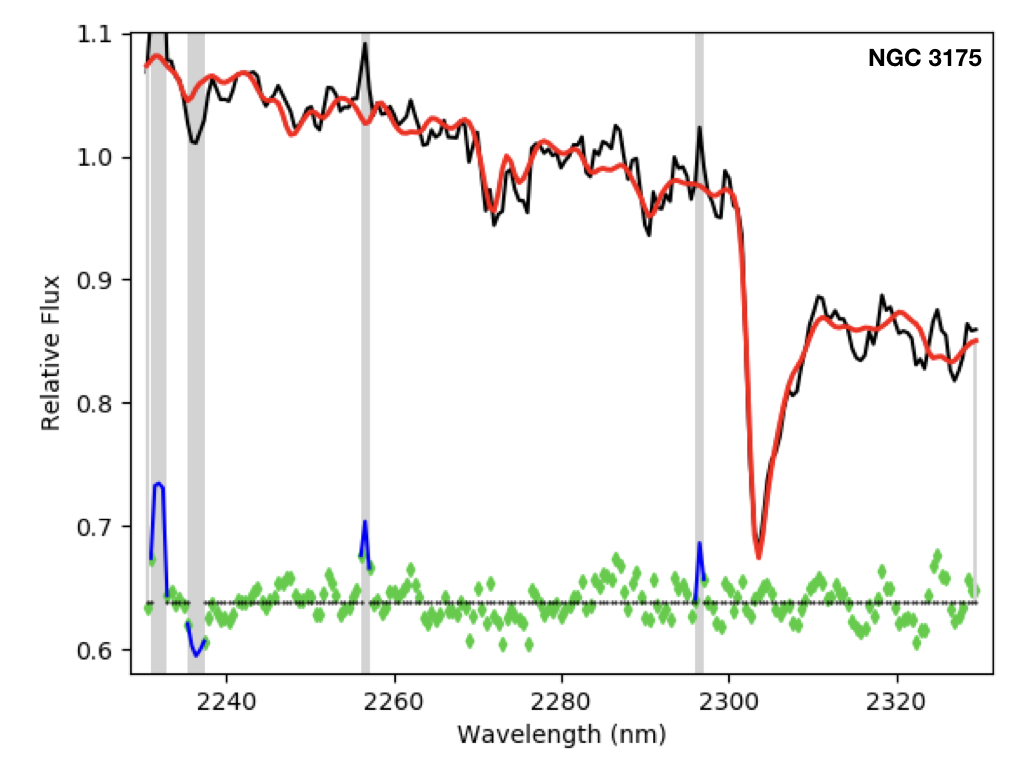}
\includegraphics[width=9.0cm]{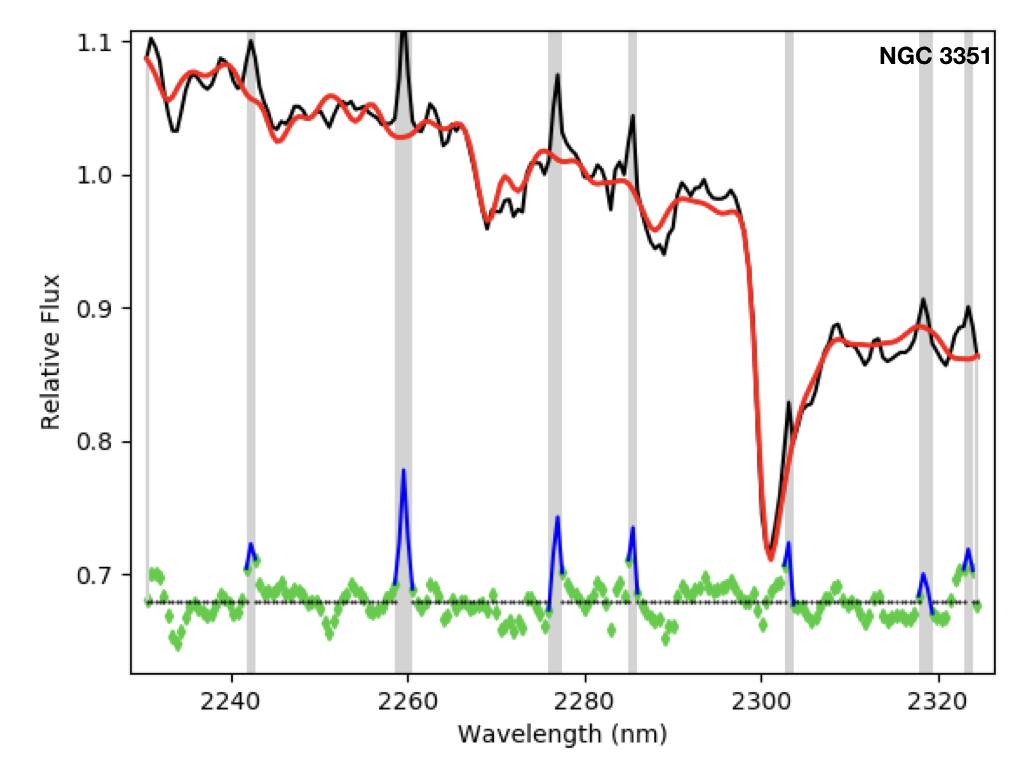}
\includegraphics[width=9.0cm]{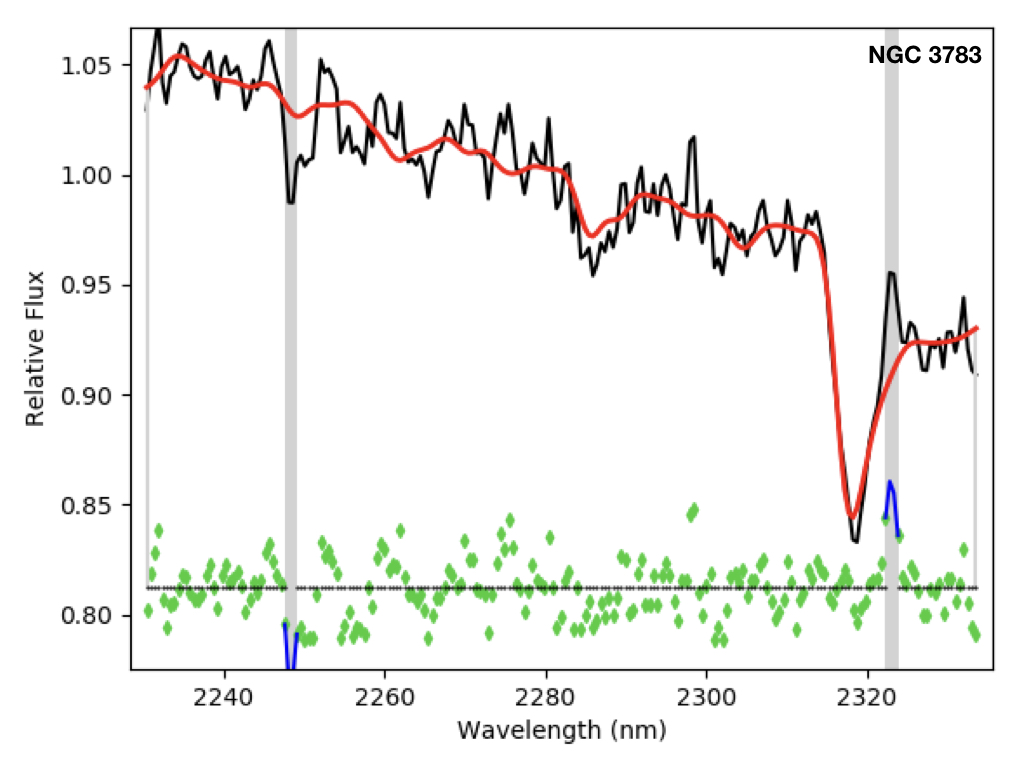}
\includegraphics[width=9.0cm]{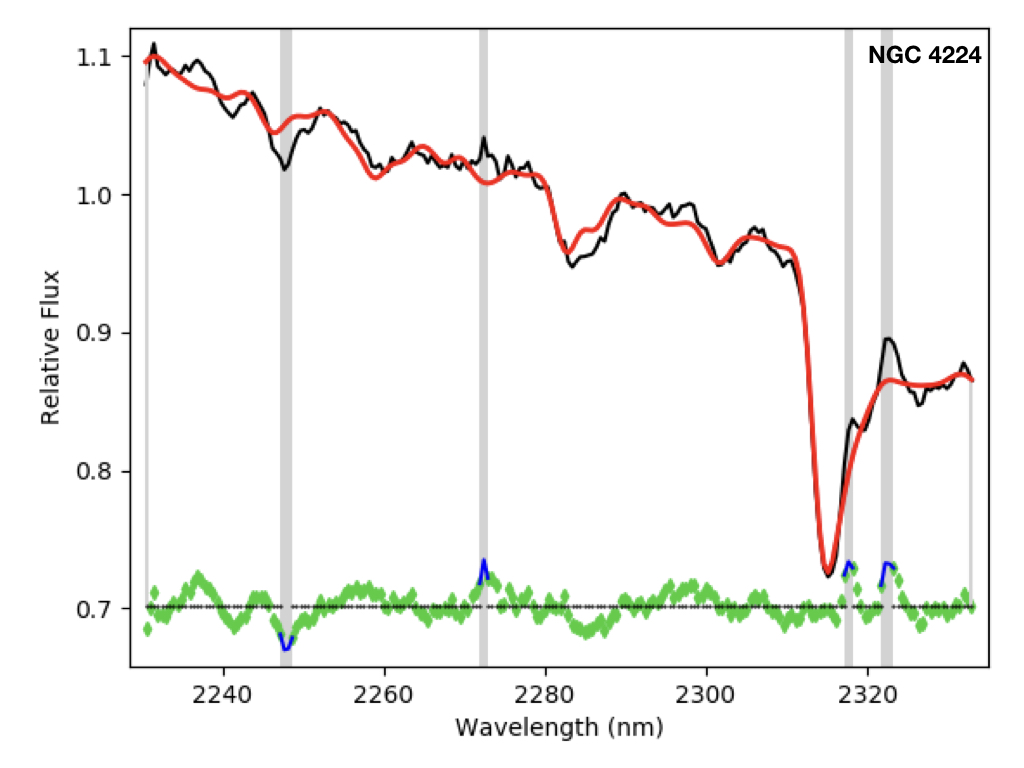}
\caption{Continued }
\label{appendixco2}%
\end{figure*}

\begin{figure*}
\centering
\includegraphics[width=9.0cm]{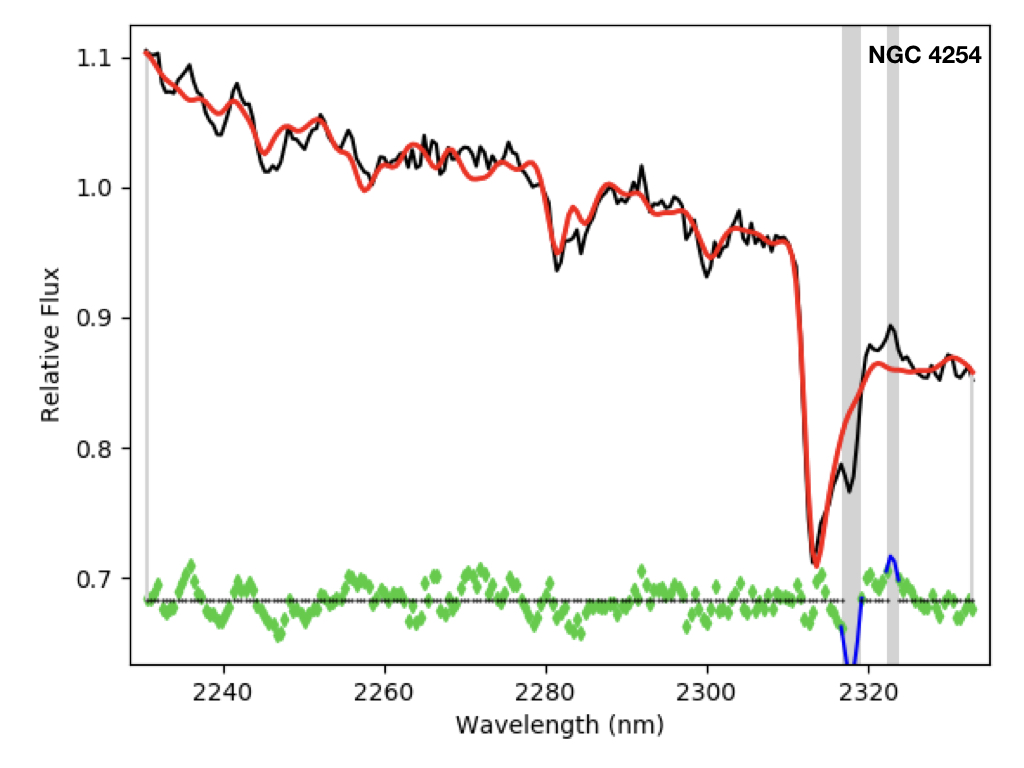}
\includegraphics[width=9.0cm]{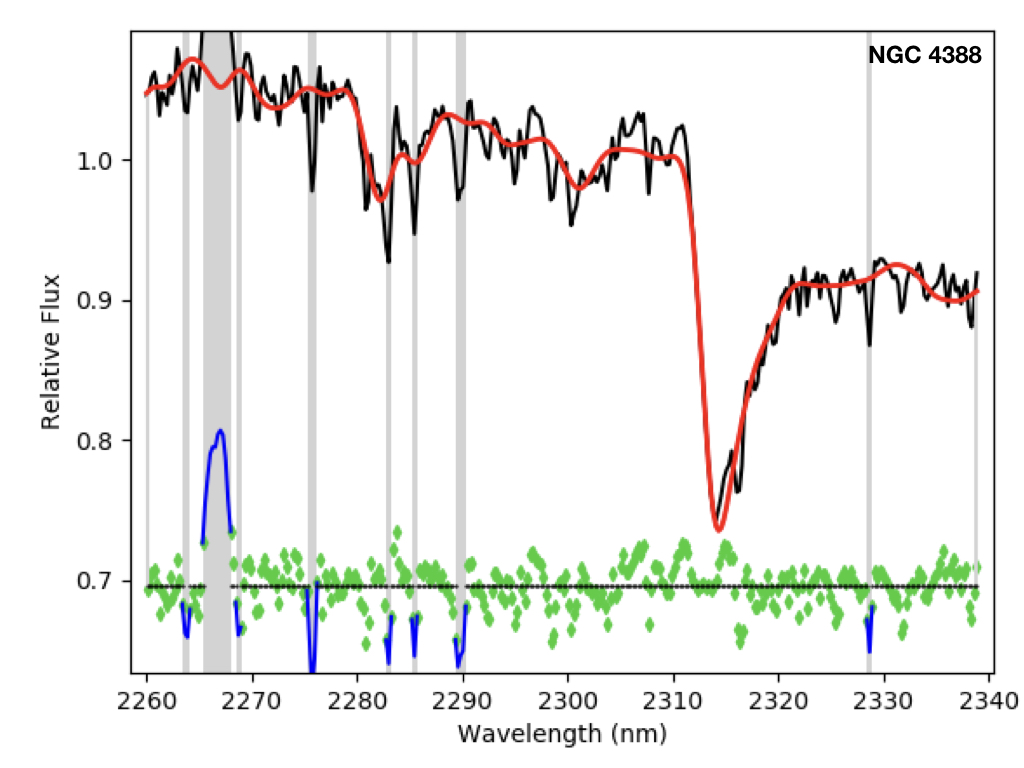}
\includegraphics[width=9.0cm]{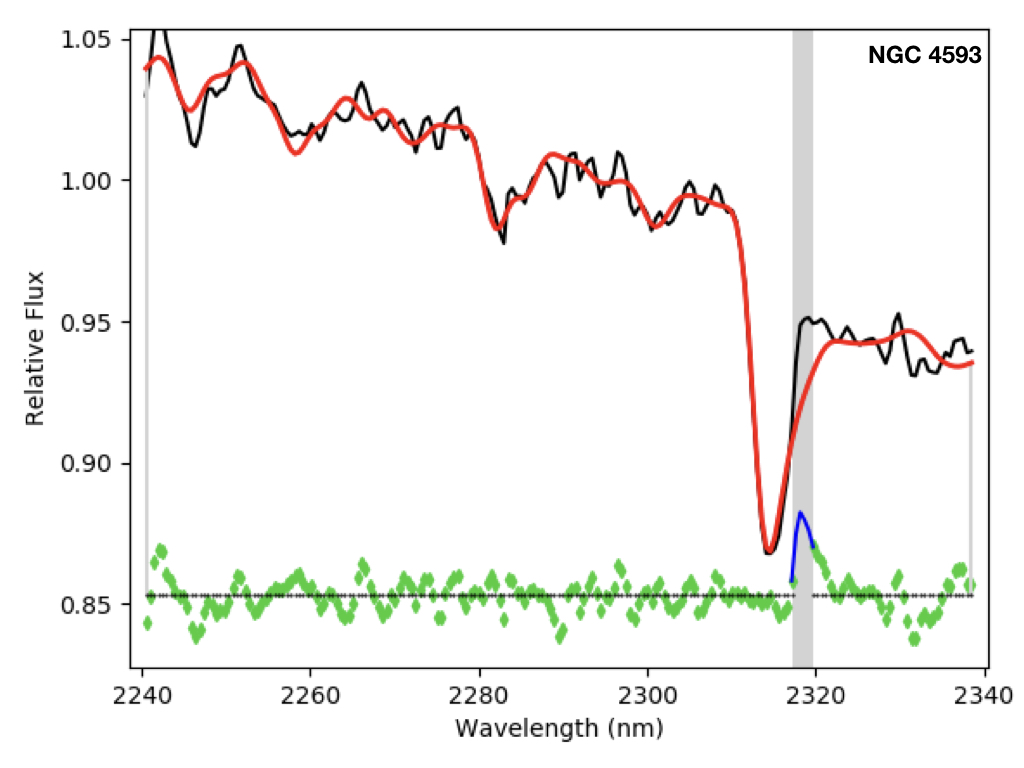}
\includegraphics[width=9.0cm]{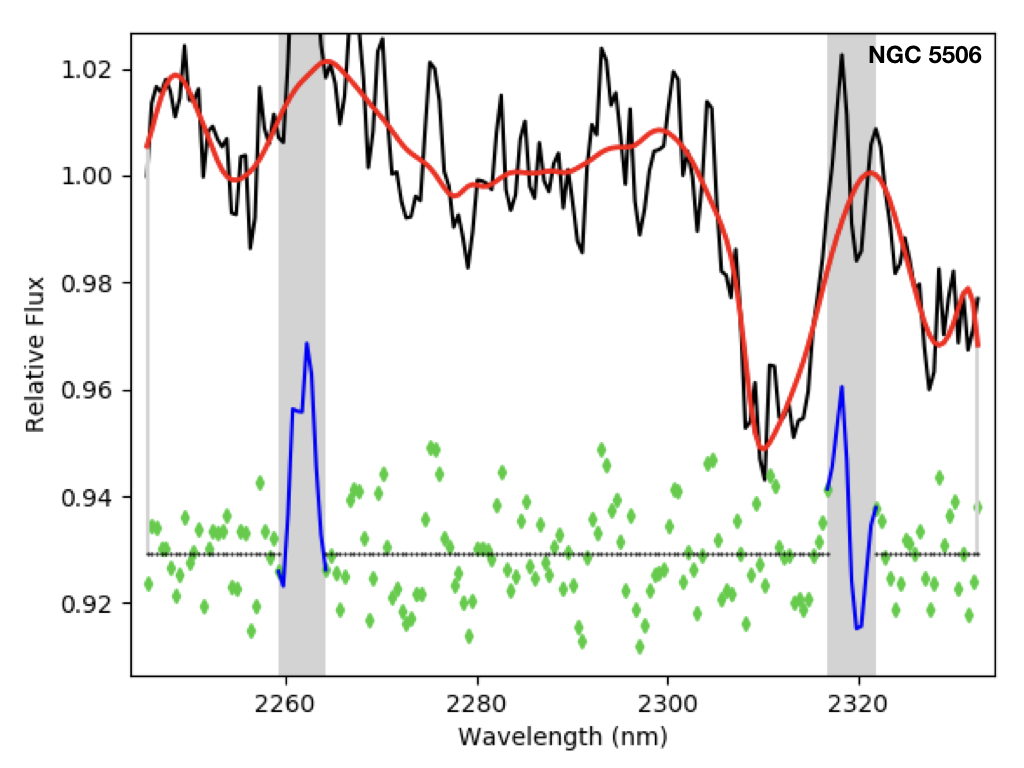}
\includegraphics[width=9.0cm]{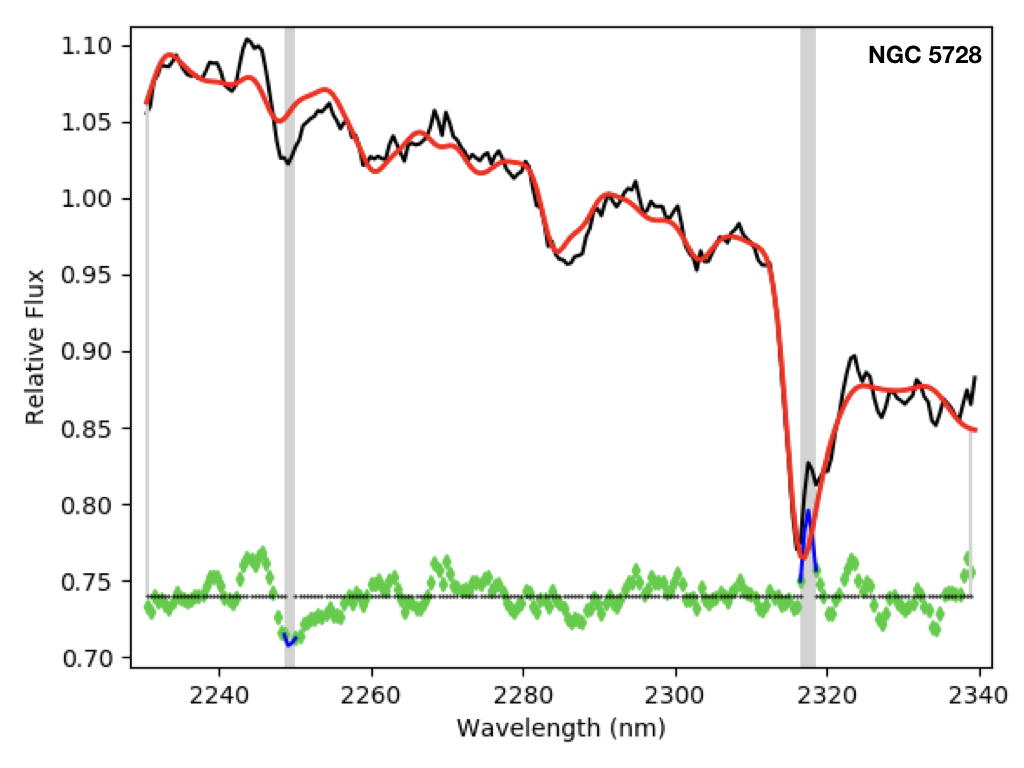}
\includegraphics[width=9.0cm]{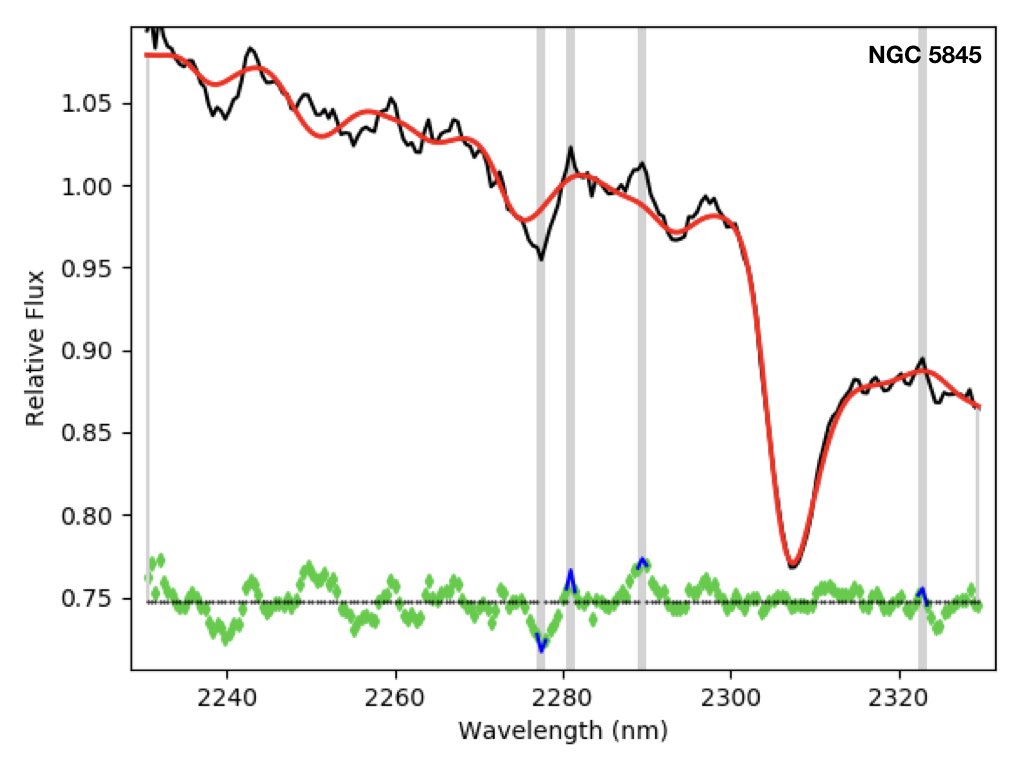}
\caption{Continued }
\label{appendixco3}%
\end{figure*}

\begin{figure*}
\centering
\includegraphics[width=9.0cm]{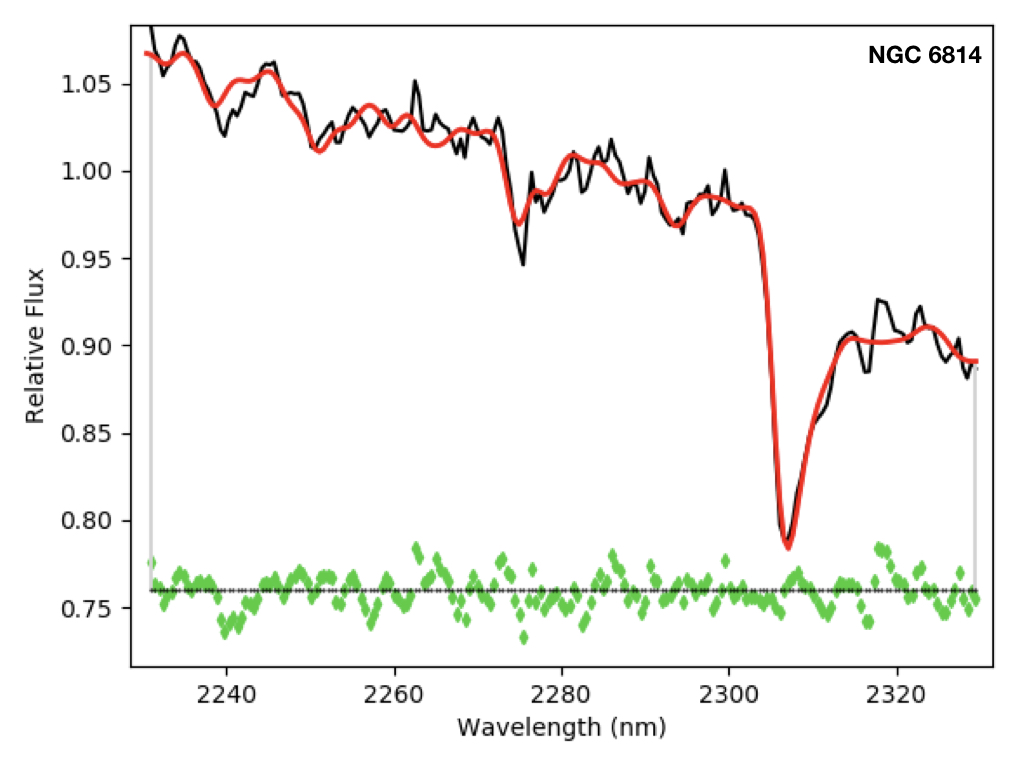}
\includegraphics[width=9.0cm]{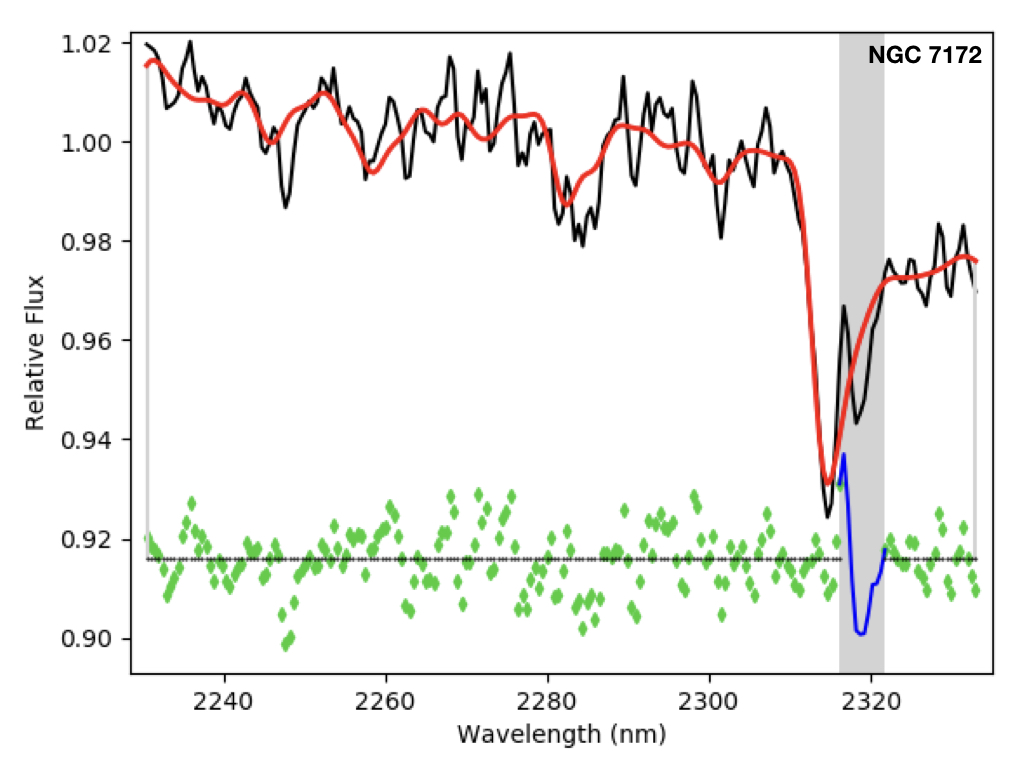}
\includegraphics[width=9.0cm]{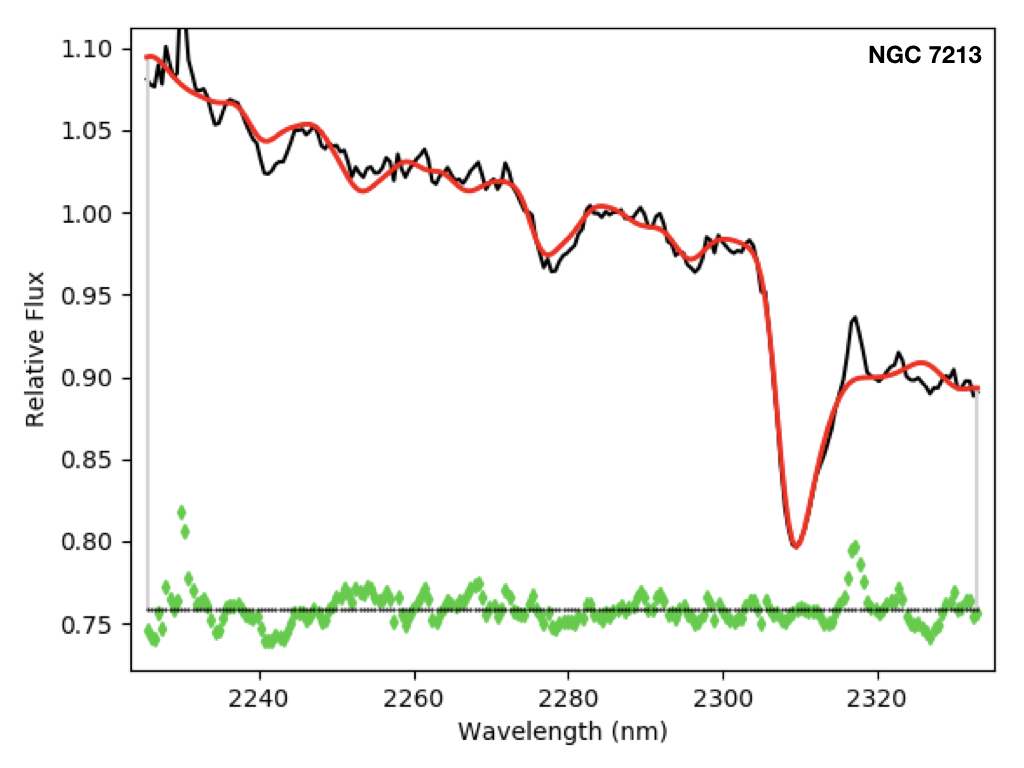}
\includegraphics[width=9.0cm]{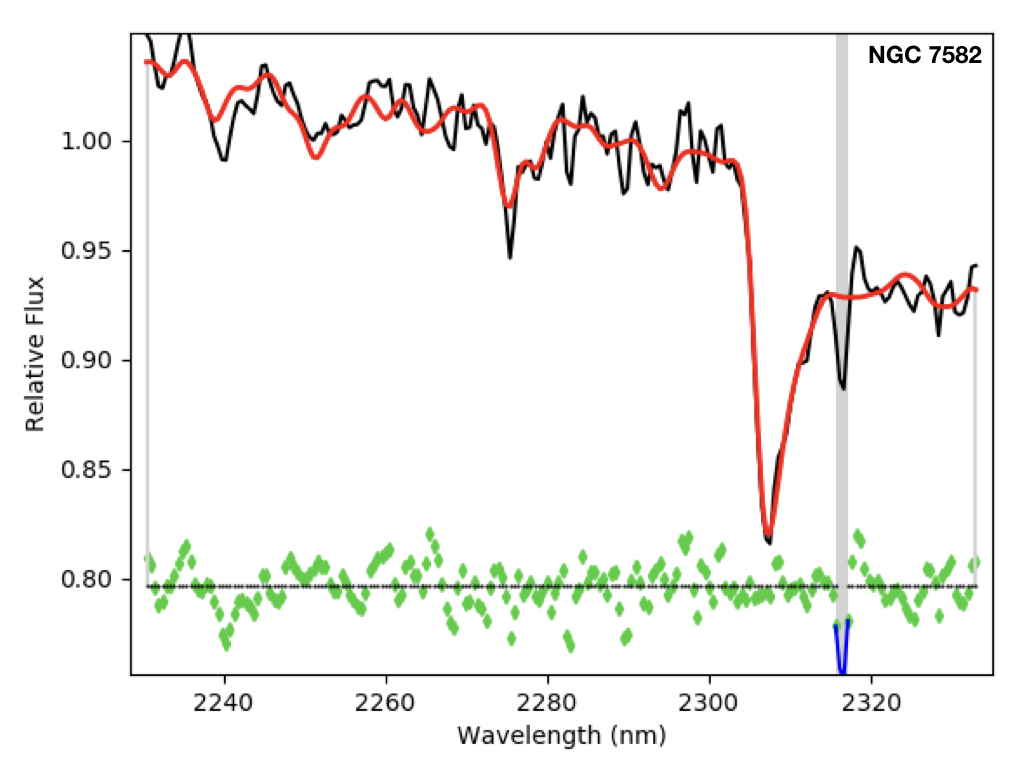}
\includegraphics[width=9.0cm]{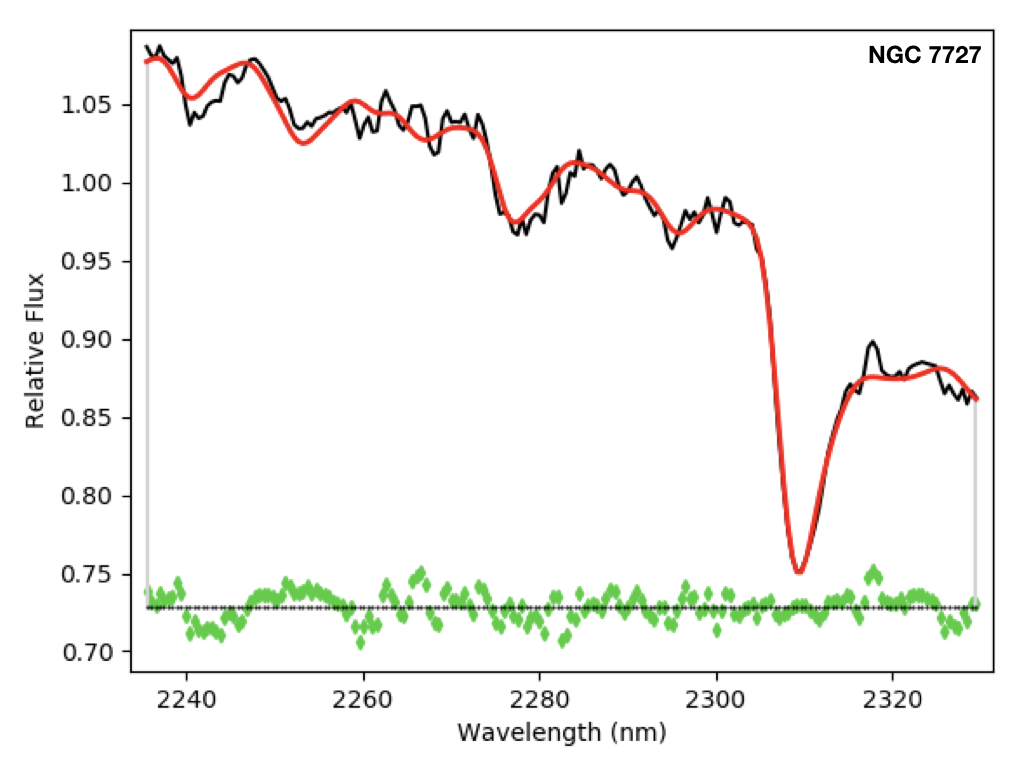}
\includegraphics[width=9.0cm]{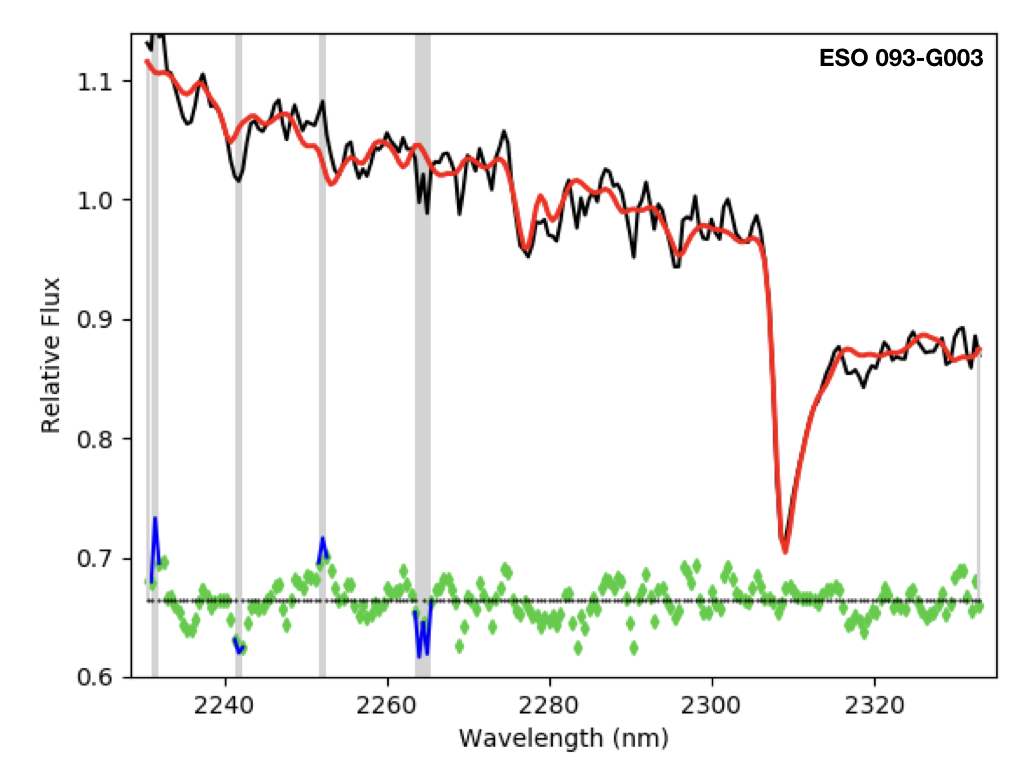}
\caption{Continued }
\label{appendixco4}%
\end{figure*}

\begin{figure*}
\centering
\includegraphics[width=9.0cm]{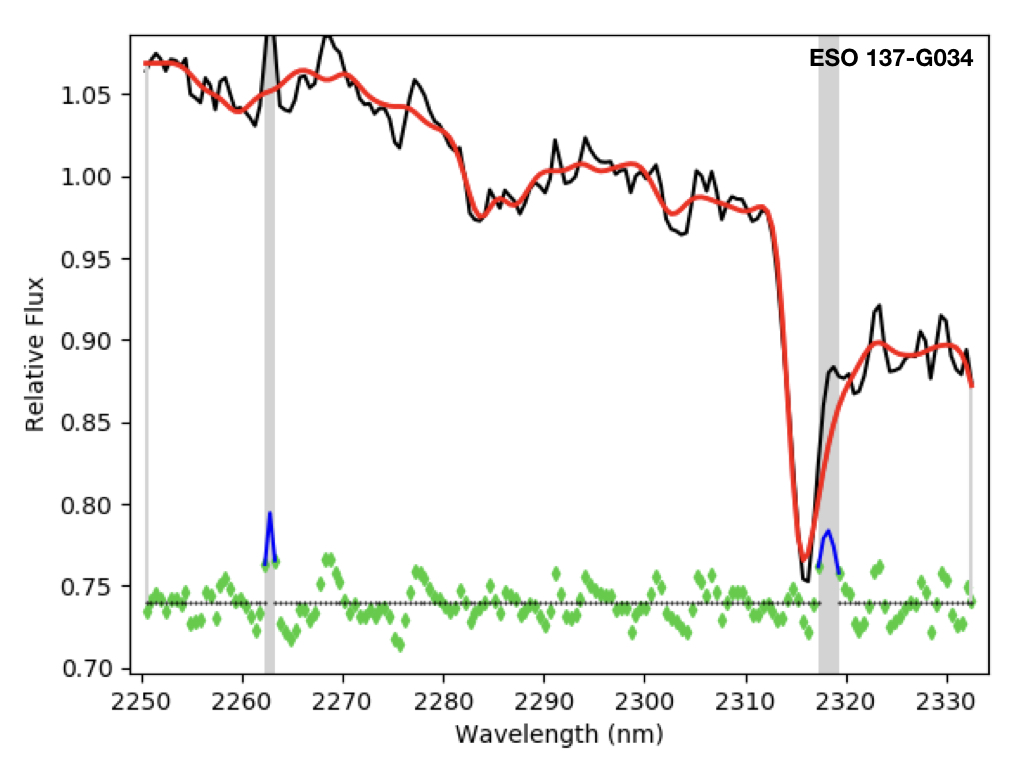}
\includegraphics[width=9.0cm]{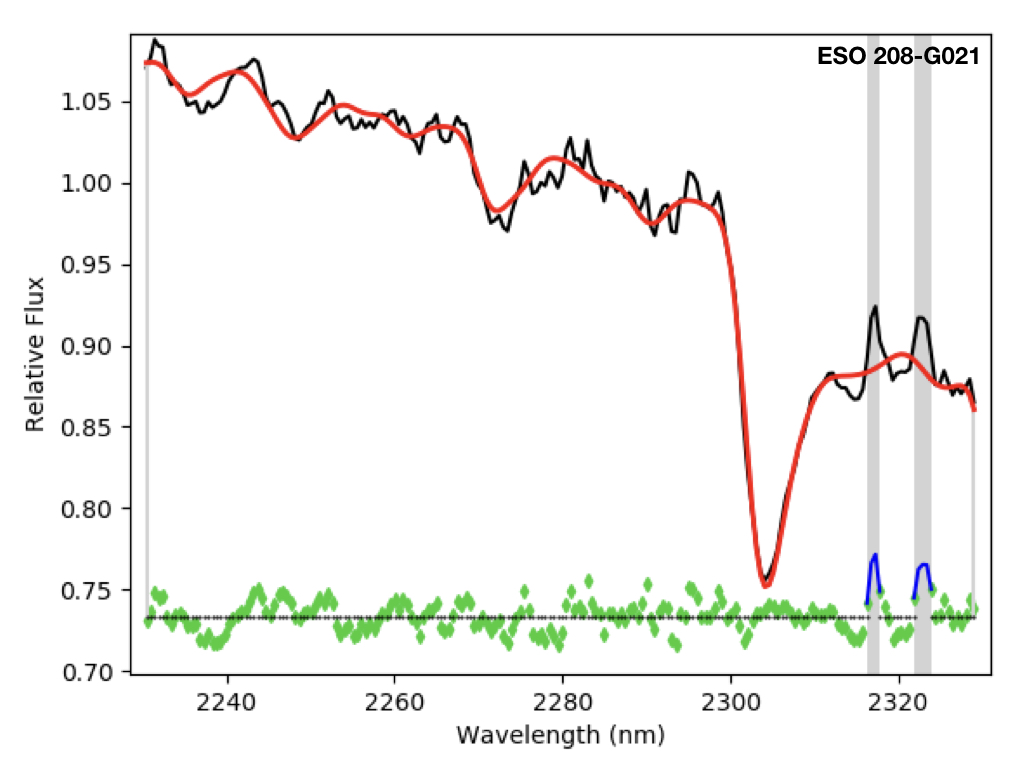}
\caption{Continued }
\label{appendixco5}%
\end{figure*}

\end{appendix}

\end{document}